\documentclass[preprint,12pt]{elsarticle}

\usepackage{amssymb}

\usepackage[cmex10]{amsmath}
\usepackage{multibib}

	\usepackage{mathtools}							
	\usepackage{amssymb,mathrsfs}
	\usepackage{pifont}
	\usepackage{algorithmic}						
	\usepackage{algorithm}							
	\usepackage{calc}										
	\usepackage{relsize}								
	\usepackage{framed} 
	\usepackage{nomencl}
	\usepackage[font={footnotesize}]{subcaption}
	\usepackage[font={footnotesize}]{caption}
	\usepackage[abbreviations = true]{siunitx}
	\DeclareSIUnit\year{yr}
	\DeclareSIUnit\dollar{\$}
	\DeclareSIUnit \VA { VA }
	\DeclareSIUnit \pu { p.u. }
	\DeclareSIUnit \MWh { MWh }
	\usepackage{longtable}							
	\usepackage{multirow}								
	\usepackage{array}									
	\usepackage{booktabs}
	\usepackage{tikz}
	\usepackage{mwe}
	\DeclareUnicodeCharacter{2212}{-}
	\usetikzlibrary{calc}

	\usepackage{pgfplots}
	\pgfplotsset{compat=1.16}
\usepackage{cite}
\setcitestyle{square}
\usepackage{stackengine}
\usepackage{enumitem}

\newcommand\addplotgraphicsnatural[2][]{%
    \begingroup
    \pgfqkeys{/pgfplots/plot graphics}{#1}%
    \setbox0=\hbox{\includegraphics{#2}}%
    \pgfmathsetmacro{\xfactor}{\wd0/(\pgfkeysvalueof{/pgfplots/plot graphics/xmax} - \pgfkeysvalueof{/pgfplots/plot graphics/xmin})}%
    \pgfmathsetmacro{\yfactor}{\ht0/(\pgfkeysvalueof{/pgfplots/plot graphics/ymax} - \pgfkeysvalueof{/pgfplots/plot graphics/ymin})}\yfactor%
    \pgfmathsetmacro{\xunit}{\xfactor<\yfactor ? 1 : \xfactor/\yfactor}
    \pgfmathsetmacro{\yunit}{\xfactor<\yfactor ? \yfactor/\xfactor : 1}
    \xdef\marshal{%
        \noexpand\pgfplotsset{unit vector ratio*={\xunit\space \yunit}}%
    }%
    \endgroup
    \marshal
    \addplot graphics[#1] {#2};
}   

	\usepackage{ifthen}
	\makenomenclature
	\renewcommand{\nomgroup}[1]{%
	\ifthenelse{\equal{#1}{A}}{\item[\textbf{\textit{Indices}}]}{%
		\ifthenelse{\equal{#1}{B}}{\item[\textbf{\textit{Sets}}]}{%
			\ifthenelse{\equal{#1}{C}}{\item[\textbf{\textit{Parameters}}]}{%
				\ifthenelse{\equal{#1}{D}}{\item[\textbf{\textit{Continuous Variables}}]}{%
					\ifthenelse{\equal{#1}{E}}{\item[\textbf{\textit{Binary Variables}}]} {} }}}}}

\journal{arXiv}

\begin{document}

\begin{frontmatter}


\title{A Machine-learning based Probabilistic Perspective on Dynamic Security Assessment}


\author{Jochen~L.~Cremer\corref{cor1}}
\ead{j.cremer@imperial.ac.uk}
\author{Goran~Strbac}
\ead{g.strbac@imperial.ac.uk}
\address{Imperial College London, South Kensington, London SW7 2BU, United Kingdom}

\cortext[cor1]{Corresponding author}

\begin{abstract}
Probabilistic security assessment and real-time dynamic security assessments (DSA) are promising to better handle the risks of system operations. The current methodologies of security assessments may require many time-domain simulations for some stability phenomena that are unpractical in real-time. Supervised machine learning is promising to predict DSA as their predictions are immediately available. Classifiers are offline trained on operating conditions and then used in real-time to identify operating conditions that are insecure. However, the predictions of classifiers can be sometimes wrong and hazardous if an alarm is missed for instance. 

A probabilistic output of the classifier is explored in more detail and proposed for probabilistic security assessment. An ensemble classifier is trained and calibrated offline by using Platt’ scaling to provide accurate probability estimates of the output. Imbalances in the training database and a cost-skewness addressing strategy are proposed for considering that missed alarms are significantly worse than false alarms. Subsequently, risk-minimised predictions can be made in real-time operation by applying cost-sensitive learning. Through case studies on a real data-set of the French transmission grid and on the IEEE 6 bus system using static security metrics, it is showcased how the proposed approach reduces inaccurate predictions and risks. The sensitivity on the likelihood of contingency is studied as well as on expected outage costs. Finally, the scalability to several contingencies and operating conditions are showcased.
\end{abstract}





\end{frontmatter}


\printnomenclature 
\nomenclature[A]{{$c$}}{{contingency}}
\nomenclature[A]{{$i$}}{{pre-fault operating condition}}
\nomenclature[A]{{$l$}}{{weak classifier, for instance a DT}}
\nomenclature[A]{{$s$}}{{scenario of a pre-fault operating condition exposed to a contingency}}

\nomenclature[B]{{$\Omega^C$}}{{set of contingencies}}

\nomenclature[B]{{$\Omega^E$}}{{set of weak classifiers in an ensemble of classifiers}}

\nomenclature[B]{{$\Omega^K$}}{{set of operating conditions for calibration}}

\nomenclature[B]{{$\Omega^S$}}{{set of scenarios}}
\nomenclature[B]{{$\Omega^{S,H}$}}{{set of high risk scenarios}}
\nomenclature[B]{{$\Omega^{S,L}$}}{{set of low risk scenarios}}
\nomenclature[B]{{$\Omega^S_n$}}{{n'th subset of the operating conditions in the calibration set $\Omega^K$}}

\nomenclature[B]{{$\Omega^T$}}{{set of operating conditions in the training database}}
\nomenclature[B]{{$\Omega^{T,1}_c$}}{{operating conditions in the training database that are secure for contingency $c$}}
\nomenclature[B]{{$\Omega^{T,0}_c$}}{{operating conditions in the training database that are insecure for contingency $c$}}

\nomenclature[B]{{$\Omega^P$}}{{set of possible future operating conditions}}

\nomenclature[C]{{a}}{{fitting parameter in sigmoid function}}

\nomenclature[C]{{b}}{{fitting parameter in sigmoid function}}

\nomenclature[C]{{B}}{{Brier score}}

\nomenclature[C]{{$C_c^{F0}$}}{{Expected outage costs for missing an alarm of contingency $c$}}
\nomenclature[C]{{$C_c^{F1}$}}{{Expected costs of preventive and corrective control for a false alarm of contingency $c$}}
\nomenclature[C]{{$CVM$}}{{ratio of scenarios assessed by conventional security assessment}}
\nomenclature[C]{{$\mathcal{C}_c$}}{{Cost ratio of missed and false alarm of contingency $c$}}

\nomenclature[C]{{$h_l(x_i)$}}{{hypothesis (prediction) of the weak classifier $l$, $0$ if the hypothesis for condition $i$ is insecure, and $1$ is hypothesis is secure}}
\nomenclature[C]{{$H^E(x_i)$}}{{prediction of the ensemble based on the feature vector $x_i$ for operating condition $i$, $1$ if prediction is secure, $1$ is insecure}}

\nomenclature[C]{{$M$}}{{maximal number of weak learners in an ensemble}}
\nomenclature[C]{{N}}{{user-specified number of sets for computing Brier score}}
\nomenclature[C]{{$N^1_c$}}{{number of missed alarms }}
\nomenclature[C]{{$N^0_c$}}{{number of false alarms }}

\nomenclature[C]{{$\hat{p}^1(x_i)$}}{{probability estimate for the secure class of the operating condition $i$}}
\nomenclature[C]{{$\hat{p}^0(x_i)$}}{{probability estimate for the insecure class of the operating condition $i$}}
\nomenclature[C]{{$p_c^C$}}{{likelihood of contingency $c$}}
\nomenclature[C]{{$p^I_i$}}{{likelihood of operating condition $i$}}
\nomenclature[C]{{$p^S_s$}}{{likelihood of scenario $s$}}

\nomenclature[C]{{$\pi^0_c$}}{{Prior of the insecure class for contingency $c$ in the training database}}
\nomenclature[C]{{$\pi^1_c$}}{{Prior of the secure class for contingency $c$ in the training database}}

\nomenclature[C]{{$R^1_c(x_i)$}}{{Risk of predicting an operating condition $i$ as secure}}
\nomenclature[C]{{$R^0_c(x_i)$}}{{Risk of predicting an operating condition $i$ as insecure}}
\nomenclature[C]{{$RISK^{ML^*}$}}{{risk of using machine learning for the assessment of the subset of low-risk scenarios $\Omega^{S,L}$}}
\nomenclature[C]{{$RISK^{ML}$}}{{risk of using machine learning for the assessment of all scenarios $\Omega^S$}}
\nomenclature[C]{{$RISK^{SA^*}$}}{{risk of using conventional security assessment for the subset of high-risk scenarios $\Omega^{S,H}$}}
\nomenclature[C]{{$RISK^{SA}$}}{{risk of using conventional security assessment for all scenarios $\Omega^{S}$}}
\nomenclature[C]{{$\mathcal{R}_s$}}{{risk relying on a machine learning prediction for scenario $s$}}

\nomenclature[C]{{$s^1(x_i)$}}{{weighted and normalized score for the secure class of the operating condition $i$}}
\nomenclature[C]{{$s^0(x_i)$}}{{weighted and normalized score for the insecure class of the operating condition $i$}}
\nomenclature[C]{{$S$}}{{computational capacity as the number of security assessments that the operator can run close to real-time operation}}
\nomenclature[C]{{$S_{i,c}$ }}{{severity of the system response when the pre-fault operating condition $i$ is subjected to contingency $c$}}

\nomenclature[C]{{$w_l$}}{{weight of hypothesis of weak learner $l$}}

\nomenclature[C]{{$x_i$}}{{Extracted/selected feature vector describing the pre-fault operating condition $i$}}

\nomenclature[C]{{$y_{i,c}$}}{{security label of operating condition $i$ subjected to contingency $c$, $1$ if secure, $0$ if insecure}}

\nomenclature[C]{{$z_c$}}{{decision threshold of the classifier for contingency $c$}}

\nomenclature[C]{{$Z^*_c$}}{{estimated residual risk of relying on the machine learning model of contingency $c$ for a testing set }}

\section{Introduction}

The operation of the power system undergoes significant changes. In the past, the operations of the power system involved to predict the demand pattern and accordingly controlling/planning conventional power supplies such that demand and supply matches (simplified description). The demand was straightforward to predict, and the operator knew with little uncertainty the next day's operations. In the future, to meet targets of reducing carbon emissions, the share of renewable energy on the overall energy supply will increase and renewables are neither controllable nor accurately predictable. Instead of controlling the supply of energy, in the future, the demand will be made flexible and controllable. Therefore, a key change for future power systems operations is that the generation and the demand are significantly more uncertain. This change underpins several challenges for the operations and a new set of operating tools is needed to address these challenges \citep{Pan12}. This increased uncertainty is specifically challenging for the management of the reliability of the power system.

\subsection{Reliability management}

\begin{figure}
\centering
\includegraphics[width=1\linewidth]{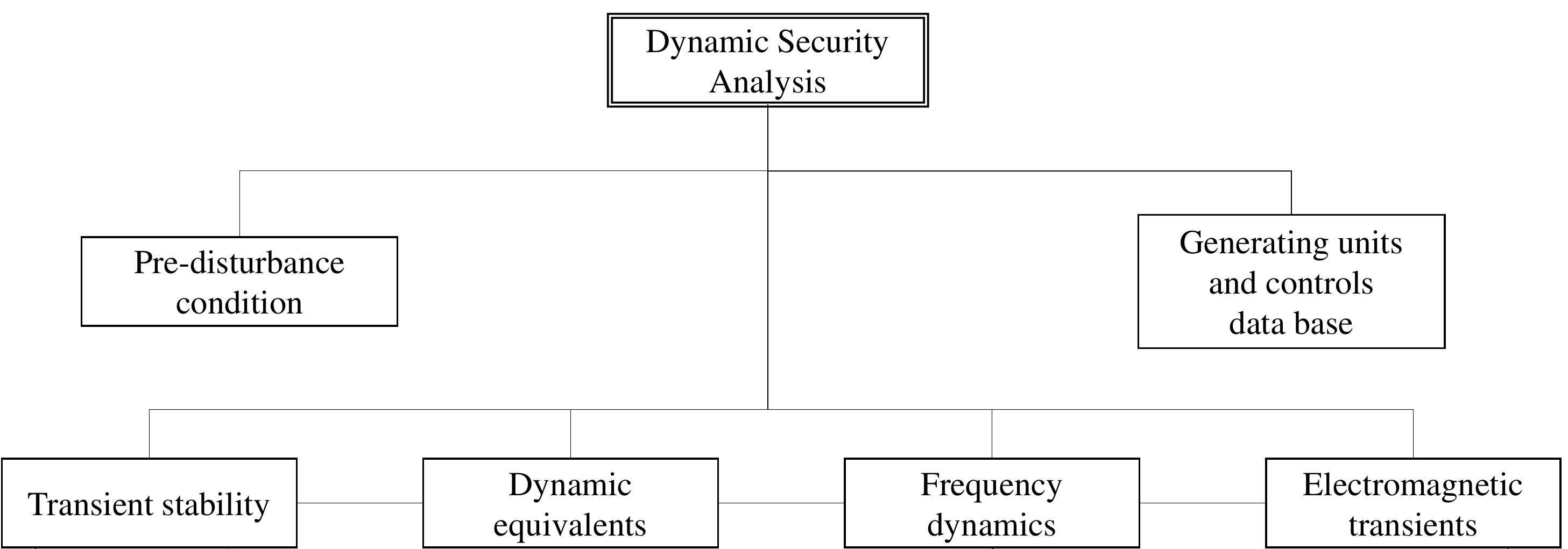}
\caption{Example of functions in the Dynamic Security Assessment, adapted from \citep{Fou88}.}\label{fig:DSA}
\end{figure}

{The reliability of the power system requires the two functions of adequacy and security \citep{Bil96}. Adequacy is the ability of the system to supply electricity to the end-users with a probability that is high enough at all times considering outages of assets in the system. Adequacy is evaluated over months and years and can be measured by computing metrics such as loss of load probability or the expected energy not supplied. The security of the power system refers to the degree of risk that the power system survives imminent disturbances/contingencies without the interruption of the service to the end customers \citep{Kun04}. Hence, security focuses on the real-time or short-term operations and involves control mechanisms, such as in \citep{Lia67}, whereas adequacy focuses on the reliability function over a longer time-span. The security of the system operation depends on the probability of the disturbances and the system operating condition. When analysing the security of the system two important components are to be distinguished: the static and dynamic security analysis. It is important to analyse for the power system subjected to a disturbance (small or large) whether the new post-disturbance operating condition, where the system settles, fulfils all physical constraints. This is considered as the static security analysis and involves the steady-state analysis of the post-disturbance operating condition and the corresponding verification of remaining within the equipment ratings and voltage constraints. Static security can be considered in real-time operation by modelling the energy balances and constraints and equipment ratings of the pre-fault and the post-fault operating conditions in the Security-Constrained Optimal Power Flow problem \citep{Cap11,Cap16}. The dynamic analysis of security refers to analyse whether the system survives the transition from the pre-disturbance to the post-disturbance condition. Considering dynamic security in operations is more difficult and often it is not considered, rather stronger (more conservative) static security limits are considered \citep{Cap16}. The analysis of dynamic security typically involves studying various types of system stability phenomena (e.g as in Fig. \ref{fig:DSA} from \citep{Fou88}), such as rotor angle stability, voltage stability, frequency stability as classified in \citep{Kun94}. Rotor angle stability refers to the ability of the synchronous machines in the power system remaining in synchronism following a disturbance and voltage stability refers to the ability of the system to maintain steady voltages at all buses \citep{Kun04}. Frequency stability refers to the ability of the system to maintain a steady frequency in a disturbance that involves a major imbalance between generation and load. Although power system stability is in itself a single problem referring to the continuation of intact operation following a disturbance, these stability phenomena are distinguished for the purpose of addressing effectively the stability analysis. For analysing each of these specific phenomena different assumptions are typically being made and the appropriate analytical techniques used. An overview of the analysis of the various types of stability can be found in \citep{Kun04,Kun94}. The analysis of some of these stability phenomena requires the analysis of nonlinear systems \citep{Blo00}, and this analysis can require time-domain simulations, for instance for voltage stability or transient stability. This simulation-based analysis of the nonlinear power system can be challenging given the size of the power system and the need to consider disturbances as event-type perturbations \citep{Kun04,Kun94,Chi01}. An example of the different functions to analyse the stability phenomena for the analysis of dynamic security is shown in Fig. \ref{fig:DSA} \citep{Fou88} and, for instance, the analysis of transient stability requires a time-domain simulation. Therefore, it is computational challenging to assess dynamic security in real-time operations for every single disturbance that can possibly occur as stability analysis is part of a full analysis of dynamic security. It is also challenging to assess dynamic security in advance to real-time operation as the operating conditions are unknown and uncertain in advance and hence multiple possible operating conditions may need to be assessed. }





{\subsection{Security Assessment}}
{In the past, the assessment of the security was deterministic, meaning that the power system was designed and operated to withstand a fixed shortlist of contingencies selected based on the significance of the likelihood of occurrence. This shortlisting minimizes the computational effort of the assessments. In practice, this list typically consists of contingencies such as a sudden loss of the functioning of a single piece of major equipment, such as a three-phase fault of a transmission line \citep{Kun04}. The study of the response of the system to losing a piece of single equipment is typically referenced as the 'N-1' criterion as it corresponds to the operation of the power system reduced by one equipment. The deterministic approach to the assessment of security has resulted in a high level of security in the past. However, several limitations of this deterministic approach exist one being that each studied scenario is assumed with the same risk. This assumption may be inaccurate as some operating conditions may be more likely and more severe than others, therefore their risk is higher than of other scenarios. Hence, it is a need for systems operations to quantify and manage the risks. This requires to account for the probabilistic nature of the operating conditions and the likelihood of the contingencies \citep{Kun04}. Such a risk-focused security assessment would use resources on high-risk scenarios rather than on some that have lower risks. }

{The probabilistic approach to security assessment is to estimate the level of security by considering the probability of the system becoming unstable and the corresponding consequences \citep{Kun04,McC99,McC04}. Considering the likelihood of contingencies allows focusing on the assessment of contingencies with high risks. The underlying idea of computing a risk rather than assessing a deterministic set of contingencies has been studied in the past extensively and several different risk metrics were proposed \citep{All00,Hey19} that compute the physical risk \citep{Kir04,Mar04}, the socio-economic risk for the end-consumer \citep{Kir07,Kir03} or the risk in terms of physical system parameters that, for instance, relate to the system security as in \citep{Mar04,Ni03}. Most of these probabilistic security assessments focus on static security, and only a small number include dynamic stability/security, such as for instance on transient stability \citep{Dis11} or rotor angle stability \citep{Pre15}. \citep{Cia17} extends this to probabilistic dynamic security by considering multiple stability phenomena. Generally, the probabilistic approach resulted already when focusing on static security only in reductions of the risks in comparison to deterministic approaches where the short-list of contingencies is fixed and assumed to have equal risks \citep{Kir07}. More reductions are expected when considering dynamic security in the probabilistic approach to security assessment as well. The use of a probabilistic approach is also needed to enable making full use of corrective control actions following a disturbance and to trade these off with preventive control actions \citep{Str16}. However, in Europe for instance, operators still seldom make use of probabilities in the short-term management of reliability \citep{Gar14}, and in the US, the development of metrics on security by the North American Electric Reliability Corporation remains deterministic in their nature \citep{Hey18,Hey19}. Operators are not eager to change from deterministic to probabilistic reliability metrics as probabilistic criteria are perceived as complex and the results from deterministic metrics are perceived as satisfactory \citep{Hey19}. Additionally, the quantification and management of risks require accurate estimations of the likelihood of contingencies \citep{McC04} and it is challenging to predict these estimates as they vary with time being dependent on the asset health and on the weather. Various techniques were proposed for predicting these estimates \citep{Xia06,Fan16,Jam20}.}

{Several challenges of probabilistic management of reliability remain. For instance, the probabilistic approach to reliability management is not broadly adopted although being studied extensively in the literature showing demonstrable benefits as in \citep{Str16,Hey18,Hey19}. This barrier has to do with the conservative operating paradigm from the past rather than being a technical challenge \citep{Hey19}. A technical challenge is on considering dynamic security in real-time operations. Currently, most operators select higher static security limits rather than considering dynamic security in real-time operations \citep{Cap16}. This technical barrier for not considering DSA in real-time operations is the computational complexity of the task at hand \citep{Kun04,Chi01,Cap16,Don18}. In this work, it is proposed to use machine learning within the probabilistic security assessment to address the aforementioned technical challenge.}

{\subsection{Machine Learning for Dynamic Security Assessment}}
\begin{figure}
\centering
\includegraphics[width=0.6\linewidth]{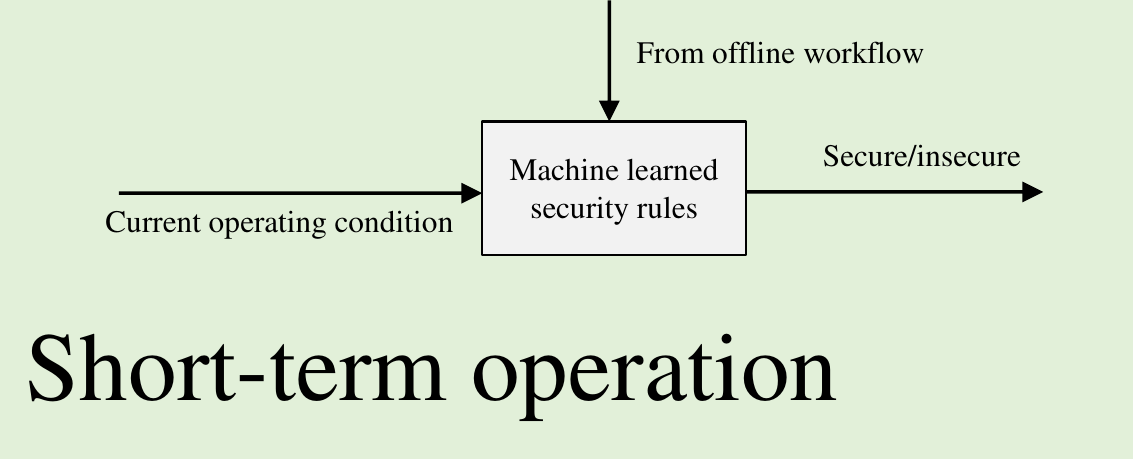}
\caption{{The real-time workflow of using machine learning in DSA.}}\label{fig:online_classic}
\end{figure}
{Machine learning has been proposed for the management of reliability \citep{Duc20} including for the assessment of dynamic security. The purpose of using machine learning in DSA is to overcome the technical barrier of real-time DSA being too computationally challenging. The idea is that a machine learning approach can predict either the outcome of a stability analysis without performing the analysis as per standard stability analysis procedures \citep{Weh98} or to directly predict the dynamic security that comprises these stability analyses. This use of machine learning is illustrated in Fig. \ref{fig:online_classic}. The feature of machine learning rendering this as a  suitable approach is that the predictions are instantaneously available in contrast to classical approaches such as time-domain simulations that are computationally intensive. The applicability of these machine learning approaches was studied (and tested) in control rooms for the real-time operations by a small set of system operators, for instance in Canada \citep{Lou10,Sam10} and in a large-scale European project, the Innovative Tools for Electrical System Security within Large Areas (iTesla) project \citep{Vas16,Kon16} showing promising results, however, several problems and challenges still exist. In order to have such a machine learning-based predictor prepared for real-time operation requires a training procedure of the machine that involves data. This training procedure is carried out well-ahead of real-time operations and typically involves four steps as illustrated in Fig.  \ref{fig:offline}. The first step is to build a training database, the second to pre-process the data, the third to learn the model and finally, the fourth step is to validate and update the machine learning model. }

\begin{figure}
\centering
\includegraphics[width=1\linewidth]{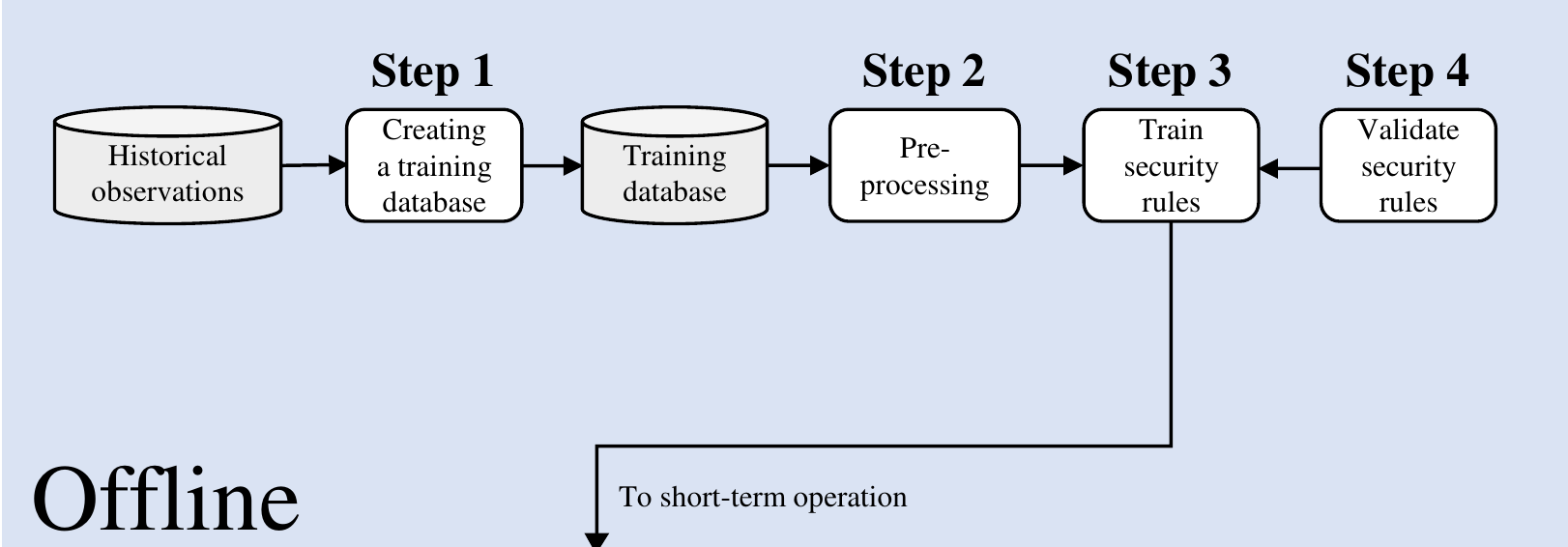}
\caption{The offline machine learning workflow to train a model/security rules for the real-time/short-term operation as in Fig \ref{fig:online_classic}. This work focuses on Step 3 and proposes a new real-time workflow.}\label{fig:offline}
\end{figure}

In the first step, the training database is built by using typically a mix of historical observations and synthetically generated data based on these observations \citep{Kon16,Kon18}. The database lists pre-fault operating conditions and their corresponding post-fault level of stability and/or security. First, the pre-fault operating conditions are generated and subsequently serve as initial conditions to analyse the stability phenomena in question for a specified list of contingencies. Subsequently, a metric for post-fault stability is computed for each operating condition. For instance, \citep{Van14} computed the post-fault level of stability for voltage stability. If the final database should consider (dynamic) security, then the analysis of multiple stability phenomena is combined as in \citep{LiuA14}, or static security can be considered as well \citep{Sev15}. Finally, the database consists of the list of pre-fault operating conditions and their corresponding security/stability labels for various faults depending on what was part of the analysis. An effective database requires to maximise the information content and to minimise the amount of redundant information. 
For instance, a common imbalance is that many more secure operating conditions than insecure conditions are in the database \citep{Duc20}. Hence, much more information is available for the secure conditions than for the insecure conditions. Therefore, the effective generation of insecure operating conditions is important as it adds information to the database and can be done by a broad variety of approaches, such as by iterative selecting relevant data \citep{Gen10}, by using feature selection and neural networks \citep{Jaf18}, by using importance sampling \citep{Kri11, LiuA14} or by a directed-walk \citep{Tha18}.

{In the second step, the data is pre-processed for improving the following training process of the machine. This step can involve the selection and extraction of relevant features for training the machine learning approach and addressing the imbalance in the training database. Feature selection aims to select the most relevant features \citep{He12,He13} and is applied to address the curse of dimensionality of the learning problem that can cause long training times and low performance of the machine. Feature selection approaches can be broadly classified into filter-based approaches (as used in \citep{Jaf18}), wrapper-based approaches, and embedded approaches \citep{Guy03,Li16}. Feature extraction aims to transform and reduce the data to represent the information in a more meaningful way, such as by using principal component analysis \citep{Moh15} or deep learning \citep{Sun18}. In addition to the selection and extraction of features, pre-processing can involve addressing the imbalance in the training database as well. The imbalance can be addressed within the pre-processing by either under-sampling the majority class (secure class) or oversampling the minority class (insecure class) \citep{Cha02, Dru03}, as applied in \citep{Zhu17,Tha18}. For instance, the Synthetic Minority Oversampling Technique (SMOTE) approach \citep{Cha02} creates new data without performing the stability/security analysis and under-sampling removes some data to balance the remaining data.}

{In the third step, the machine learning model (also called security rules) is trained. The final features and the training database are used as input for the training and the output is the machine learning model. The type of the model is typically a binary classifier that considers the two classes, the secure/stable and the insecure/unstable operating conditions. If the classifier aims to predict the result of specific stability analysis, then the predictions correspond to stability/instability respectively, such as for transient stability \citep{Zho19} or voltage stability \citep{Zha18}. If the classifier aims to predict directly the (dynamic) security, then the classifier predicts security/insecurity, as for instance in \citep{LiuA14}. Across the different stability phenomena or security criteria, several machine learning approaches were proposed to learning the model \citep{Bal18,Duc20}, such as Artificial Neural Networks \citep{Sun18,Tan17}, Support Vector Machines \citep{Wan16,Roz17}, Decision Trees (DTs) \citep{Weh98,Moh15,Kri11,LiuA14,Kon16,Vas16,Cre19}, Ensemble methods \citep{Sam10, Cre18b, Liu18,Maj18}, \textit{k}-Nearest Neighbour \citep{Hou95,Fan18}, or fine operating rules \citep{Sun16}. The best suitable learning approach may change dependent on the characteristics of the application and a broader overview can be found in \citep{Duc20}. For instance, some consider one model per contingency (e.g. \citep{Weh98,LiuA14}) and others consider multiple contingencies in a single model, as in \citep{Cos16,Sun18,Cre18a}. Often this choice depends on the accuracy, required training time, and frequency of data and system changes such as topology changes that can vary with the application \citep{Duc20}. For instance, whether the accuracy of the prediction of the final model is higher when training a single- or a multiple-contingency model depends on the selected machine learning approach. Also, in terms of training time, the overall training time may be higher in a single-contingency approach as one model need to be trained per contingency and in a multiple-contingency approach, only $1$ complex model needs to be trained. However, whether the training of many simple models or $1$ complex model is faster depends on the application and how often changes occur that trigger retraining of the models. All the aforementioned differences in the application are important to consider when designing the right machine learning workflow. The final trained machine learning model can subsequently be used in real-time operation to predict for a pre-fault operating condition (input to the model) whether this condition belongs to the secure/stable or insecure/instable class (output of the model). }

{In the fourth step, the machine learning model is validated and updated. The majority of these machine learning approaches typically aim to provide a model that is accurate in predictions. To validate the accuracy of the model a testing set of the database is used that was not part of the training process. For this testing data set the actual output of the stability/security analysis is known and can be compared against the prediction from the machine learning model, and the average testing accuracy (or the testing error, the complement to the accuracy) can be computed. However, using the testing accuracy assumes that prediction errors in the two classes have equal severeness. Missing an insecure operating condition is more severe than missing a secure operating condition. Another measure for the testing accuracy is the $F1$ score that allows balancing the precision and recall for different errors \citep{Pow11}, as done in \citep{Sun18,Roz17,Maj18}. Considering different impacts of errors is particularly important if the training database is imbalanced, hence, the accuracy is biased toward predicting the majority class with higher accuracy than the minority class \citep{Elk01,Duc20}. As pointed out earlier, it is important to address this bias as the minority class is typically the insecure/unstable class and identifying/predicting these is the goal of the operator with the dynamic security assessment. As strategies to address this imbalance by re-balancing the database through over- or under-sampling, several approaches were proposed to address this imbalance in the training and validation of the model. A first approach is to account for the imbalance within the training of the model and the second to account for it after the model was learned. The first approach is to weight within the training the data of the minority class more than of the majority class, such as done in \citep{LiuA14,Tha18}. The second, cost-sensitive learning approach is to adjust the predicting decisions by using the probability estimates from the model \citep{Elk01}. 
For instance, a cost-sensitive approach was used in combination with ensemble DTs in \citep{Han17} and with deep learning in \citep{Zho18}. It was also combined with the oversampling approach SMOTE in \citep{Zhu17}. In these works, the costs are used to bias the predictions and accurate predictions remain the objective, similar to weighting the classes within the training. The aforementioned challenges and strategies to address imbalances and the inaccuracies are important to consider when using machine learning in security assessments. Ultimately in security assessments, specifically in probabilistic security assessment the risk from contingencies should be considered and ideally directly the expected outage cost to the end-customer \citep{Kir07}.}

{\subsection{Probabilistic Security Assessment with Machine Learning}}
{This work proposes to use machine learning for probabilistic security assessment including dynamic security. A probabilistic framework is proposed that combines the strength of conventional security assessment methods with the strength of machine learning for real-time operation. This overcomes the technical challenge of the computational complexity of assessing dynamic security in real-time operation, typically resulting in probabilistic security assessments mainly focusing on static security and not on dynamic security.}

\begin{figure}
\centering
\includegraphics[width=1\linewidth]{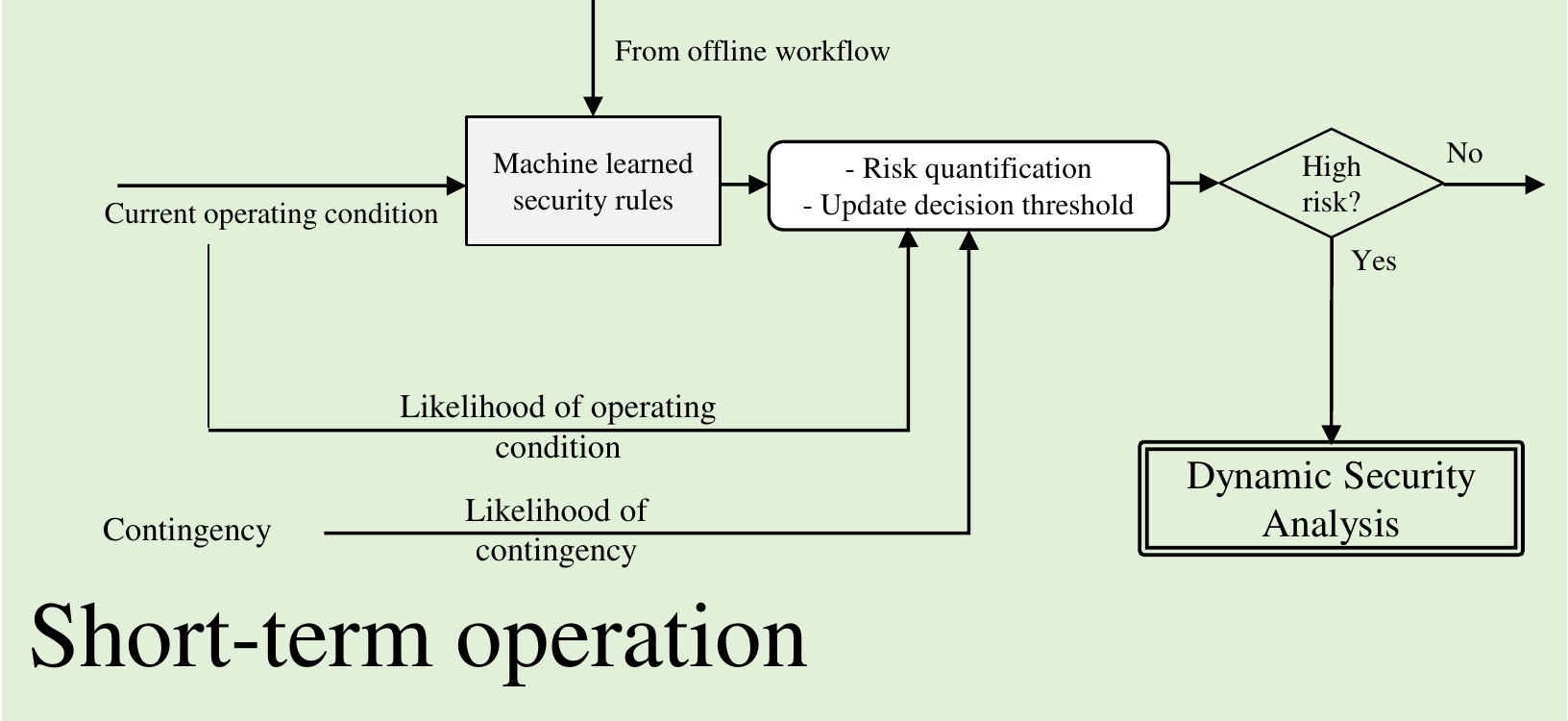}
\caption{{The proposed real-time workflow of using machine learning within a probabilistic security assessment. This workflow combines the trained machine learning model (security rules) from Fig. \ref{fig:offline} and the conventional way of dynamic security assessment Fig. \ref{fig:DSA}}}\label{fig:online}
\end{figure}

This work proposes to extend the risks discussed in the context of probabilistic security assessment by the risk when using machine learning within these security assessments. This extension is necessary when machine learning is used for security assessments as a key difference to other probabilistic security assessments, as in \citep{Dis11,Cia17,Pre15}, is that at least the output of the stability analysis can be relied on, however, when using machine learning these can be inaccurate. Hence, in this work, it is proposed to consider this risk of inaccuracy within probabilistic security assessments. In addition, this work proposes a framework that allows using at the same time two ways of assessing the dynamic security: the conventional method to assess the different stability phenomena by using time-domain simulations and by the machine learning approach. The closest work is \citep{Don18} where the risks of contingencies were ranked using a deep neural network. This proposed work extends \citep{Don18} by including many operating conditions and applying cost-sensitive learning to address the imbalance and differences in costs. In this proposed workflow, as illustrated in Fig. \ref{fig:online}, initially, all operating conditions and contingencies are assessed with the machine learning model. Subsequently, the risk-metric is used to rank these and make corrections to the assessment with the conventional assessment methods if the risk is high. The same risk-metric was also used in \citep{Cre18b}, however, there the objective was to find corrective control decisions and not the probabilistic assessment of security. This proposed work can be used for individual dynamic stability phenomena and various metrics; however, it is demonstrated with a static metric. It proposes a general methodology of how to include machine learning in probabilistic dynamic security assessments.

\begin{figure}
\centering
\includegraphics[width=0.8\linewidth]{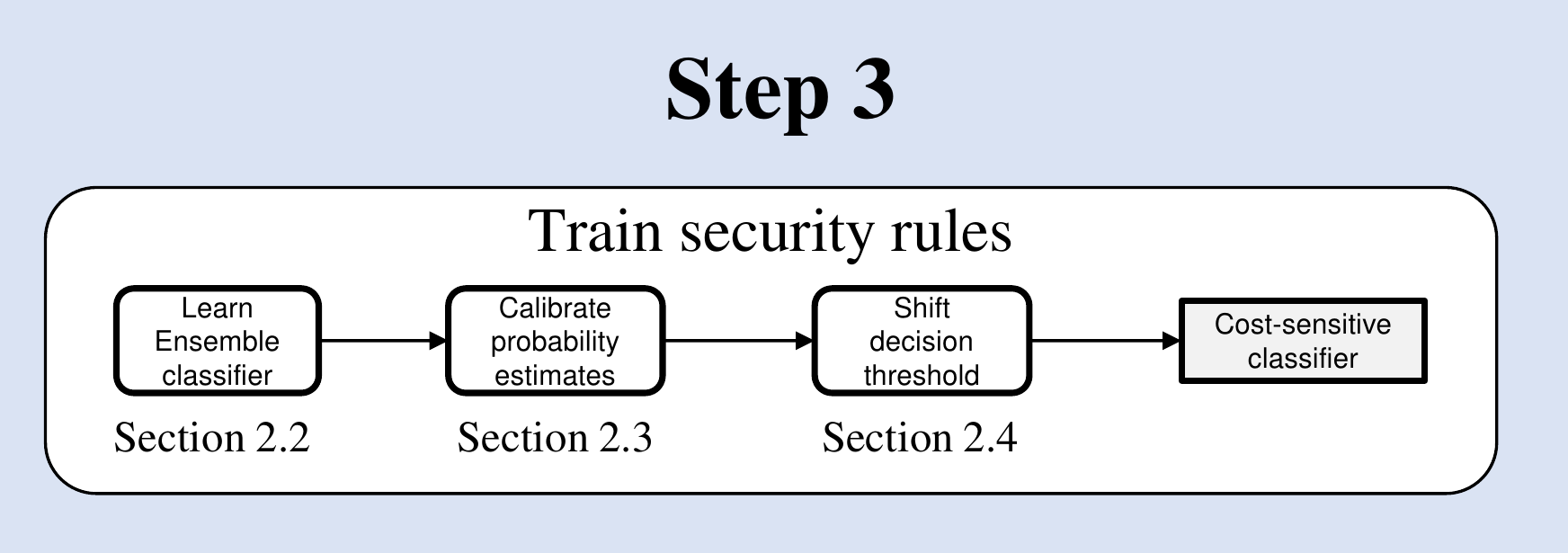}
\caption{{The proposed training of the machine learning model. This is Step 3 in Fig. \ref{fig:offline}.}}\label{fig:step3}
\end{figure}

{The risk of using machine learning within probabilistic security assessments is computed as follows. Initially, a cost-sensitive machine learning approach is selected that allows addressing the imbalance of the training database, a common problem. Adaptive boosting (AdaBoost) \citep{Fre97, Has09} is selected to train an ensemble machine learning model of many DTs \citep{Die00}. In large problems, this approach is well-suited to address the problem of imbalance \citep{He09,Car05}. Subsequently, Platt calibration \citep{Pla99} is used to calibrate the probabilistic output of the trained machine learning model, such that the outputs of the machine learning model are accurate probability estimates whether the operating condition is secure/insecure. Subsequently, a shifted decision threshold is used to make cost-sensitive predicting decisions. This specific workflow is selected as illustrated in Fig. \ref{fig:step3} from Adaboost over Platt calibration to shifted decision thresholds as is simple, flexible and has the best performance among a wide-range of other boosting variants \citep{Nik16}. The decision threshold can be adjusted in real-time to accommodate frequent changes in the likelihood of contingencies from approaches as in \citep{Xia06,Fan16,Jam20}. Finally, the risk is computed based on the theory from cost-sensitive learning. }

{The contribution of this work is to extend other machine learning-based approaches for DSA by proposing
\begin{itemize}
	\item calibration for accurate probability estimates of predictions
	\item a risk-metric for using machine learning in probabilistic security assessment
	\item a framework to efficiently balance machine learning with conventional security assessment within probabilistic security assessment
\end{itemize}}

{This allows operators to better focus their security assessment on operating conditions and contingencies that are of high-risk. }

The proposed approach is studied on a small and a large system using static security metrics. A case study on the IEEE 6-bus system is used to illustrate the challenges of imbalances in the training data and how these and probabilistic estimated being calibrated with the proposed approach. Then, the French transmission system is used to showcase that the proposed framework can be used for multiple contingencies and the performance to inaccurate estimations of the likelihood of contingencies.

{The rest of the paper is structured as follows. In Section \ref{sec:ml}, the cost-sensitive machine learning workflow is proposed. Thereafter, the framework for using machine learning in probabilistic security assessment is proposed in Section \ref{sec:pr}. Subsequently, the case study is presented in Section \ref{sec:cs}. Finally, Section \ref{sec:conc} is the conclusion.}

\section{Training Machine Learning model}\label{sec:ml}
To use machine learning in DSA typically involves a binary classification problem. {An operating condition $i$ of the system subjected to a disturbance/contingency $c$ can either in the post-fault (a) fulfil security criteria and is considered as secure $y_{i,c}=1$ or (b) the operating condition does not fulfil the criteria and is considered as insecure $y_{i,c}=0$ against this contingency.} Binary classification is used to predict this binary label $y_{i,c}$ based on a feature vector $x_i$ consisting out of steady-state real values that describe the operating {condition}, such as power injections, loads, phase angles and voltages of the buses. {This work is to use the score output of a binary classifier to predict probability estimates for the classes. The probability estimate $\hat{p}^1(x_i)$ corresponds to the likelihood that the operating condition $x_i$ belongs to the secure class and $\hat{p}^0(x_i)$ that $i$ belongs to the insecure class. }To learn a classifier requires a training database: a population of operating conditions $\Omega^T$, where each operating condition $i \in \Omega^T$ corresponds to a feature vector $x_i$ and a label $y_{i,c}$. The population of insecure operating conditions at contingency $c$ is $\Omega^{T,0}_c = \{i \in \Omega_T | y_{i,c} = 0 \}$ and of secure operating conditions $\Omega^{T,1}_c = \{i \in \Omega_T | y_{i,c} = 1 \}$, thus $\Omega^T = \Omega^{T,0}_c \cup \Omega^{T,1}_c$. 

\subsection{Cost-sensitive Learning}\label{sec:asm} 
{Cost-sensitive learning under imbalances is typically twofold challenging for machine learning applications \citep{Elk01,Duc20}. }

{The first challenge is the class imbalance of the training database. $|\Omega^{T,1}_c| > |\Omega^{T,0}_c|$ can be typically observed \citep{Kon16,Duc20}.} $|\cdot|$ denotes the cardinality of a set. This class imbalance can be captured by the two different class priors $\pi^0_c = |\Omega^{T,0}_c| / |\Omega^{T}|$ and $\pi^1_c = |\Omega^{T,1}_c| / |\Omega^{T}|$. 

The second challenge is the cost-skewness as the cost/severeness of inaccurate predictions differs for the two classes. Missing an insecure operating condition {(missing an alarm)} is more severe than missing a secure operating condition {(false alarm), that is why often other metrics for the testing accuracy are used as in \citep{Sun18,Roz17,Maj18}. Hence,} the costs of predicting an insecure operating condition as secure $C_c^{F1}>0$ is typically greater than the costs corresponding to predicting a secure operating condition as insecure $C_c^{F0}>0$, thus $C_c^{F1} \gg C_c^{F0}$. {Ideally the cost for missing an alarm $C_c^{F1}$ are directly the expected cost to the end-customer \citep{Kir07}. The cost $C_c^{F0}$ considers unnecessarily preventive or corrective control actions.} These costs may differ for each contingency $c$ as well.

In fact, the two challenges of class imbalances ($\pi^0_c \neq \pi^1_c$) and skewed costs ($C_c^{F0} \neq C_c^{F1}$), although typically discussed and addressed in different contexts, can be addressed very similarly \citep{Elk01,Fla12}. For example, assume $C_c^{F1} = 3 C_c^{F0}$, then this is equivalent to adjusting the class priors, which means, in this example, to use each insecure operating condition $i \in \Omega^{T,0}_c$ three times in the training database. In general, if the cost ratio is defined as

\begin{equation}
\mathcal{C}_c = \frac{C_c^{F1}}{C_c^{F1} + C_c^{F0} },
\label{eq:costratio}
\end{equation}
then, the adjusted class distributions would be

\begin{subequations}\label{eq:classdis}
\begin{align}
& \hat{\pi}^0 = \frac{\mathcal{C} |\Omega^{T,0}|} {\mathcal{C} |\Omega^{T,0}| + (1-\mathcal{C}) |\Omega^{T,1}| } \\
& \hat{\pi}^1 = \frac{ ( 1 - \mathcal{C}) |\Omega^{T,1}|} { \mathcal{C} |\Omega^{T,0}| + (1-\mathcal{C}) |\Omega^{T,1}| }.
\end{align}
\end{subequations}

Note that the index for the contingency $c$ was dropped for simplicity reasons in Eq. \eqref{eq:classdis}. Hence, there exist a duality of the problems of skewed cost and class imbalances \citep{Nik16}, thus, the first two challenges can be similarly interpreted and addressed in classification. 

\subsection{Classifier training}\label{sec:ab}
An ensemble classifier consists out of many weak classifiers $\Omega^E$. Each of these weak classifiers $ l \in \Omega^E$ corresponds to a hypothesis $h_l(x_i) = \{ 0,1 \}$, which predicts the binary label for operating condition $i$ based on the corresponding feature vector $x_i$. Subsequently, these hypotheses are weighted by $w_l$ and combined as a weighted majority vote to obtain the binary label
\begin{equation}
H^E(x_i) = \begin{cases}
             0  & \text{if } \mathop{\mathlarger{\sum}} \limits _{l \in \Omega^E} w_l h_l (x_i) < 0.5 \\
             1  & \text{if } \mathop{\mathlarger{\sum}} \limits _{l \in \Omega^E} w_l h_l (x_i) \geq 0.5.
       \end{cases}\label{eq:majvot}
\end{equation}
Adaptive boosting (AdaBoost) is used to learn this ensemble classifier from a training population of operating conditions $\Omega^T$. AdaBoost is an iterative process \citep{Fre97, Nik16}, which is here not further discussed. Standard \textit{k}-fold cross-validation was used for trading-off over- and under-fitting by stopping the learning to result in $|\Omega^E|$ weak learners. 

The output of an ensemble classifier can be either the predicted binary label $H^E(x_i)$, as introduced in Eq. \eqref{eq:majvot}, or a score such as 
\begin{equation}\label{eq:score}
s^1(x_i) = {\mathop{\mathlarger{\sum}} \limits _{ j \in \Omega^E |  h_j(x_i) = 1  } w_j} \bigg/ {\mathop{\mathlarger{\sum}} \limits _{ j \in \Omega^E } w_j }.
\end{equation}
$s^1(x_i) \in [0,1]$ quantifies the weighted and normalised vote for the secure class $1$ of the operating condition $i$ (thus, the $1$ as a superscript) by accounting for the votes of all weak classifiers $\Omega^E$. The score vote of the insecure class can be calculated through $s^1(x_i) + s^0(x_i) =1$. These scores can be used to obtain accurate probability estimates after calibrating.

\subsection{Calibration for accurate probability estimates}\label{sec:cal}
{Calibration is used to compute a probability estimate from the score output of the classifier using the methodology of Platt \citep{Pla99}} as this method resulted in the best probability estimates across several calibration methods for boosting classifiers \citep{Nic052}. {A sigmoid function is used to map the score $s^1(x_i)$ to the probability estimate }
\begin{equation}\label{eq:calprob}
\hat{p}^1(x_i) = \frac{1}{1+ e^{a s^1(x_i) + b} },
\end{equation}
where parameters $a$ and $b$ are fitted by maximising the likelihood of a separate calibration data-set $\Omega^K$ of operating conditions. {The superscript $1$ in the probability estimate of the secure class} and accordingly the probability estimate for the insecure class can be calculated through $\hat{p}^1(x_i)+\hat{p}^0(x_i)=1$. 

{The performance of this calibration can be quantified by using the Brier score. First, the following sequence of operating conditions in the calibration set $\Omega^K$ is defined as $k=1,2,\dots |\Omega^K|$, where $\hat{p}^1(x_k) \leq \hat{p}^1(x_{k+1})$. Then, this sequence is split into $N$ subsets of operating conditions, where each subset $\Omega^S_n$ for $n = 1,2,3,... N$ has the same size $|\Omega^S_n|=\frac{|\Omega^K|}{N}$ and $N$ is specified by the user. For instance, the first subset is $\Omega^S_1=1,2,3,...\frac{|\Omega^K|}{N}$. Subsequently, the average probability estimates $\bar{p}^{1}_n$ and the fraction of secure operating conditions $\pi^{1}_n$ are calculated for each of the subsets $n = 1,2,3,... N$. Finally, the Brier score is the squared sum of the differences between $\bar{p}^{1}_n$ and $\pi^{1}_n$ as follows
\begin{equation} \label{eq:brier}
B = \frac{1}{N} \sum_{n \in 1,2,3,...N} (\bar{p}^{1}_n - \pi^{1}_n)^2.
\end{equation}
The score $\bar{s}^{1}$ replaces $\bar{p}^{1}$ in Eq. \eqref{eq:brier} if the Brier score is computed for an uncalibrated classifier.}

\subsection{Shifting the Decision Threshold}\label{sec:shi}
{A shifted decision threshold $z_c$ on the probability estimates is proposed to make risk- and cost-optimal predicting decisions.} 
The risk of predicting an operating condition $i$ can be described as a disjunction $R_c = R^1_c \vee R^0_c$ of the two risks 
\begin{subequations}\label{eq:risks}
\begin{align}
& R^1_c(x_i) =   C^{F1}_c \, p_c^C \, (1 - \hat{p}^1(x_i) ) \label{eq:risk1}\\
& R^0_c(x_i) = C^{F0}_c \, (1 - p_c^C ) \,  \hat{p}^1(x_i) \label{eq:risk0}, 
\end{align}
\end{subequations}
where $R^1_c(x_i)$ and $R^0_c(x_i)$ are the risks of predicting operating condition $i$ as secure or insecure. $\hat{p}^1(x_i)$ is the probability estimate obtained from the classifier learned for contingency $c$ and corresponds to Eq. \eqref{eq:calprob}. $p_c^C$ is the likelihood that contingency $c$ occurs. 

For making risk-optimal predicting decisions the operating condition $i$ should be predicted with lowest residual risk: this means to predict as secure if (and only if) the risk of predicting as secure $R^1_c$ is lower than the risk of predicting as insecure $R^0_c$, thus if (and only if) $R^1_c<R^0_c$. Given this rationale and Eq. \eqref{eq:risks}, an operating condition should be predicted as secure if (and only if)
\begin{equation}\label{eq:threshold}
\hat{p}^1(x_i) > \frac{C^{F1}_c \, p_c^C }{C^{F1}_c \, p_c^C + C^{F0}_c \, (1 - p_c^C ) } \coloneqq z_c,
\end{equation}
where $z_c$ is the shifted decision threshold for the model trained for contingency $c$. 
{Therefore, the computation of the decision threshold requires estimates for three parameters:
\begin{enumerate}[label={(\arabic*)}]
   \item estimates for the costs of missing an alarm and contingency $c$ occurring $C_c^{F1}$ that is ideally the expected outage cost to the end-customer \citep{Kir07} 
   \item estimates for the costs of a false alarm $C_c^{F0}$ that can be estimates of economic costs for unnecessarily planning for preventive and corrective control actions
   \item estimates for the likelihood of the contingency $p_c^C$ that can be forecasted based on weather data or asset health \citep{Xia06,Fan16,Jam20}
\end{enumerate}
The cost ratio $\mathcal{C}_c$ may be used instead of the individual costs. The above parameters are difficult to estimate as discussed in the introduction. The inaccuracy of these parameter estimates is studied in the case study section.} Note how $z_c$ simplifies Eq. \eqref{eq:threshold} if there exist no cost skew $C^{F0}_c = C^{F1}_c$ or/and no class imbalance $p_c^C =0.5$. Subsequently, the risk of a prediction can be summarised:
\begin{equation}
R_c(x_i) = \begin{cases}
             R^1_c(x_i)  & \text{if } R^1_c(x_i) < R^0_c(x_i) \\
             R^0_c(x_i)  & \text{if } R^1_c(x_i) \geq R^0_c(x_i)
       \end{cases}\label{eq:resrisk}
\end{equation}

\section{Probabilistic Security Assessment with Machine Learning}\label{sec:pr}
{The probabilistic security assessment is to use a risk index for assessing the future security of the power system. The results from the assessment can subsequently be used such that operating decisions can be made on future conditions. However, such a future assessment of the future power system requires forecasts of possible future operating conditions $\Omega^P$. The full state $x_i$ of the operation conditions $i \in \Omega^P$ can for instance be obtained by running a load flow analysis combining future load forecasts with the latest results from state estimations. These forecast models need to appropriately describe the forecast uncertainty such that the likelihood of each operating condition $p^I_i$ can be computed. In this assessment, the time dependency is neglected due to simplicity reasons. The future assessment of security requires as well as forecasts of the likelihood of contingencies $p^C_c$ for each contingency $c$. These can be obtained through asset health predictions or inferred from weather forecasts \citep{Xia06,Fan16,Jam20}. These two likelihoods $p^I_i$ and $p^C_c$ change over time. } 

{The probabilistic assessment of security is to compute a risk metric 
\begin{equation}\label{eq:probsec}
RISK^{SA} = \mathop{\mathlarger{\sum}} \limits _{ i \in \Omega^P} \mathop{\mathlarger{\sum}} \limits _{ c \in \Omega^C} p^I_i \, \, p^C_c \, \, S_{i,c}
\end{equation}
and, subsequently, to make operating decisions that minimise this risk \citep{McC04}. The risk is the summation of all risks that each contingency $\Omega^C$ can cause to all possible operating conditions $\Omega^P$, where $S_{i,c}$ is the severity of the contingency. The severity can be for instance the expected outage cost to the end-customer \citep{Kir07}. The challenge is to calculate the severity function $S_{i,c}$ as the performance of the post-fault operation depends on the contingency $c$ and the operating condition $i$, hence this function needs to be evaluated individually for each of these scenarios $\Omega^S = \{ (i,c) \} | \, i \in \Omega^P, c \in \Omega^C \}$ \citep{McC04}. The likelihood of each scenario $s$ is $p^S_s = p^I_i \, \, p^C_c$ assuming these are not correlated. For static security this may be feasible with conventional assessments as their computation is moderately computational demanding. For instance, a linear severity function was selected that increase with the violation of physical limits in \citep{McC04}, or others select a binary severity function that is for violating/not-violating the physical limits. For dynamic security this is more challenging, as pointed out in the introduction, the analysis of the various phenomena of stability requires time-domain simulations and these are computational challenging and given they need to be executed for each scenario $s$ it is computational infeasible. Others use short-cut methods to compute severity functions for some specific stability phenomena. For instance, fuzzy inference systems were used to estimate the severity for rotor angle instability \citep{Pre15} or sensitivities of the severity to the operating conditions were computed for transient stability in \citep{Dis11}. These approaches are binary or linear approximations of the relation severity/operating conditions and follow similar assumptions as in probabilistic static security. Machine learning is an alternative approach to learn advanced models based on data and these models allow fast evaluations of the severity function that outperform conventional security assessments.}

\subsection{Risk from using Machine Learning}

{When using machine learning in probabilistic security assessments it is important to consider the uncertainty/inaccuracy of the predictions of the severity. This uncertainty has an additional risk and hence needs to be considered. This work is to compute this risk from machine learning and consider this risk in probabilistic dynamic security assessments as it is further described. First, the risk function of Eq. \eqref{eq:probsec} needs to be extended as it does not consider that the severity function can be sometimes inaccurate. Although the severity $\hat{S}_{i,c}$ of the operating condition $i$ subjected to contingency $i$ can be probabilistic predicted with machine learning as
\begin{equation} \label{eq:severityML}
\hat{S}_{i,c} =  C^{F1}_c \, (1 - \hat{p}^1(x_i) ), 
\end{equation}
this does not include and consider risks of inaccuracies. The risk of using machine learning in probabilistic security assessment is 
\begin{equation}\label{eq:probsecML}
RISK^{ML} = \mathop{\mathlarger{\sum}} \limits _{ i \in \Omega^P} \mathop{\mathlarger{\sum}} \limits _{ c \in \Omega^C} p^I_i \, \, R_c(x_i),
\end{equation}
where $R_c(x_i)$ is the disjunction of risks as in Eq. \eqref{eq:resrisk}.}

\subsection{Combining Machine Learning and Conventional Security Assessment}\label{sec:combML}

\begin{figure}
\centering
\includegraphics[width=0.8\linewidth]{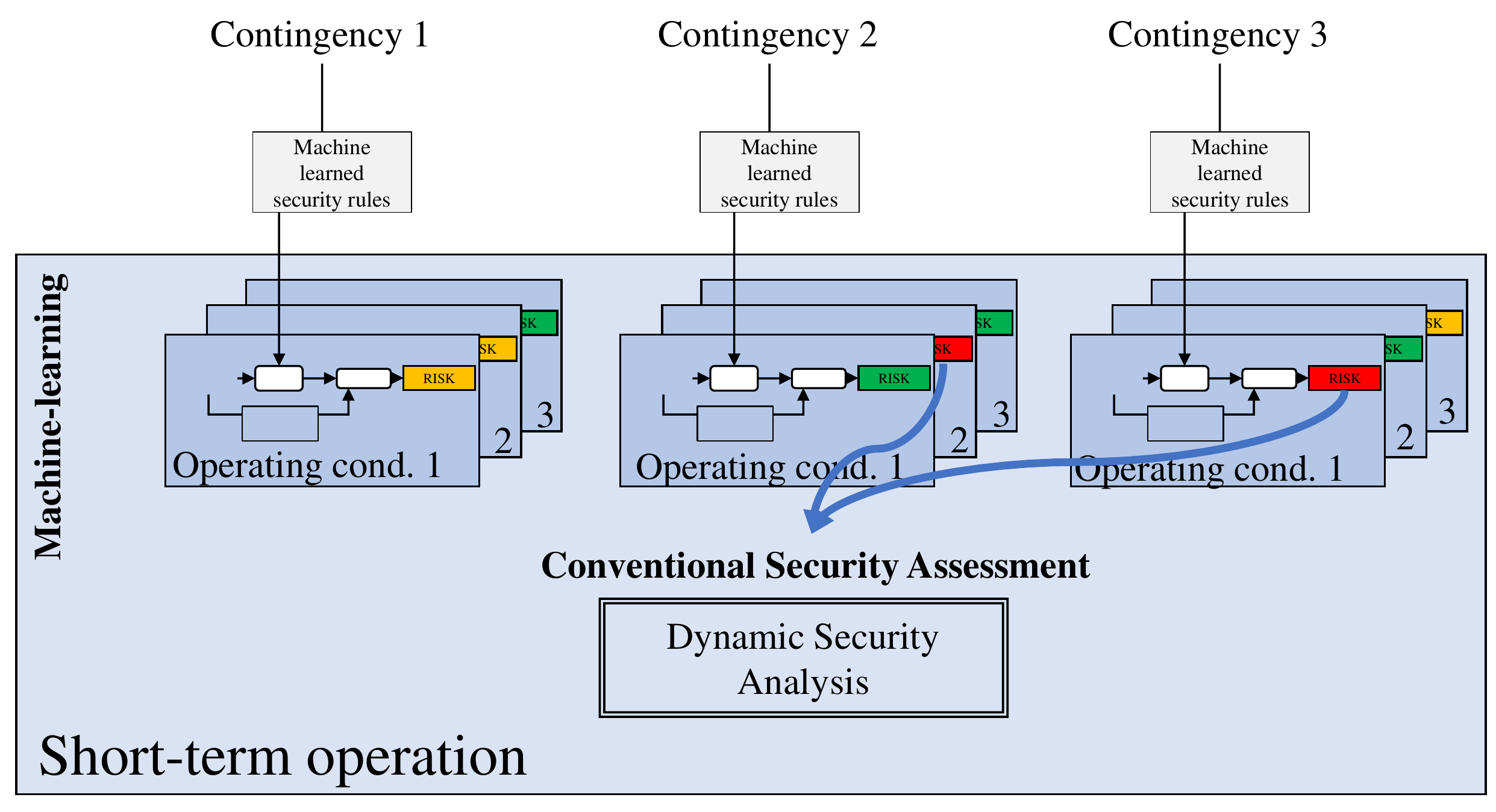}
\caption{{Probabilistic Security Assessment with machine learning and conventional Security Analysis. For each contingency one machine learning model is used to predict risks for many possible operating conditions. Subsequently, the ones with high risk \protect\includegraphics[height=0.8em]{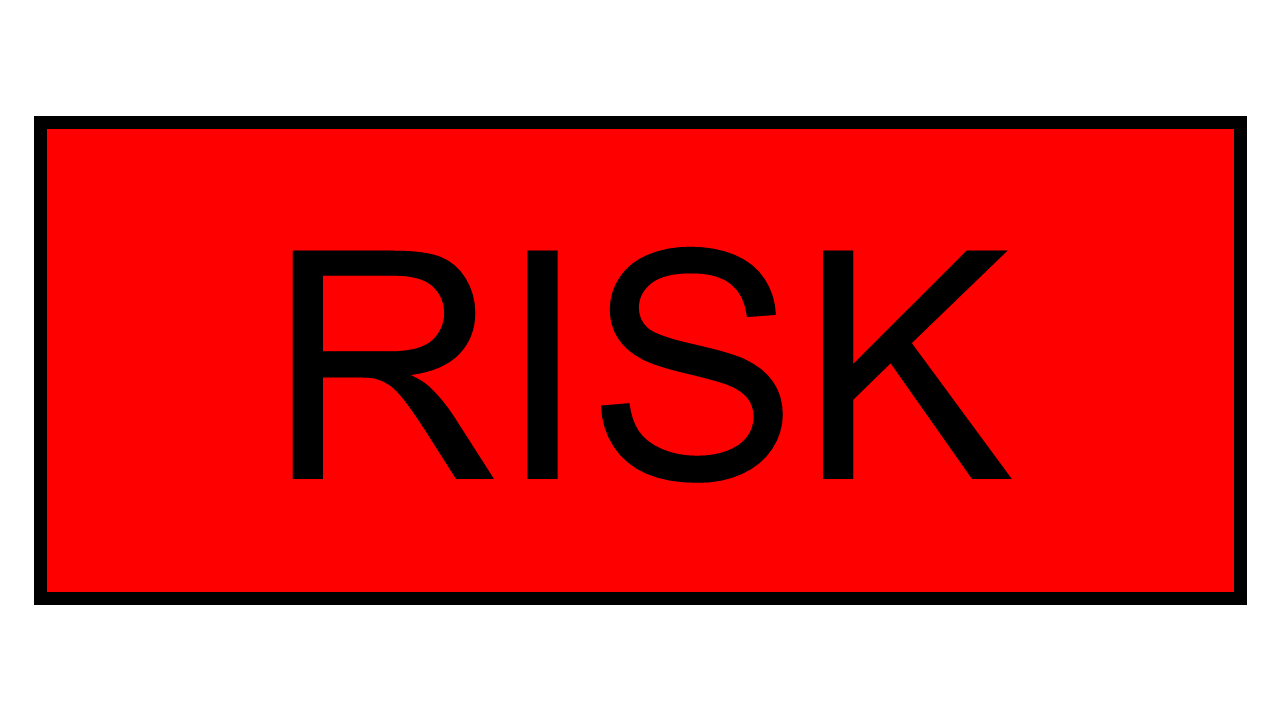} are assessed with conventional Security Assessments followed by medium \protect\includegraphics[height=0.8em]{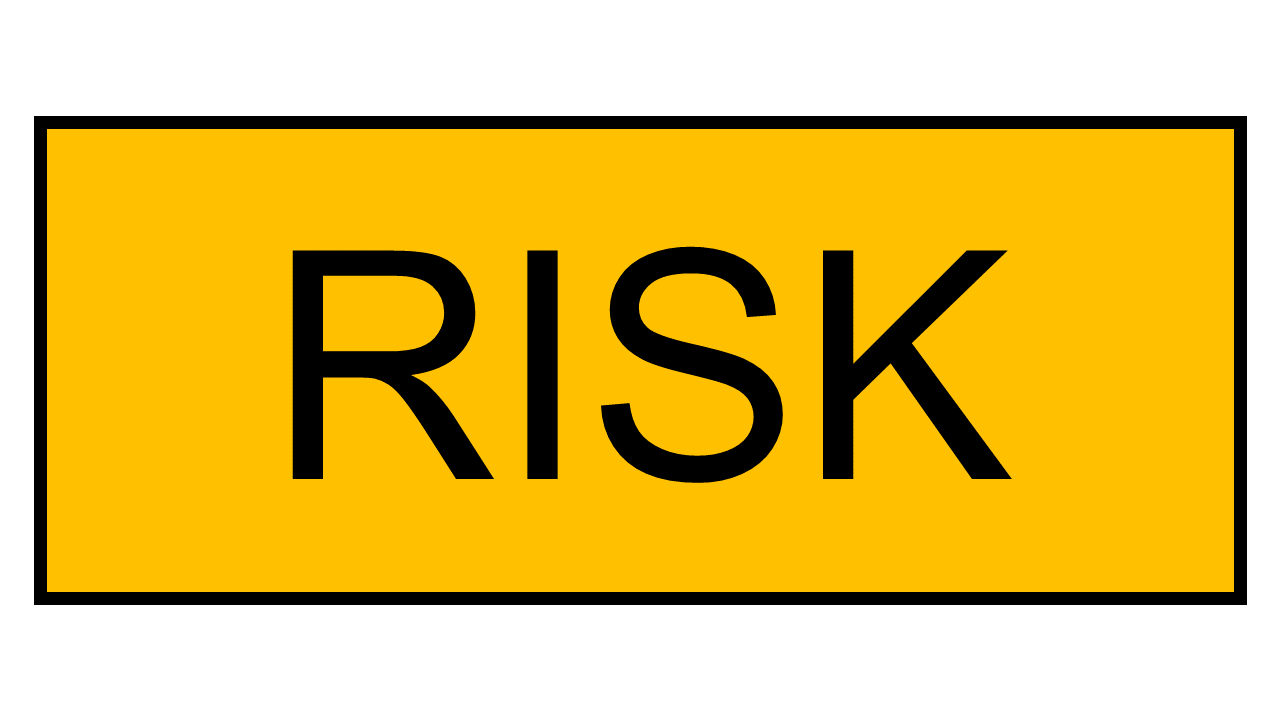} and low risks \protect\includegraphics[height=0.8em]{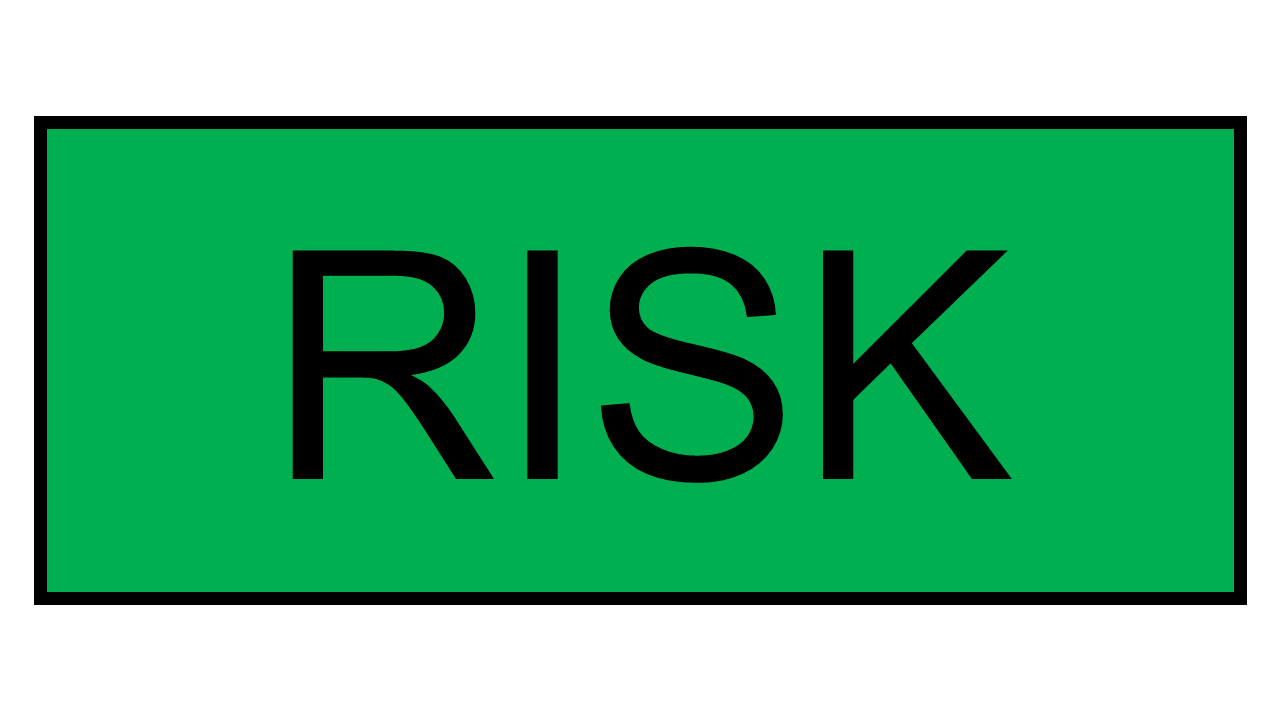}, until the computational budget is used.}}\label{fig:pp}
\end{figure}

{In this work it is proposed to combine machine learning and conventional security assessments in real-time probabilistic security assessment. The idea is to rely on machine learning on some operating conditions that have low risks $\Omega^{S,L}$ and to assess other scenarios with high risk $\Omega^{S,H}$ with conventional security assessments. Before real-time operation a different machine learning model was trained for each contingency $c$ as described in Sec. \ref{sec:ml}.}

{In close to real-time operation the following procedure is executed as illustrated in Fig. \ref{fig:pp}: firstly, risk-optimal predicting decisions are made as described in Sec. \ref{sec:shi} for all scenarios $\Omega^S$. Subsequently, the risks of relying on these predictions are computed
\begin{equation}\label{eq:riskscenario}
\mathcal{R}_s = p^I_i \, \, R_c(x_i)
\end{equation}
for all scenarios $s \in \Omega^S$ corresponding to the operating condition $i$ subjected to contingency $c$. $R_c(x_i)$ is the disjunctive risk of relying on machine learning  from Eq. \eqref{eq:resrisk} dependent on the risk-optimal predicting decisions. Then, these risks are sorted across all scenarios $\mathcal{R}_s^{(j)} \, \forall s \in \Omega^S$ with descending risks, such that $\mathcal{R}_s^{(j)}$ is the $j$th largest value. Dependent on the computational capability of the operator the scenarios are split into high risk $\Omega^{S,H}$, and low risk $\Omega^{S,L}$ scenarios according to the sorted list,
where $\Omega^{S} = \Omega^{S,H} \cup \Omega^{S,L}$. The computational capability $S$ determines the split and is denoted as how many security assessments the operator can run close to real-time operation, hence the number of high-risk scenarios is kept below $|\Omega^{S,H}| <= S$. Finally, only these high-risk scenarios are assessed in real-time operation with conventional security assessment. Or, in other words, the scenarios are assessed with conventional security assessments in the order of $j = 1, 2, \dots \min \{|\Omega^S|, S \}$ until the maximal computational budget of $S$ security assessments is used.
The ratio of scenarios assessed with conventional security assessments is 
\begin{equation}\label{eq:riskscenario2}
CVM = \frac{|\Omega^{S,H}|}{|\Omega^{S}|}.
\end{equation}}

{The residual estimated risk of combining machine learning and conventional security assessment is
\begin{equation}\label{eq:probsecOverall}
RISK^{TOT} = RISK^{SA^*} + RISK^{ML^*} ,
\end{equation}
where the risk from conventional from Eq. \eqref{eq:probsec} is now expressed for only a subset for high-risk scenarios that is denoted by the asterisk symbol $*$ representing the the partial risk contributions dependent on the splits in low-risk and high-risk sets. The risk of conventional security assessment is
\begin{equation}\label{eq:probsec2}
RISK^{SA^*} = \mathop{\mathlarger{\sum}} \limits _{ s \in \Omega^{S,H}}  p^S_s \, \, S_{s},
\end{equation}
where $S_{s}$ is the corresponding severity computed from conventional security assessment. The risk of machine learning from \eqref{eq:probsecML} is expressed as subset of low risk scenarios is 
\begin{equation}\label{eq:probsecML2}
RISK^{ML^*} = \mathop{\mathlarger{\sum}} \limits _{ s \in \Omega^{S,L}}  p^S_s \, \, R_s,
\end{equation}
where $R_s = R_c(x_i)$ corresponds to the correct mapping of the scenarios to operating conditions and contingencies.}

{This proposed approach effectively accounts for real-time changes in the input parameters $p_c^C$, $p_i^I$, $C_c^{F1}$ and $C_c^{F0}$. Particularly the likelihood of contingencies $p_c^C$ and operating conditions $p_i^I$ can change frequently and are considered in real-time operation as illustrated in Fig. \ref{fig:online}. Changes in these parameters require the instant update of the decision threshold $z_c$. Note that no update of the machine learning models is required. Subsequently, some risk-optimal predictions change in response to the changes of the decision thresholds as per Eq.\eqref{eq:threshold}. For instance, if the decision threshold increases in response to an increased risk from a particular contingency $c$, then some scenarios earlier predicted as secure are now predicted as insecure with the objective to minimise the risks. Then, this results in an updated risk ranking of the scenarios and subsequently a new grouping in low-risk $\Omega^{S,L}$, and high-risk scenarios $\Omega^{S,H}$. Finally, different scenarios are assessed with conventional security assessment as the set of high scenarios $\Omega^{S,H}$ changed.}

\subsection{Evaluation of risk-metric}
{The probabilistic security assessment with machine learning can be evaluated by using an independent test-set of operating conditions $\Omega^P$. The evaluation is to estimate the risk of machine learning $RISK^{ML^*}$ by using the test-set where the actual security assessments are known. }

{The following procedure is executed to compute ${Z}^*_c$ as an estimate for $RISK^{ML^*}$. First, all steps in Sec. \ref{sec:pr} are executed: the machine learning models are used for predicting the security assessments for the test set, the high-risk and low-risk scenarios are identified, and the high-risk scenarios are assessed with security assessments. Subsequently, for all scenarios within the low-risk scenarios $\Omega^{S,L}$, the actual security assessment is compared with the risk-optimal prediction decisions of the machine learning models. This comparison provides the number of missed alarms $N^1_c$ and false alarms $N^0_c$ within the low-risk scenarios. Finally, the residual risk of using the machine learning model for contingency $c$ can be estimated as}
\begin{equation} \label{eq:costpred}
{Z}^*_c = \frac{1}{|\Omega^P|}\left(N^1_c \, \mathcal{C}_c \, p_c^C  + N^0_c \, (1-\mathcal{C}_c) \, (1-p_c^C)\right),
\end{equation}
where equal probability of each operating conditions $p^I_i = \frac{1}{|\Omega^P|} \, \forall i \in \Omega^P$
is assumed to be equal across the test-set that is the case once a Monte Carlo method is used to perform sampling. {This is the evaluation metric used in the case study to investigate the performance of the proposed approach and estimates the reduction of $RISK$.}

\section{Case Study}\label{sec:cs}
The proposed approach is studied on the IEEE 6-bus system and on the French grid using static security metrics. Initially, the challenges of class imbalances and cost skewness are illustrated, then it is showcases how calibration can be used to obtain accurate probability estimates and how to make cost-optimal predictions. The proposed approach is compared against a standard classifier and when not considering machine learning. The studies consider several contingencies and the sensitivity of parameter estimations such as the likelihood of contingencies. Finally, the computational reductions in terms of time-domain simulations and relevant aspects of this work are discussed.
\subsection{Assumptions}

\subsubsection{Test systems}
The first test system was the IEEE 6-bus system from \citep{Wod84} considering static security for illustrative purposes of the challenges and the proposed approach. The pre-fault variables $x_i$ of the operating conditions $i$ were the three loads, the three generator power outputs, six phase angles and eleven line-flows. Corrective actions were considered in the form of $\pm \SI{20}{\MW}$ (re-)dispatches of the generator powers for computing post-fault security labels. The loads were drawn from a multivariate Gaussian distribution and a Pearson's correlation coefficient of $0.75$ was used between all three load pairs. Then, the loads were converted to a Kumaraswamy($1.6$,$2.8$) distribution by applying the inverse transformation method. Subsequently, the loads were scaled such that all loads were in the range of $[50,150] \, \si{MW}$. The generator powers were dispatched by a DC optimal power flow (DCOPF) with a linear cost function with coefficients $\{12,10,8\}$ for the three generators at buses $\{1,2,3\}$. {The post-fault security label was computed by validating whether the post-fault steady-state operating condition fulfils all physical limits including voltages and overloads. More specifically, a DCOPF was solved where the generators were set to the pre-fault set values including allowing for the corrective re-dispatches mentioned above. Subsequently, line flow and voltage limits were checked.} 

The second test system was the French network for testing the scalability. The system had $1955$ transmission lines, $798$ transformers, $1886$ buses, $411$ generators and $127$ shunt elements. The data-set was also used in \citep{Kon16} and has $16722$ operating conditions, $35873$ features and $1980$ different contingencies were simulated in the time-domain as well as $9$ different reliability metrics were computed. The static metric for overloads, $5000$ operating conditions ($3500$ training and $1500$ testing) and $|\Omega^C|=11$ contingencies were randomly selected. The metric overload was used as defined in \citep{Sev15}. Although the generation of the data-set involved time-domain simulations, the static metric 'overload' was computed instead of considering the fluctuations within the time-domain. The metric overload considers the mean post-fault value across the time-domain and provides an indicator of the time-domain dynamics. The estimates for the probabilities $p_c$ and costs $\mathcal{C}_c$ of contingencies $c \in \Omega^C$ were randomly selected from $\{0.00001,0.00005,0.0001,0.0005\}$ and from $\{ \frac{500}{501},\frac{1000}{1001},\frac{5000}{5001}, \frac{10000}{10001} \}$, respectively. 

\subsubsection{Machine Learning}
The single DTs were learned via CART \citep{Bre84} by using the package \textit{scikit-klearn} 0.18.1 \citep{Ped11} in Python 3.5.2. with default settings, such as minimising gini impurity; an exception was the restriction of DT depth at $3$ to avoid over-fitting. The AdaBoost ensembles were learned by using the algorithm SAMME.R \citep{Has09} with default parameters, except the number of weak learners was increased to $M = 100$. The weak learners were DTs and pre-fitted with CART with maximal depths of $1$. A set of $|\Omega^K|=875$ operating conditions and $3$-fold cross-validation was used to calibrate.

\subsection{Class and cost imbalances}
The impact of class imbalances was studied on the IEEE 6 bus system for a three-fault contingency on line $5$ connecting buses $2$ and $4$. The class imbalance is large at $\pi^1_5 = 0.89$ versus $\pi^0_5 = 0.11$. A DT was trained on $|\Omega^T|=3500$ operating conditions, and the test error was computed using $|\Omega^P|=1500$ tests; and repeated 10 times for different combinations. Overall the test error was only $\SI{0.9}{\percent}$, however, when disaggregating the test error into classes, $\SI{0.3}{\percent}$ were false alarms and $\SI{5.4}{\percent}$ missed alarms. This imbalance showed that predictions were more accurate on the majority than on the minority class.

The cost skewness was studied on a fault at line $6$ that connected buses $2$ and $5$. Here, the classes were balanced ($\pi^1_6 = 0.52$ and $\pi^0_6 = 0.48$) and interference with the class-imbalance challenge was avoided. The average test error of $10$ DTs was on average $\SI{1.4}{\percent}$ and split into $\SI{1.4}{\percent}$ false alarms and $\SI{1.3}{\percent}$ missed alarms. Although this split of errors is balanced, when assuming the skewness of costs of $C_6^{F1} \gg C_6^{F0}$, then the overall cost of inaccurate predictions was not minimised as it would correspond to imbalances in the error.

\begin{figure}
\centering
\begin{subfigure}[b]{0.46\textwidth}
\begin{tikzpicture}
\small
\begin{axis}[
        axis on top,
        width=0.7\textwidth,
        scale only axis,
        ytick={0,0.5,1},     
        xtick={0,0.5,1},    
        ylabel={$\pi^{1}$},
        xlabel={$\bar{s}^{1}$},
        xmin=0,
        xmax=1,
        ymin=0,
        ymax=1,
        ]        
	\addplot graphics[xmin=0,ymin=0,xmax=1,ymax=1] {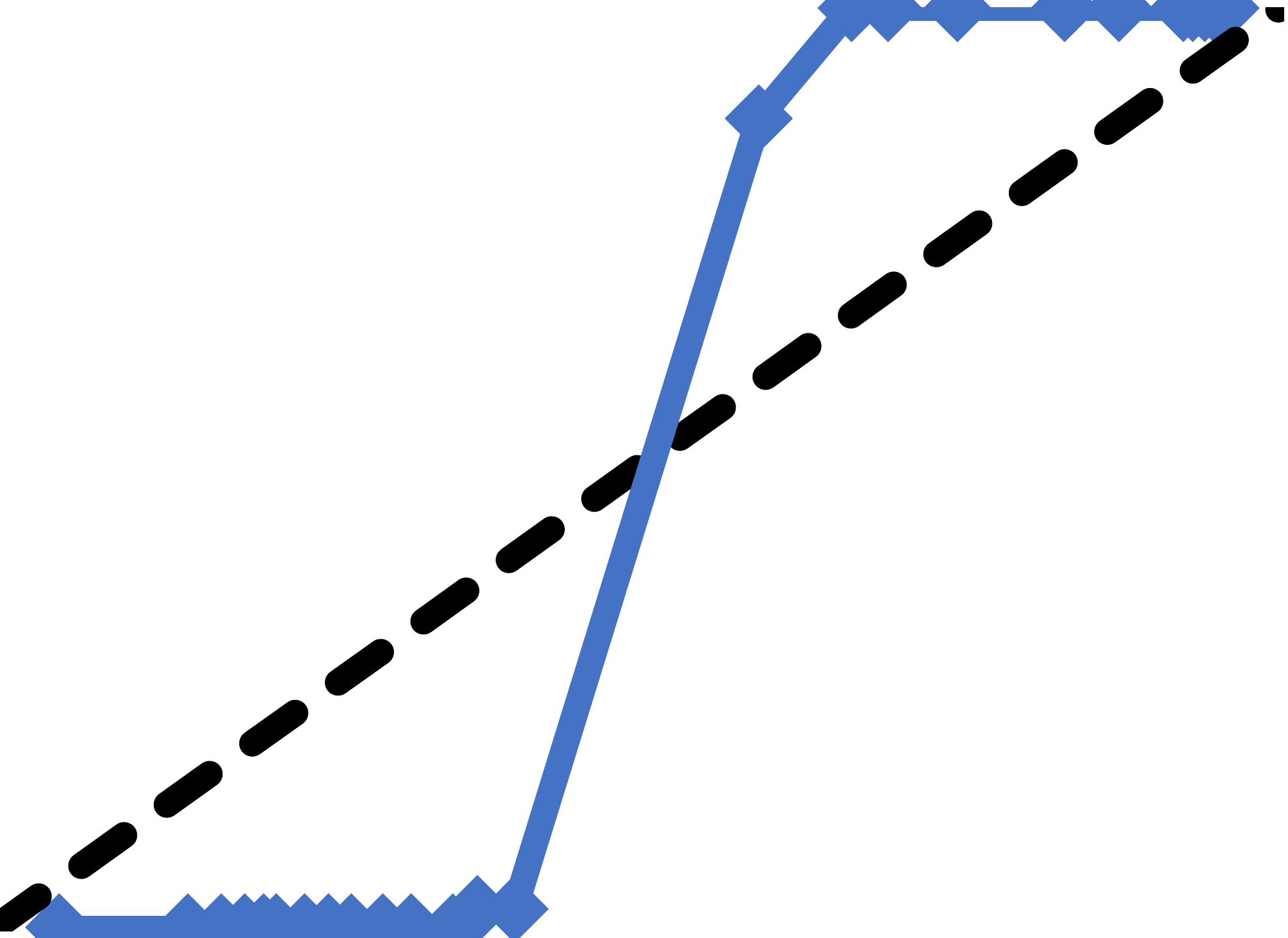};
  \end{axis}
\end{tikzpicture}
   \caption{}
   \label{fig:peuncal}
\end{subfigure}
\begin{subfigure}[b]{0.46\textwidth}
\begin{tikzpicture}
\small
\begin{axis}[
        axis on top,
        width=0.7\textwidth,
        scale only axis,
        ytick={0,0.5,1},     
        xtick={0,0.5,1},    
        ylabel={$\pi^{1}$},
        xlabel={$\bar{p}^1$},
        xmin=0,
        xmax=1,
        ymin=0,
        ymax=1,
        ]        
	\addplot graphics[xmin=0,ymin=0,xmax=1,ymax=1] {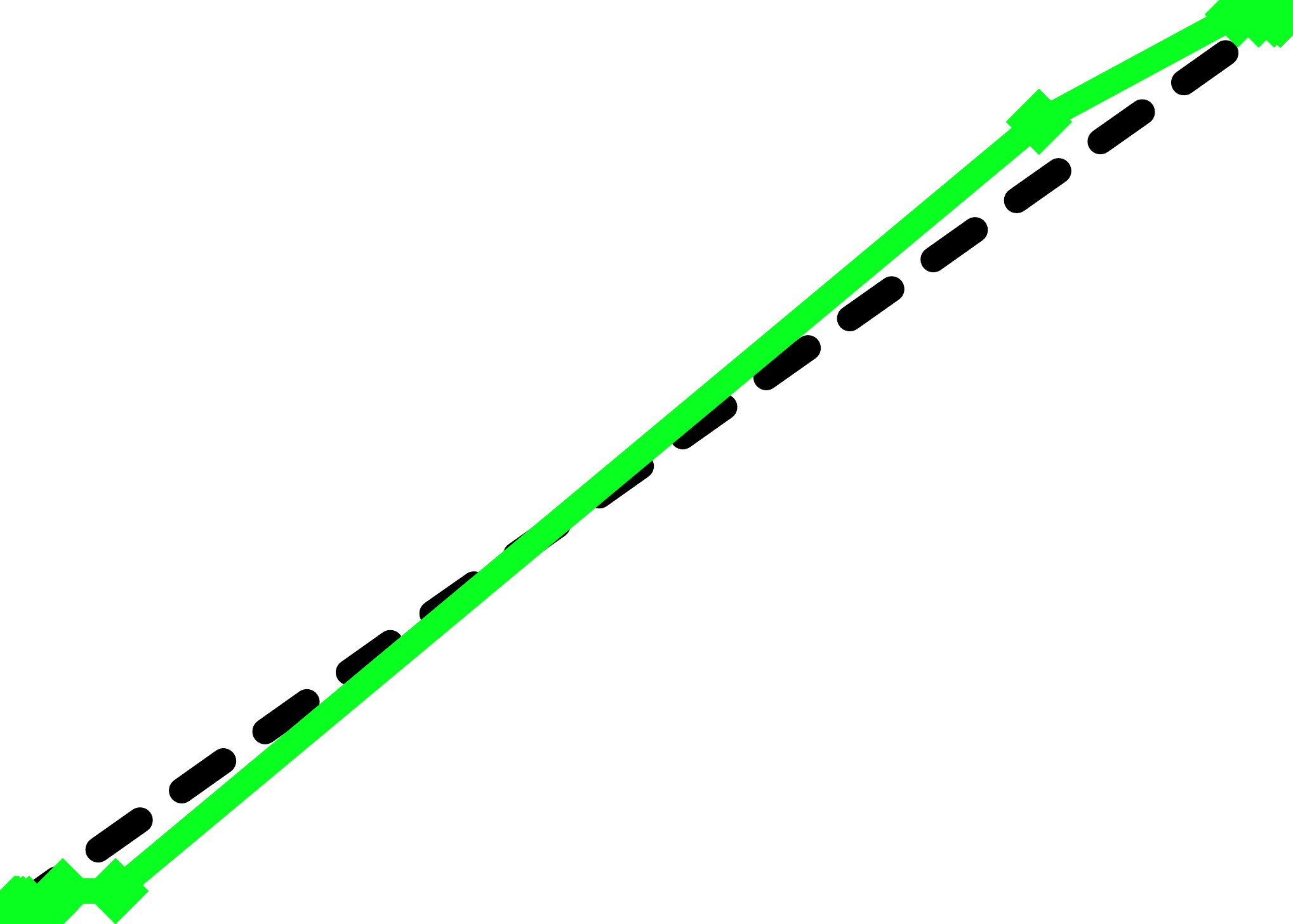};
  \end{axis}
\end{tikzpicture}
   \caption{}
   \label{fig:pecal}
   \end{subfigure}\caption{Probability estimation requires calibration: In (a) is the result of a classifier using an uncalibrated score $\bar{s}^{1}$ (\protect\includegraphics[height=0.5em]{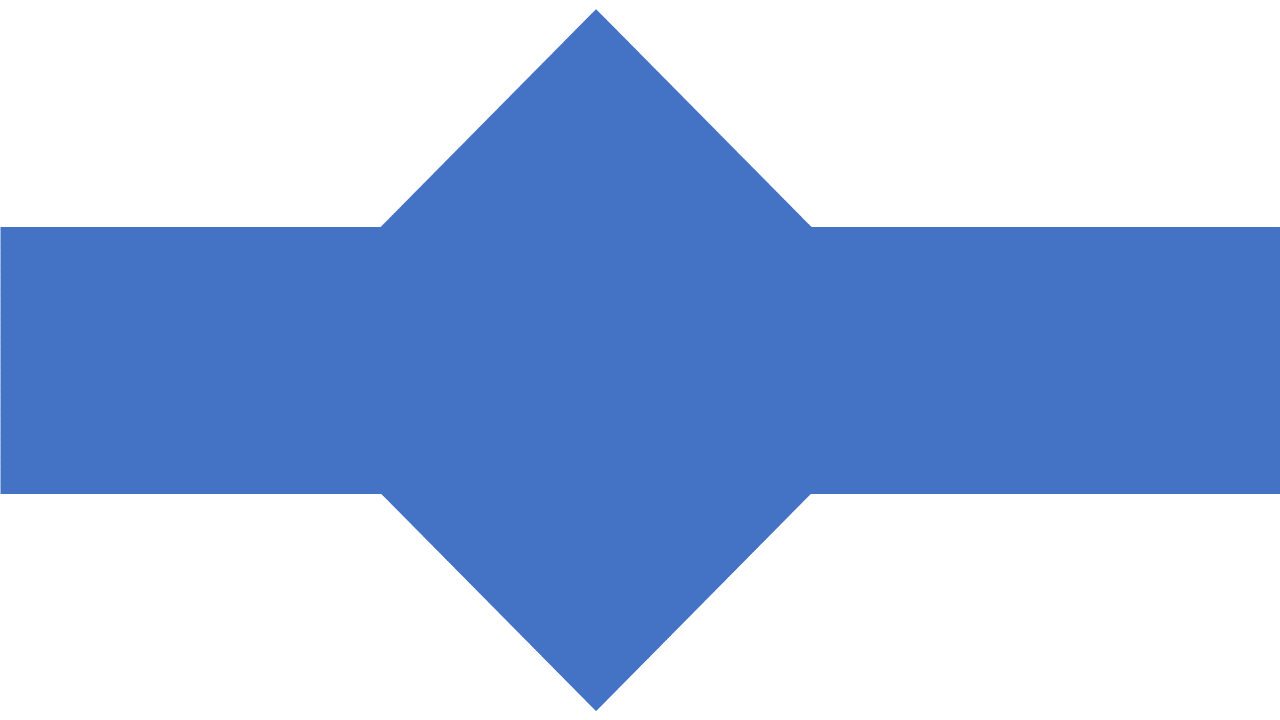}). In (b) is a Platt' calibrated classifier resulting in acceptable probability estimates $\bar{p}^1$ (\protect\includegraphics[height=0.5em]{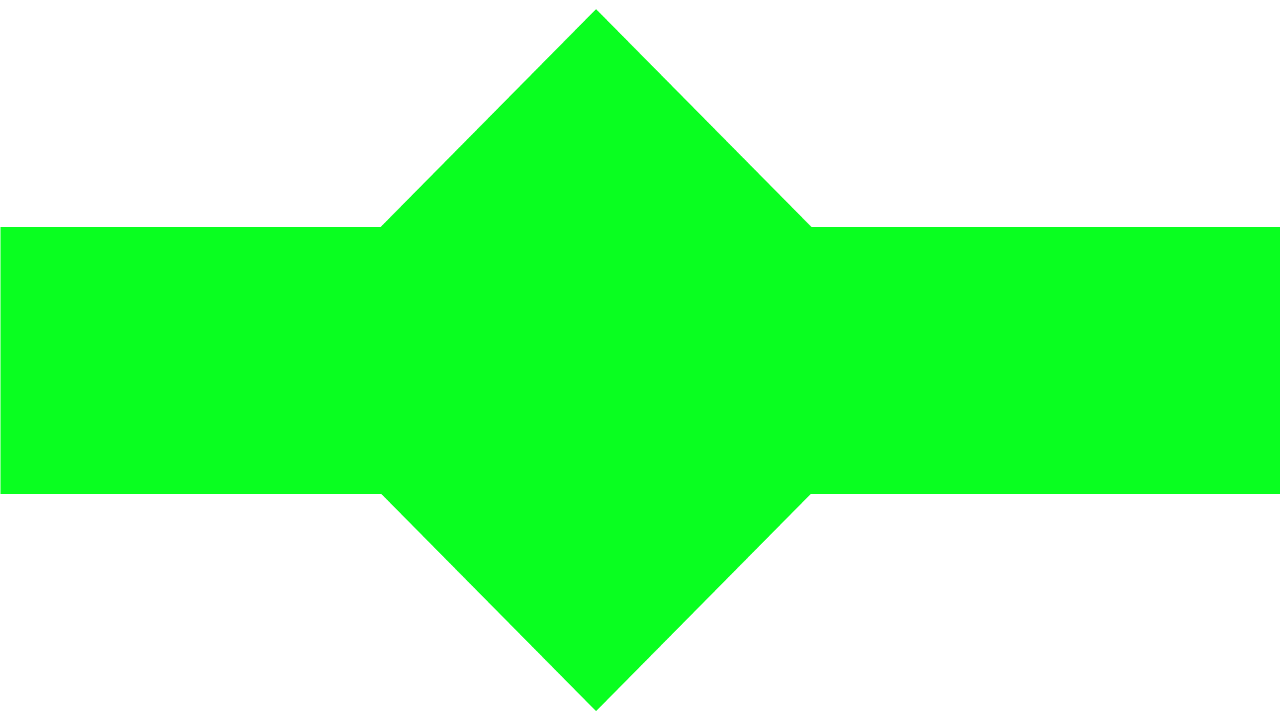}). Optimal probability estimates would follow the diagonal (\protect\includegraphics[height=0.5em]{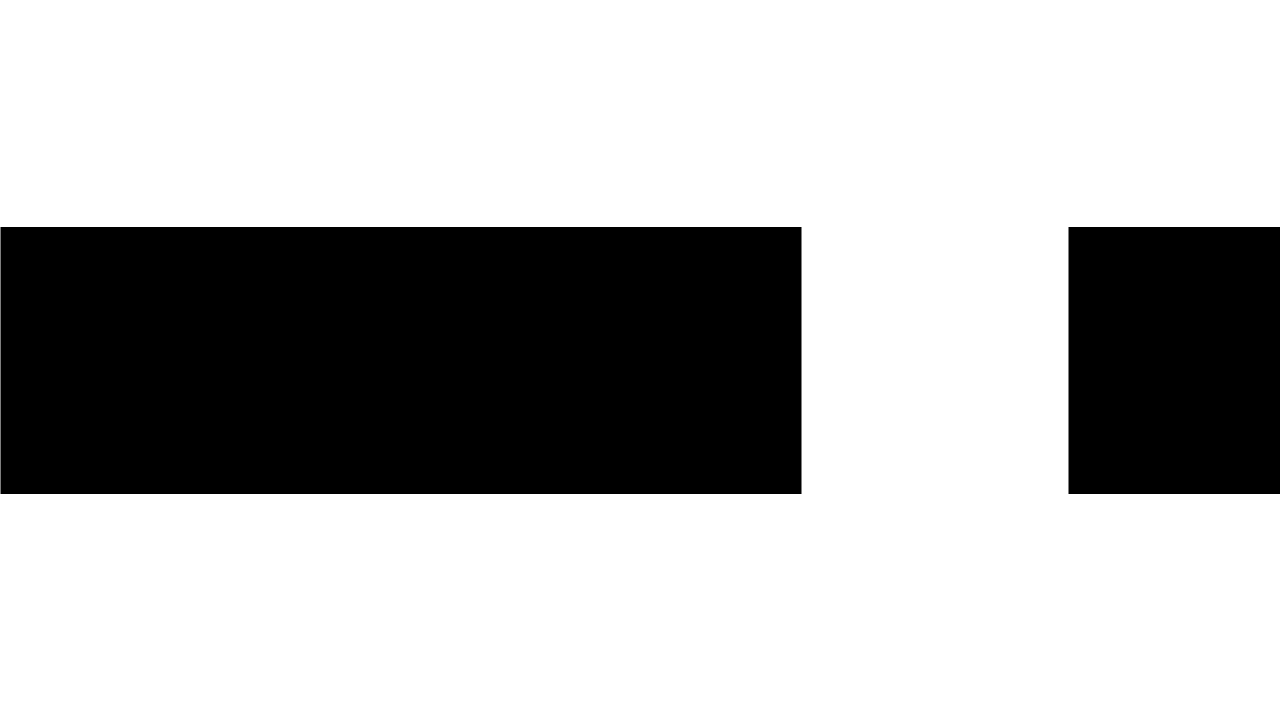}).
  }\label{fig:pe}
\end{figure}

\subsection{Calibrating imbalances}
An uncalibrated and calibrated AdaBoost classifiers were compared for contingency $c=6$ to show how to address imbalances. First, the uncalibrated scores $s^1(x_i)$ were obtained for the testing set $i \in \Omega^P$, then sorted and separated into $N$ subsets as described in Section \ref{sec:cal}. $\bar{s}^{1}_n$ was the average score in each subset $n = 1,2,3,... N$. The reliability diagram was computed based on $\bar{s}^{1}_n$ and the fraction of secure operating conditions $\pi^{1}_n$ and subsequently plotted in Fig. \ref{fig:peuncal}, where the index $n$ was dropped for simplicity reasons. Then, calibration was performed using $|\Omega^K|=875$ operating conditions and the average probability estimates $\bar{p}^1_n$ were computed for each subset $n$. Accordingly, the reliability diagram for the calibrated case was plotted in Fig. \ref{fig:pecal}. The calibrated reliability diagram aligns significantly stronger with the diagonal corresponding to the perfect probability estimates. The Brier scores for the two cases were computed $10$ times using Eq. \eqref{eq:brier}. The average Brier score of the uncalibrated case was $\bar{B}= 0.077$ and of the calibrated case $\bar{B}= 0.003$. This reduction in Brier score of $\SI{96}{\percent}$ showed the improvement of computing accurate probability estimates when using calibration.

\subsection{Cost-effective predictions}
In this study, the decision threshold $z_c$ from Eq. \eqref{eq:threshold} was used to make cost-effective predictions accordingly to the imbalances classes and the cost skewness; different cost imbalances were studied. The probabilities were equalled across the operating conditions and the probability of the contingency equalled the class prior. The actual security assessments were compared with the predictions to compute the missed ${N}^1_6$ and false alarms ${N}^0_6$. These were then used to compute the residual risk of using machine learning ${Z}^*_6$ for each cost imbalance according to Eq. \eqref{eq:costpred}. {Note that in this study machine learning is not combined with conventional security assessment, hence $\Omega^{S,L}=\Omega^{S}$ and $\Omega^{S,H}=\{\}$ is empty. }

\begin{figure}
\centering
\begin{tikzpicture}
\small
\begin{axis}[
        axis on top,
        width=0.74\textwidth,
        axis equal image,      
        enlargelimits=false, 
        ytick={0,1,2,3,4},
        yticklabels={$0.00001$,$0.0001$,$0.001$,$0.01$,$0.1$},  
        xtick={0,1,2,3,4,5,6},
        xticklabels={$\frac{2}{3}$, $\frac{5}{6}$, $\frac{10}{11}$, $\frac{50}{51}$, $\frac{100}{101}$, $\frac{500}{501}$, $\frac{1000}{1001}$},
        ylabel={${Z}^*_6$},
        xlabel={$\mathcal{C}_6$},
        ]
        \addplotgraphicsnatural [xmin=0, xmax=6, ymin=0, ymax=4] {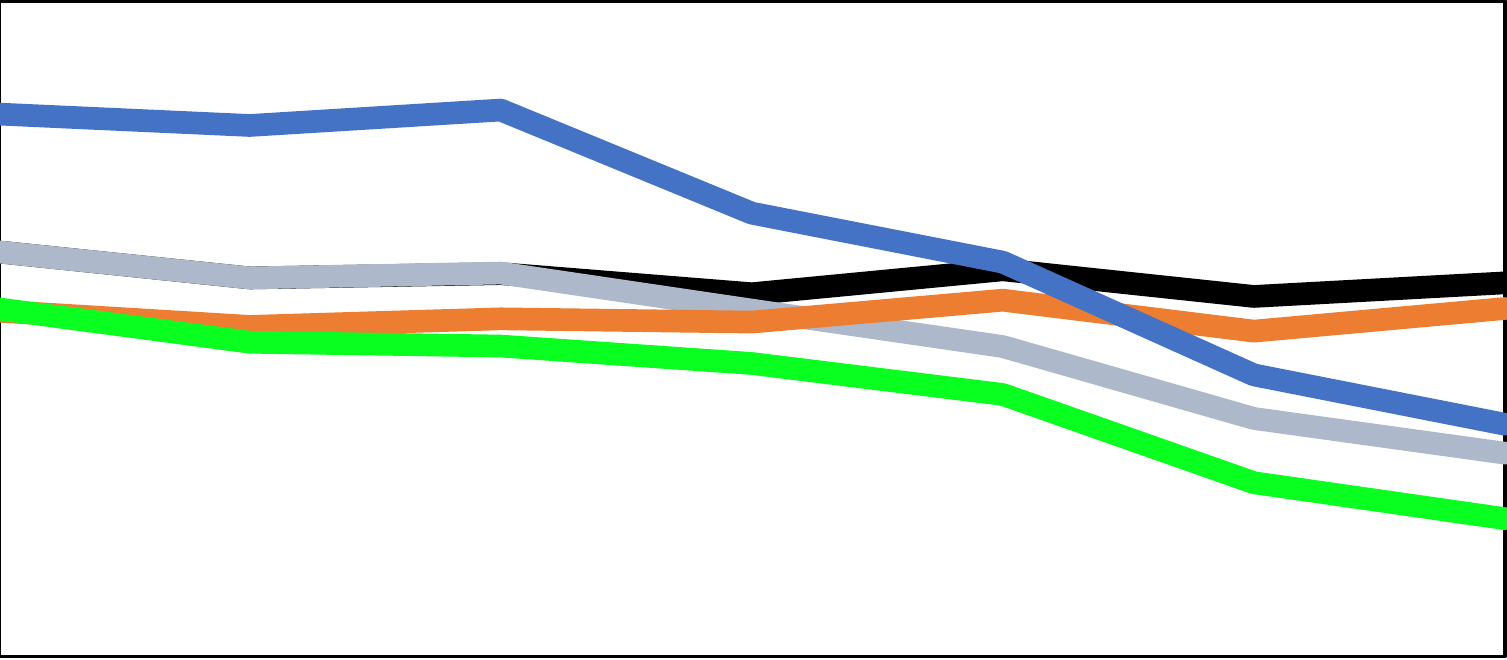};
	 \end{axis}
\end{tikzpicture} 
\caption{Using the decision threshold $z$ reduces the risk ${Z}^*_6$ if used correctly on a calibrated AdaBoost classifier (\protect\includegraphics[height=0.5em]{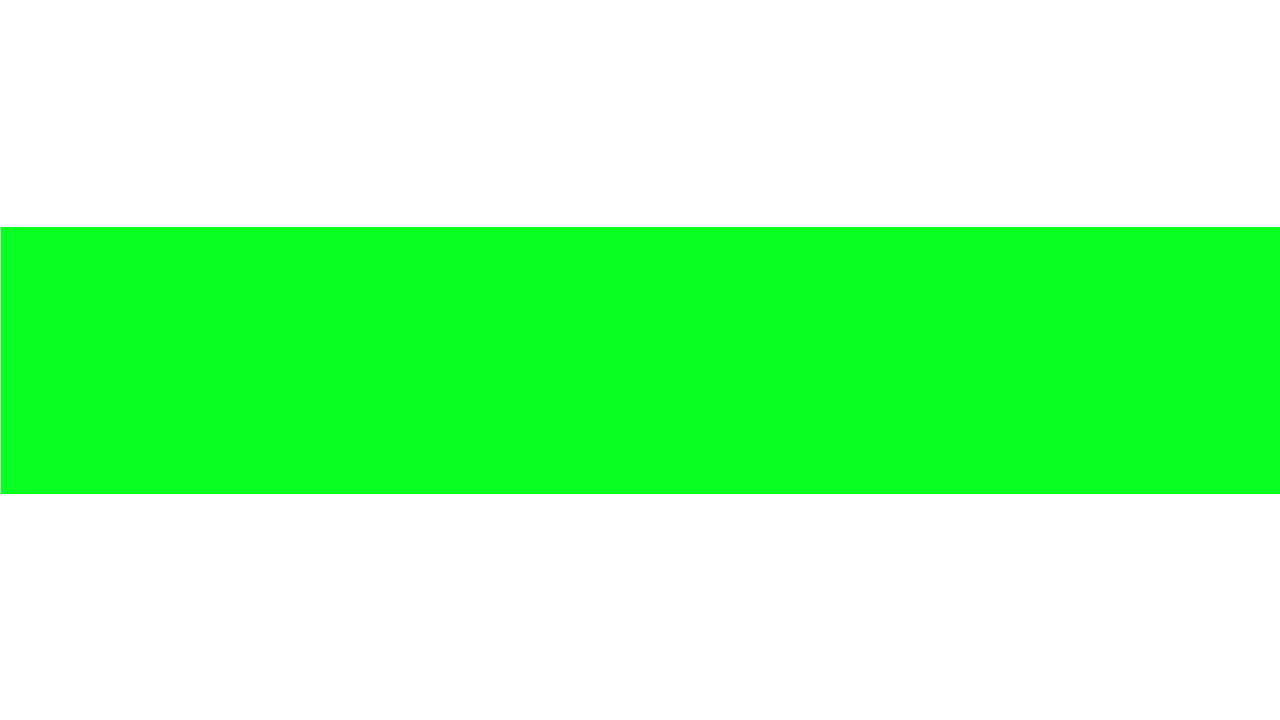}). Different cost ratios $\mathcal{C}_6$ and classifiers were used: DT without (\protect\includegraphics[height=0.5em]{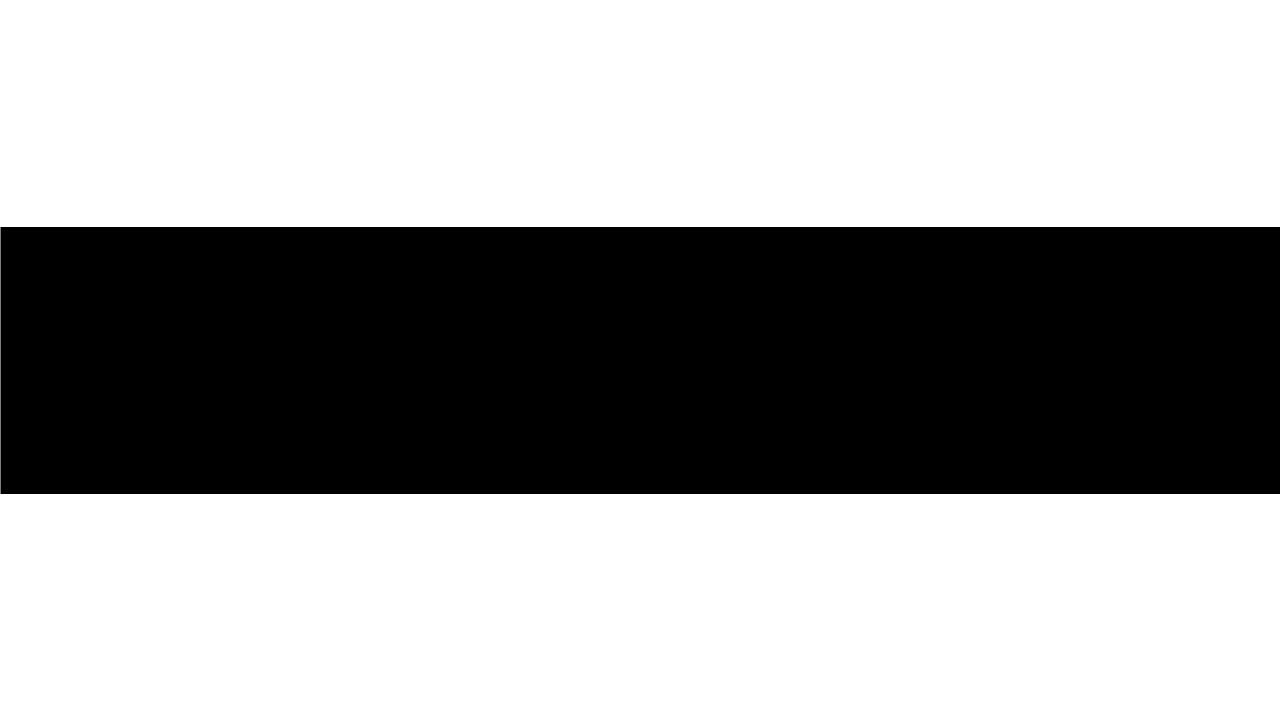}) and DT with threshold $z$ (\protect\includegraphics[height=0.5em]{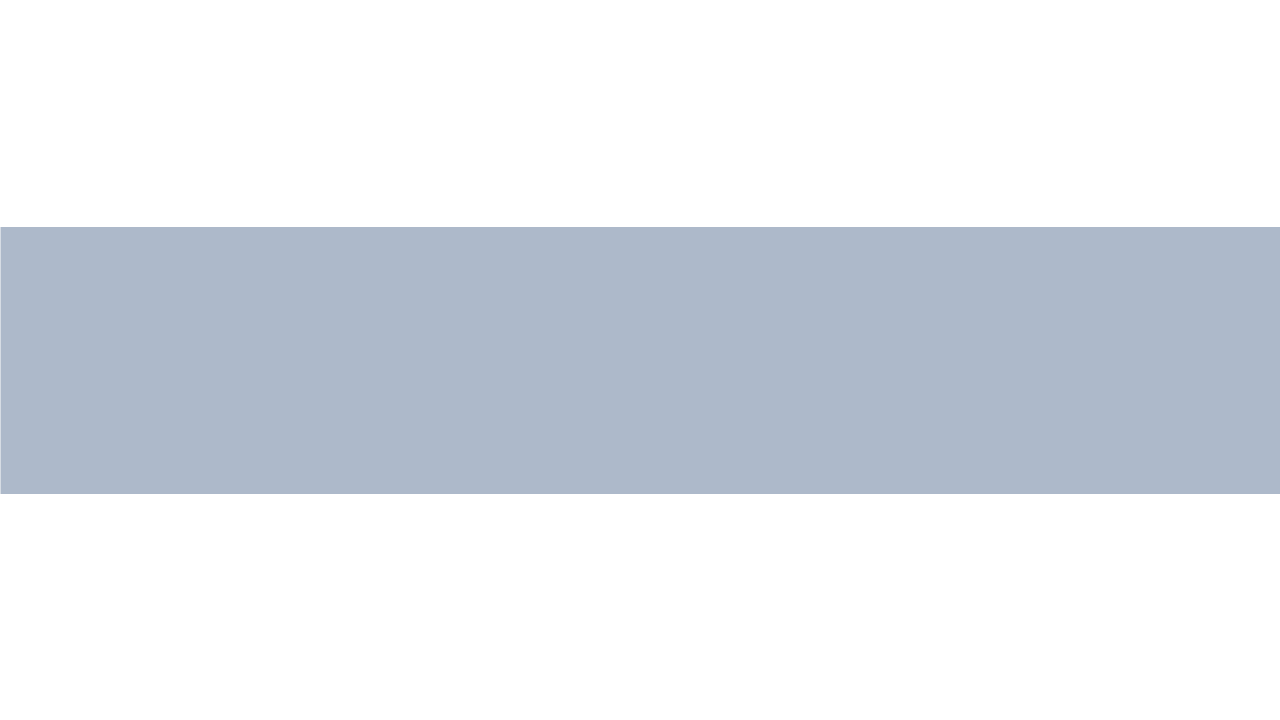}), uncalibrated AdaBoost without (\protect\includegraphics[height=0.5em]{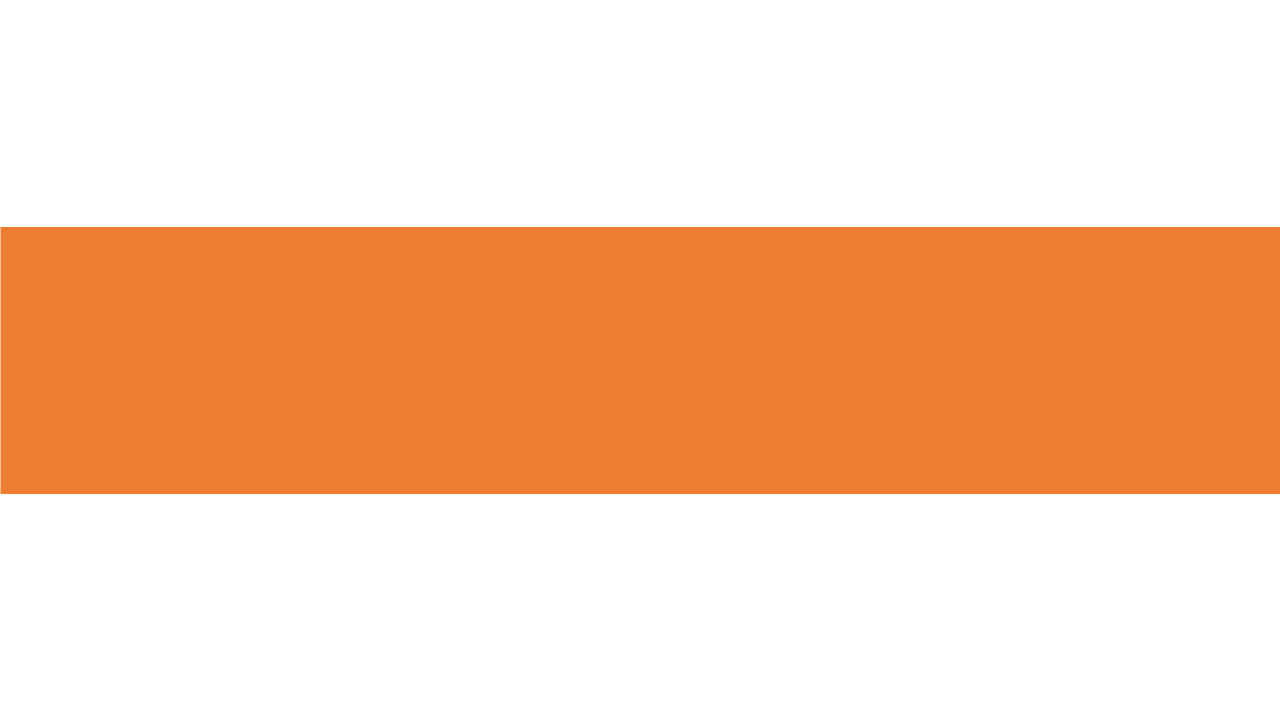}) and with $z_6$ (\protect\includegraphics[height=0.5em]{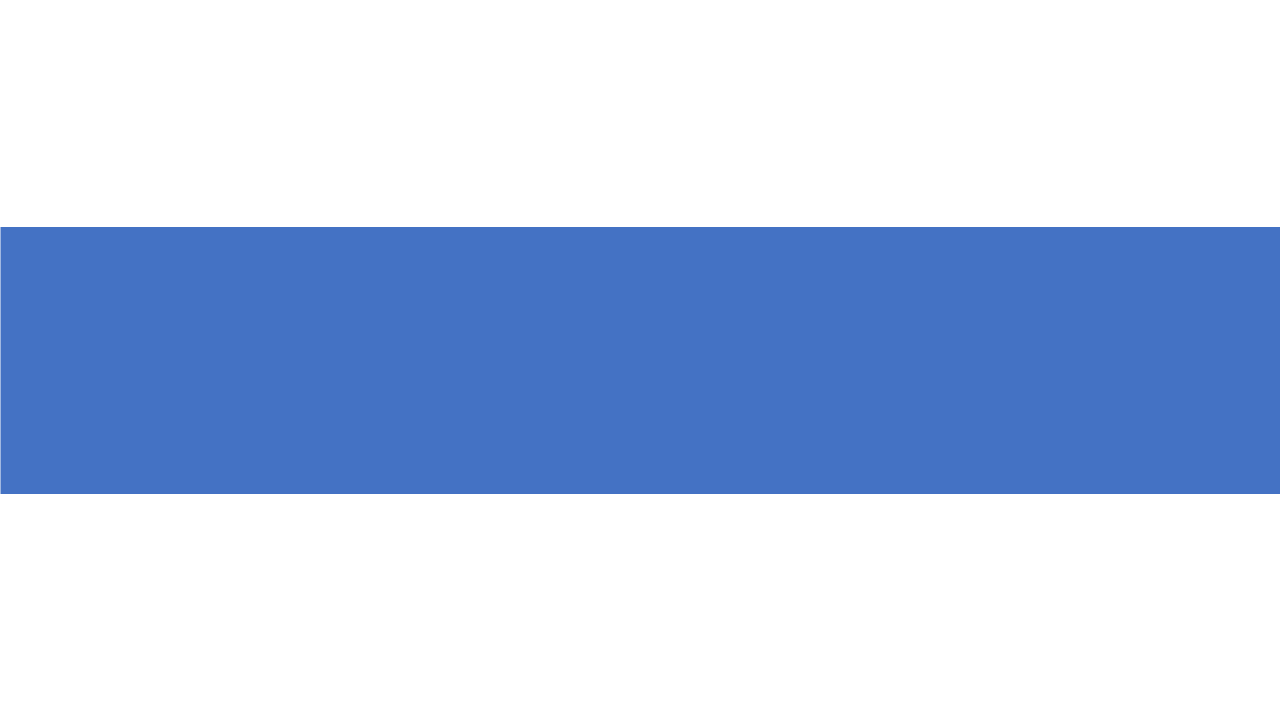}).}
\label{fig:decthres}
\end{figure}

This procedure was repeated $10$ times and the risk was averaged $\overline{{Z}^*_6}$. The results in Fig. \ref{fig:decthres} showed that combining a decision threshold with a calibrated classifier reduced risks; however, when using the threshold on the distorted score $s^1$ of an uncalibrated classifier the risk is high. The risk is particularly high for high-cost ratios of $\mathcal{C}$. Hence, the larger the imbalance the more important it becomes to calibrate and us the decision thresholds.

\begin{figure}
\centering
\begin{subfigure}[b]{0.48\textwidth}
\begin{tikzpicture}
\small
\begin{axis}[
        axis on top,
        width=0.75\textwidth,
        scale only axis,
        enlargelimits=false, 
        ytick={0,100,200,300},     
        xtick={0,500,1000,1500},    
        ylabel={$N^1_3 + N^0_3$},
        xlabel={$S$},
        xmin=0,
        xmax=1500,
        ymin=0,
        ymax=300,
        ]        
	\addplot graphics[xmin=0,ymin=0,xmax=1500,ymax=300] {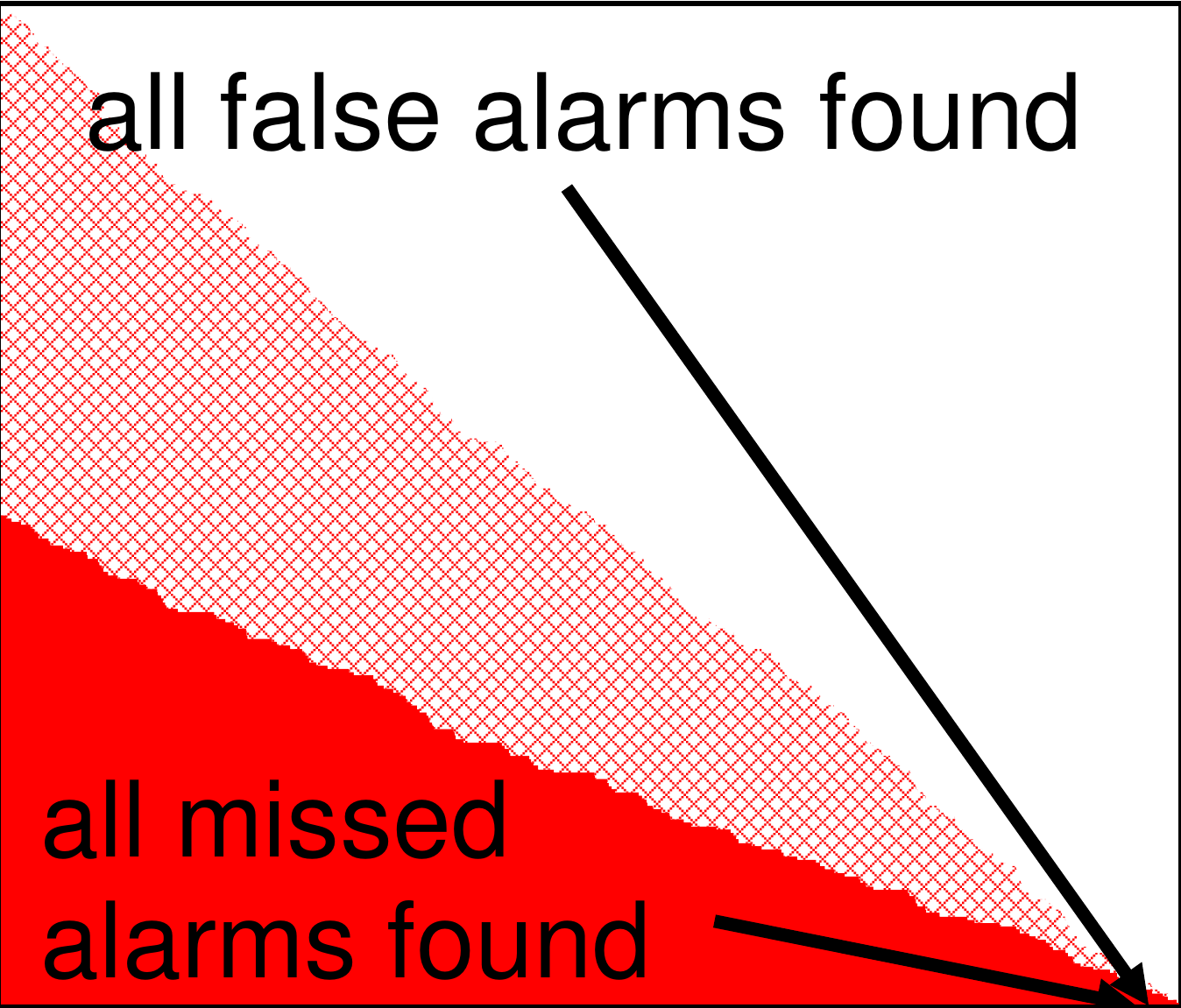};
  \end{axis}
\end{tikzpicture}
	\caption{Missed (\includegraphics[height=0.5em]{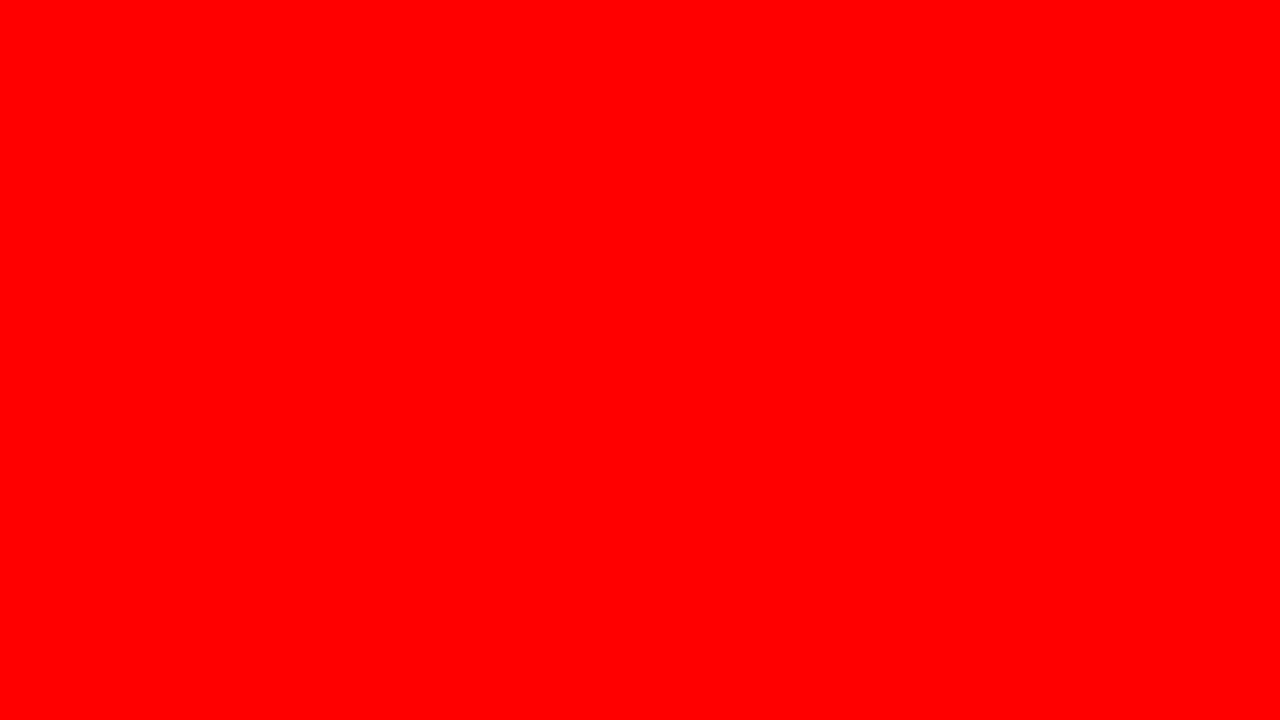}) and false (\includegraphics[height=0.5em]{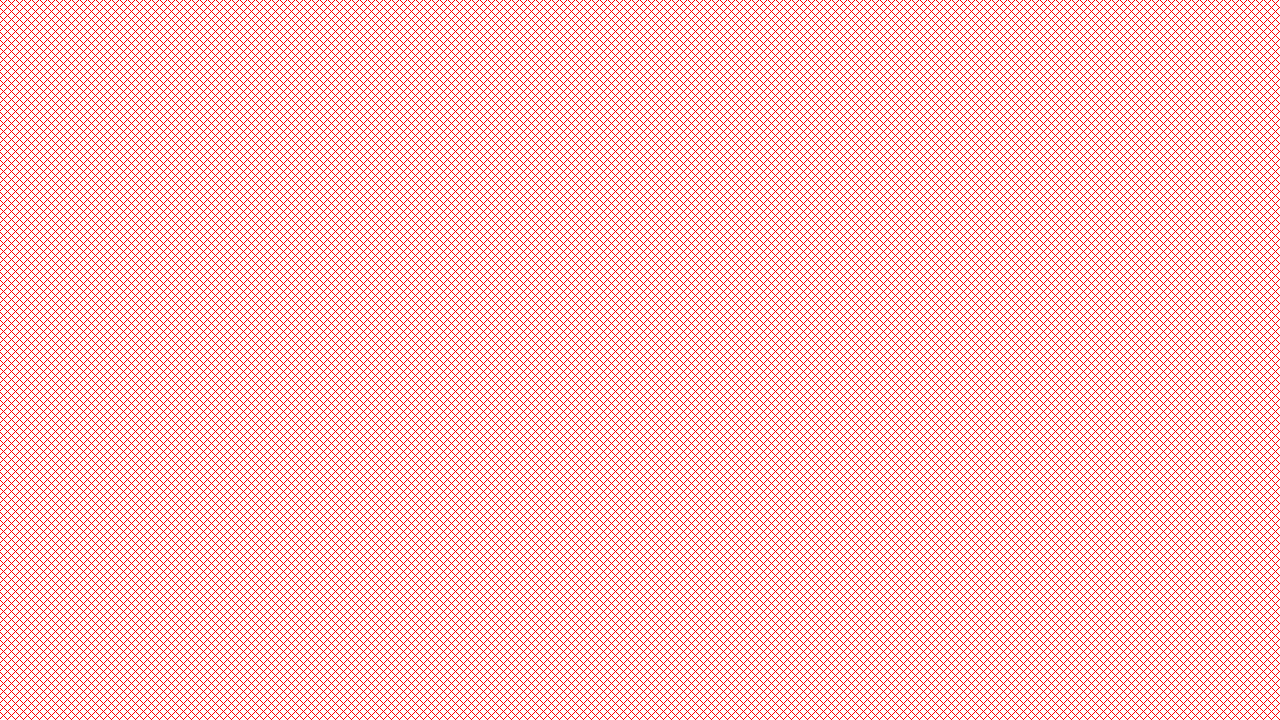}) alarms.}
   \label{fig:noml_alarms} 
\end{subfigure}\hspace{1em}
\begin{subfigure}[b]{0.48\textwidth}
\begin{tikzpicture}
\small
\begin{axis}[
        axis on top,
        width=0.75\textwidth,
        scale only axis,
        enlargelimits=false, 
        ytick={0,0.00001,0.00002,0.00003,0.00004},     
        xtick={0,500,1000,1500},    
        ylabel={${Z}_3$},
        xlabel={$S$},
        xmin=0,
        xmax=1500,
        ymin=0,
        ymax=0.00004,
    	]
	\addplot graphics[xmin=0,ymin=0,xmax=1500,ymax=0.00004] {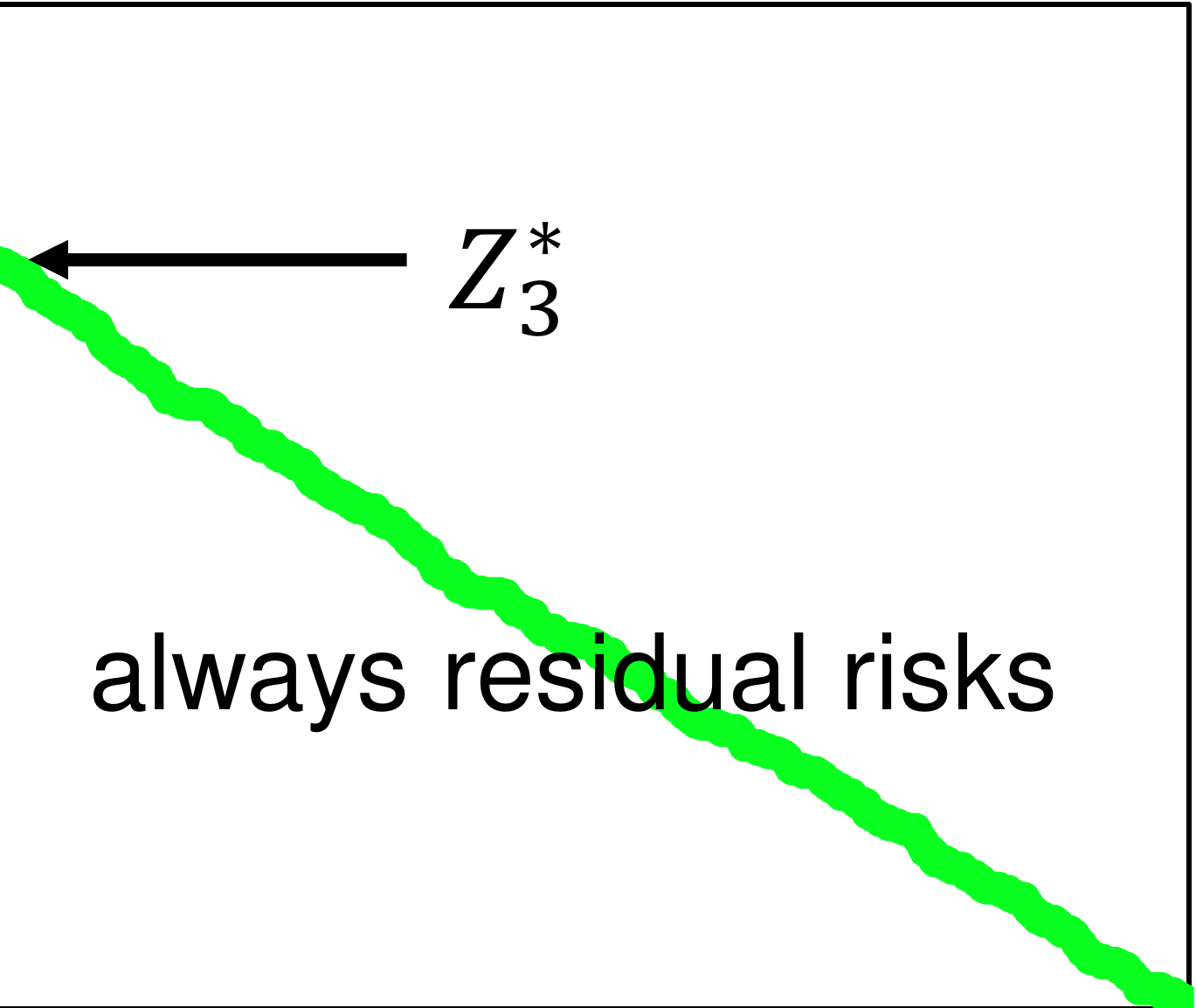};
  \end{axis}
\end{tikzpicture}]
   \caption{The residual risk (\includegraphics[height=0.5em]{grl.png}).} 
   \label{fig:noml_risk} 
\end{subfigure}
\caption{{The sole use of conventional security assessment on a subset of operating conditions. However, not all operating conditions can be assessed due to limited computational budgets. The classes of the remaining conditions are assigned based on class distributions.} (a) are the inaccurately assigned classes and in (b) is the corresponding risk.}\label{fig:plot_proper1}
\end{figure}

\begin{figure}
\centering
\begin{subfigure}[b]{0.48\textwidth}
\begin{tikzpicture}
\small
\begin{axis}[
        axis on top,
        width=0.75\textwidth,
        scale only axis,
        enlargelimits=false, 
        ytick={0,10,20,30},     
        xtick={0,500,1000,1500},    
        ylabel={$N^1_3 +N^0_3$},
        xlabel={$S$},
        xmin=0,
        xmax=1500,
        ymin=0,
        ymax=30,
        ]        
	\addplot graphics[xmin=0,ymin=0,xmax=1500,ymax=30] {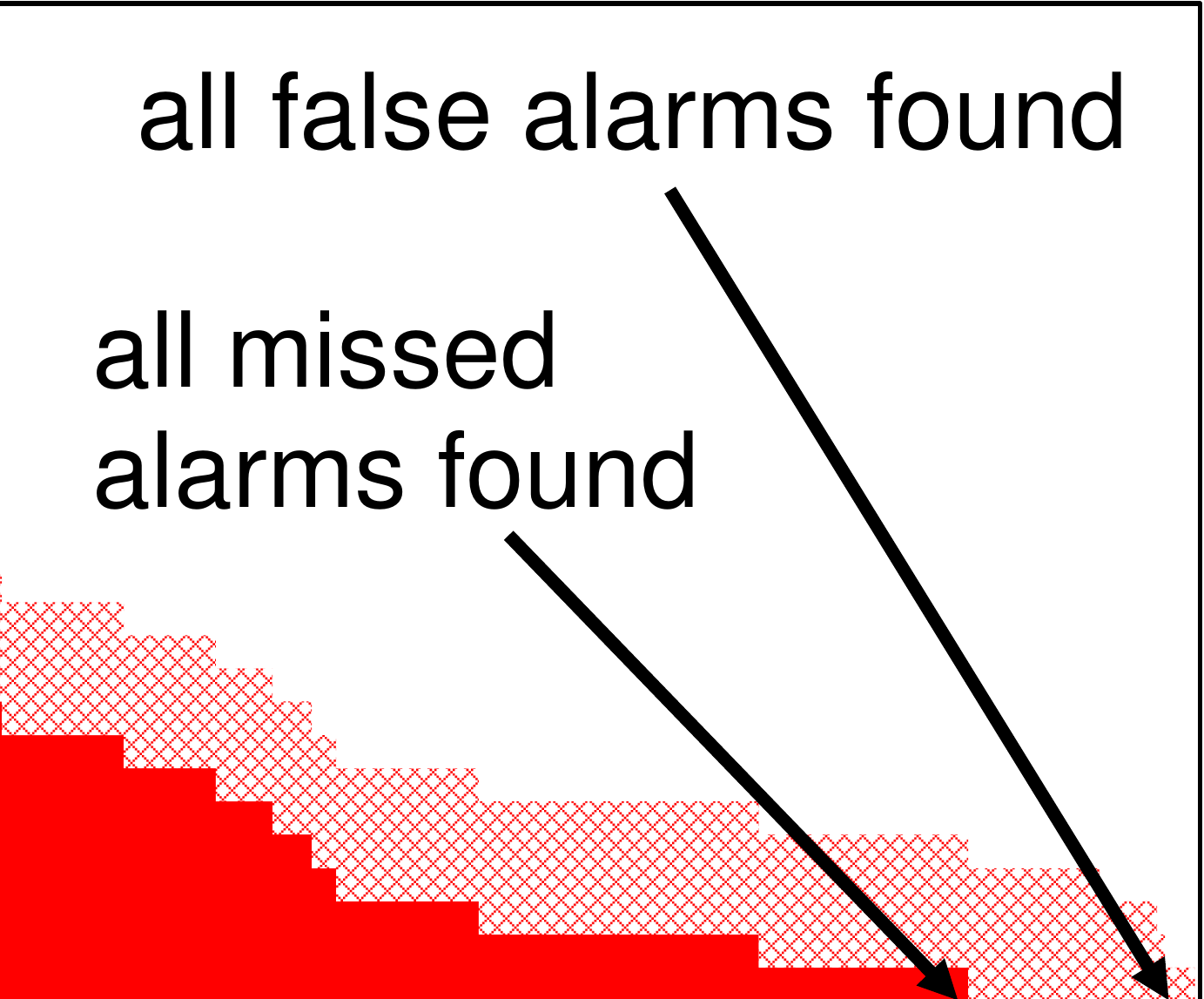};
  \end{axis}
\end{tikzpicture}
	\caption{Missed (\includegraphics[height=0.5em]{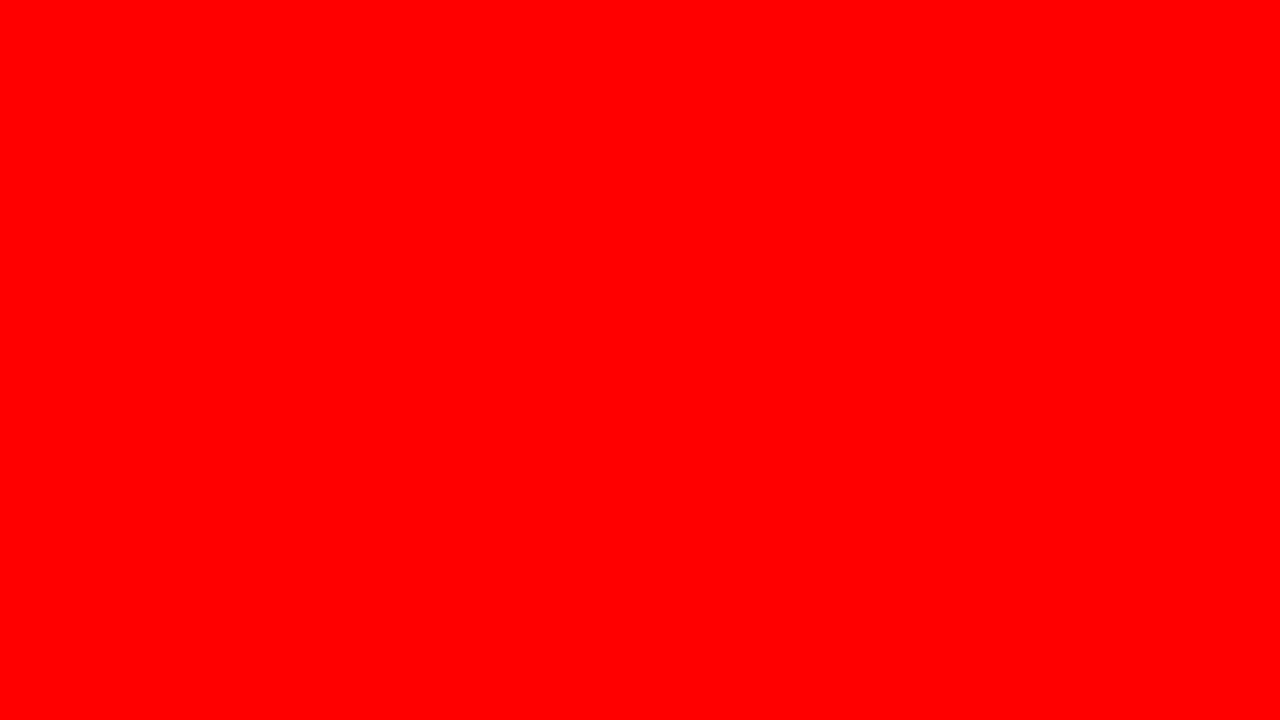}) and false (\includegraphics[height=0.5em]{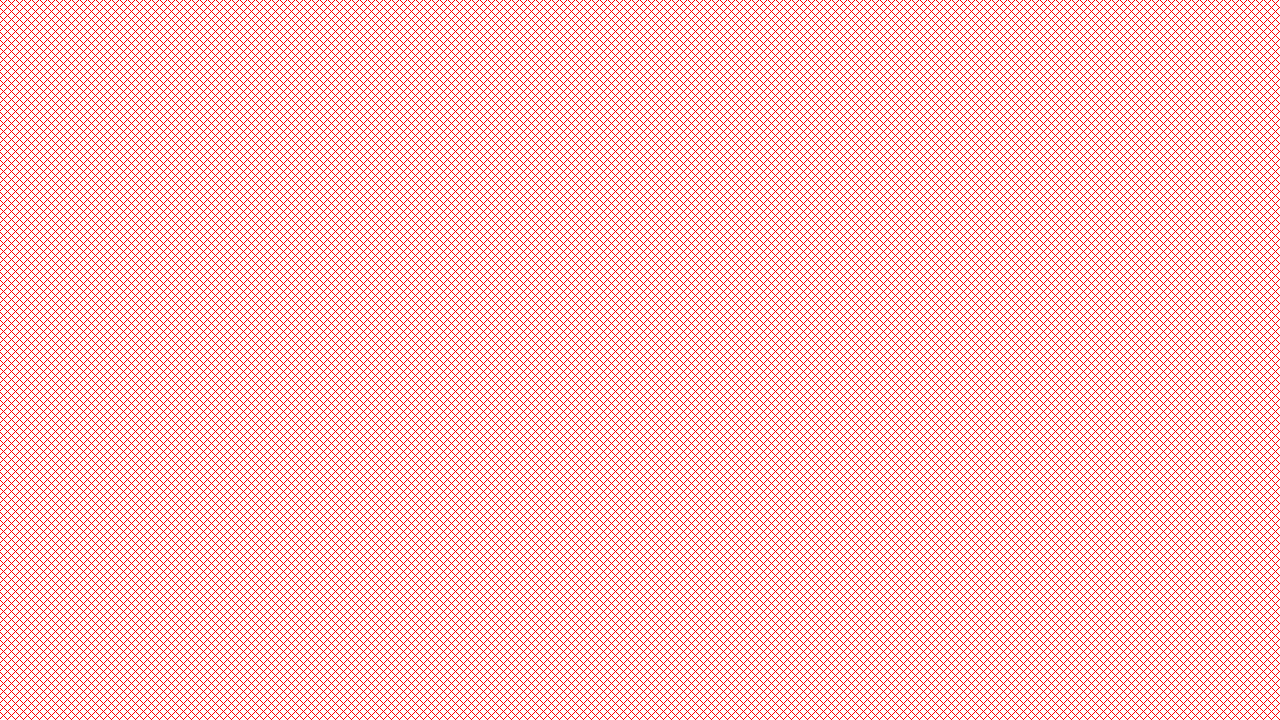}) alarms.}
   \label{fig:proper1_alarms_alt} 
\end{subfigure}\hspace{1em}
\begin{subfigure}[b]{0.48\textwidth}
\begin{tikzpicture}
\small
\begin{axis}[
        axis on top,
        width=0.75\textwidth,
        scale only axis,
        enlargelimits=false, 
        ytick={0,0.000001,0.000002,0.000003,0.000004},     
        xtick={0,500,1000,1500},    
        ylabel={${Z}_3$},
        xlabel={$S$},
        xmin=0,
        xmax=1500,
        ymin=0,
        ymax=0.000004,
        ]        
	\addplot graphics[xmin=0,ymin=0,xmax=1500,ymax=0.000004] {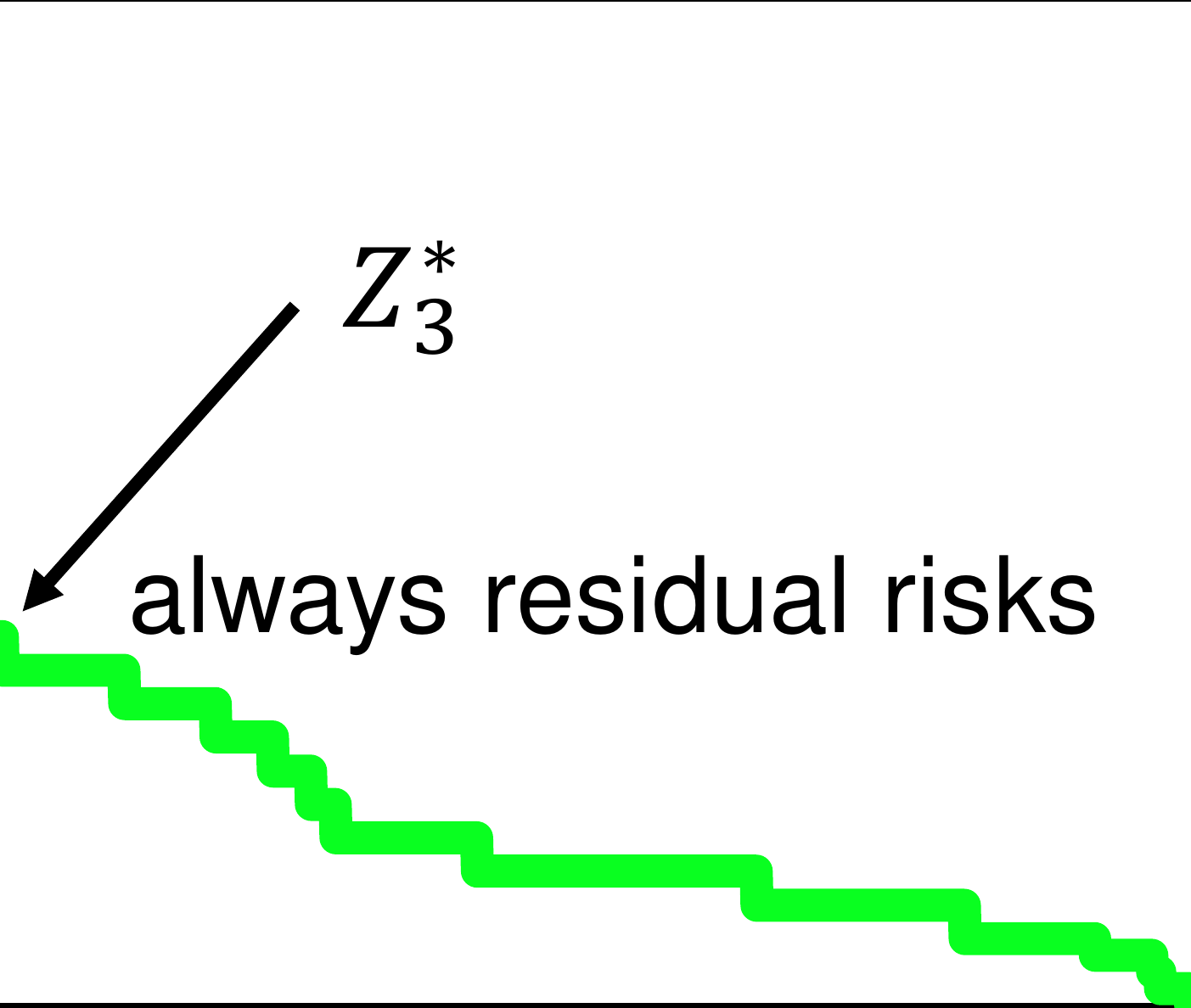};
  \end{axis}
\end{tikzpicture}
   \caption{The residual risk (\includegraphics[height=0.5em]{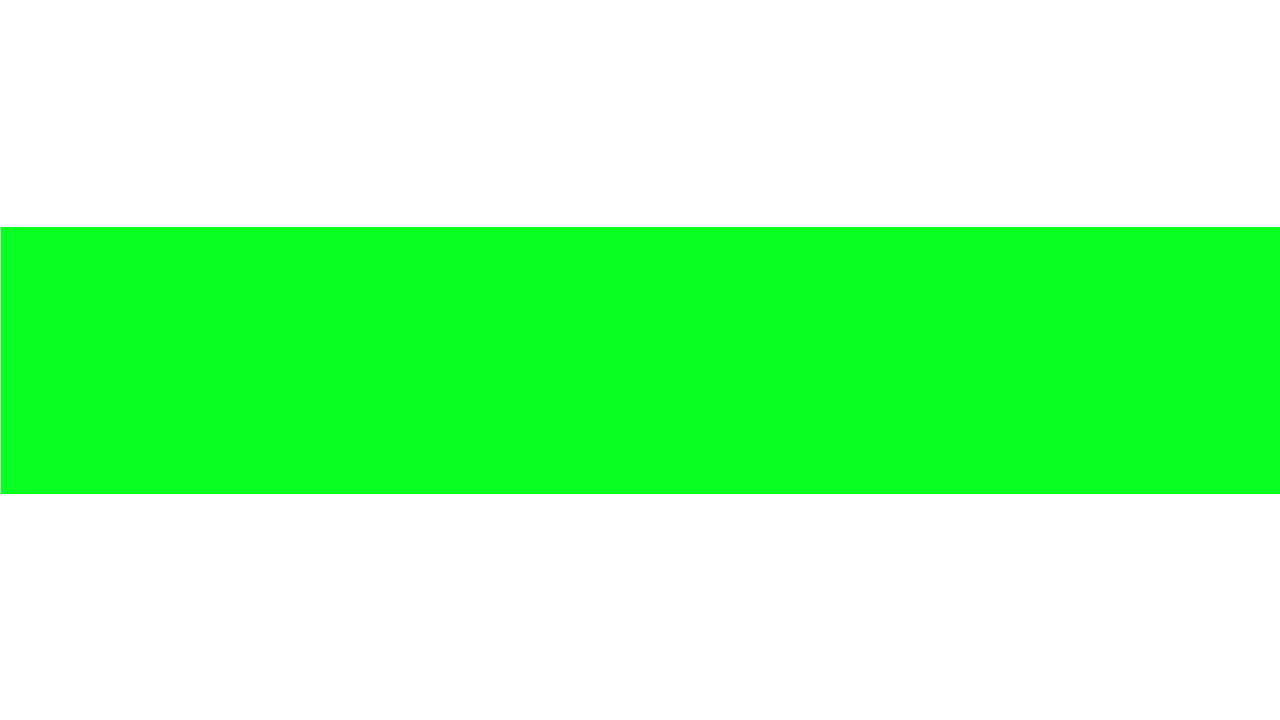}).}
   \label{fig:proper1_risk_alt} 
\end{subfigure}
\begin{subfigure}[b]{0.48\textwidth}
\begin{tikzpicture}
\small
\begin{axis}[
        axis on top,
        width=0.75\textwidth,
        scale only axis,
        enlargelimits=false, 
        ytick={0,10,20,30},     
        xtick={0,500,1000,1500},    
        ylabel={$N^1_3 +N^0_3$},
        xlabel={$S$},
        xmin=0,
        xmax=1500,
        ymin=0,
        ymax=30,
        ]          
	\addplot graphics[xmin=0,ymin=0,xmax=1500,ymax=30] {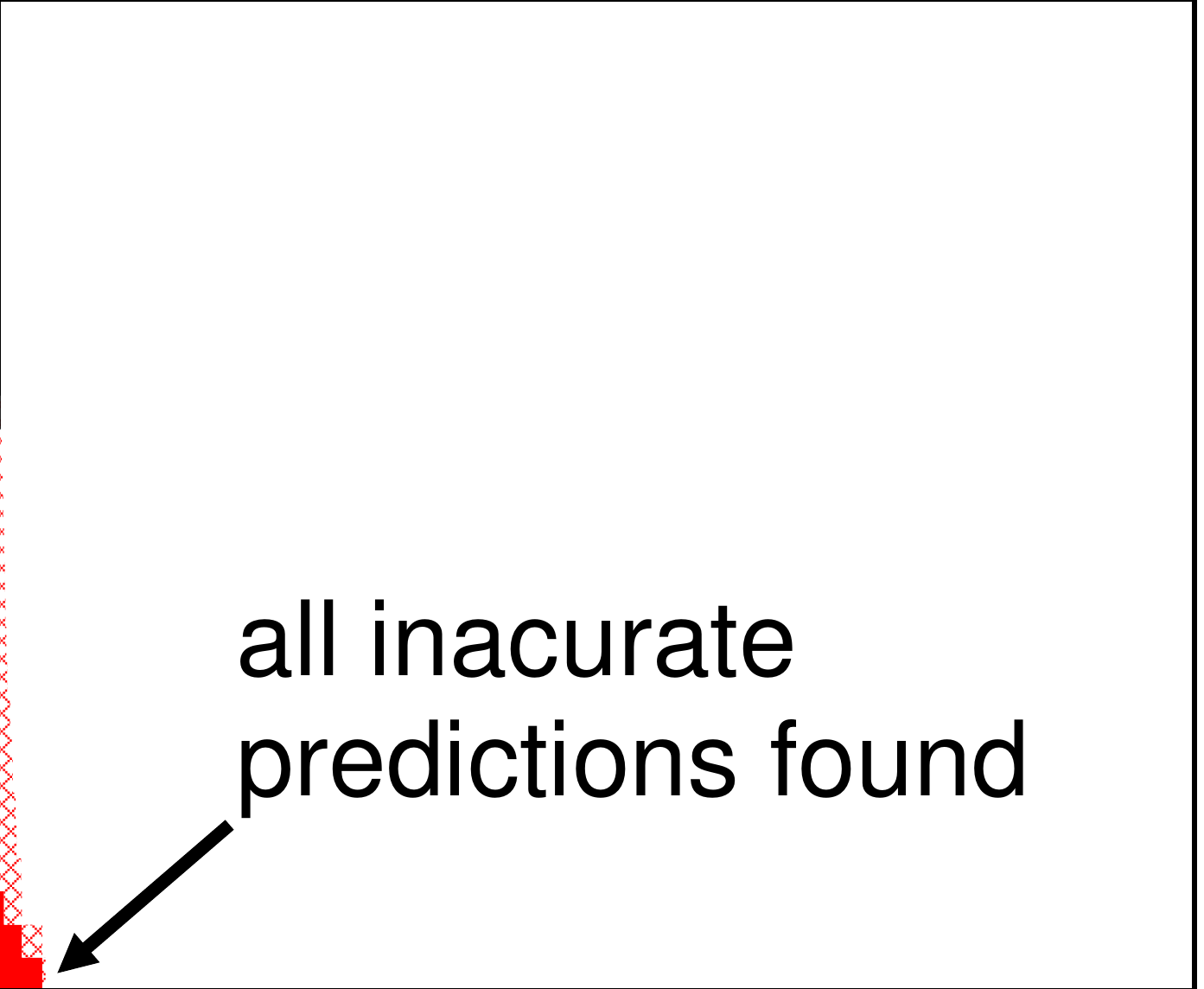};
  \end{axis}
\end{tikzpicture}
   \caption{Missed (\includegraphics[height=0.5em]{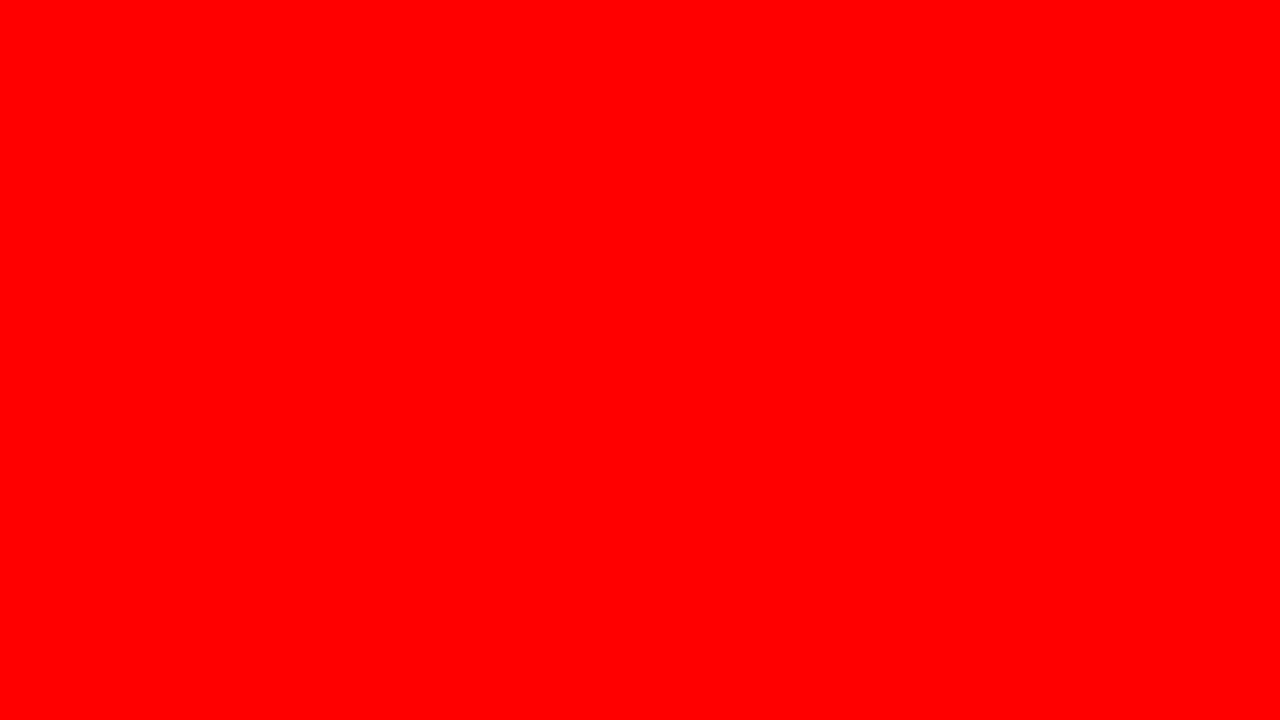}) and false (\includegraphics[height=0.5em]{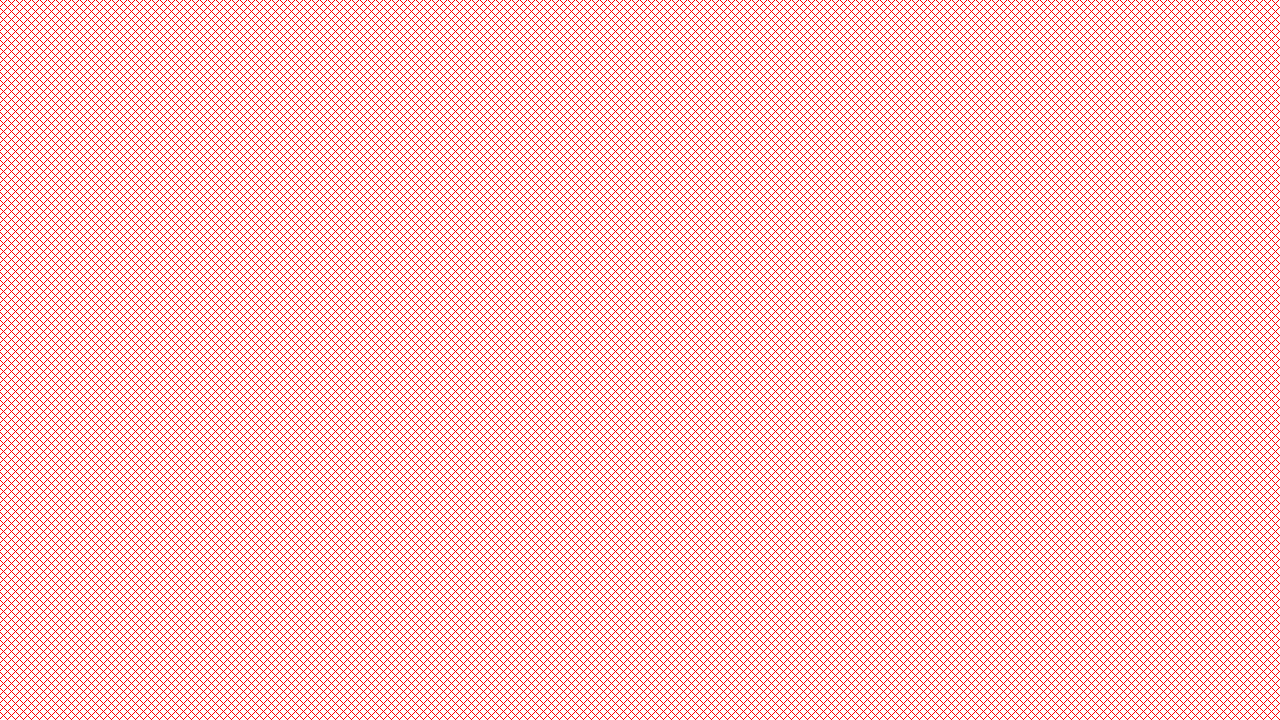}) alarms.}
   \label{fig:proper1_alarms}
\end{subfigure}\hspace{1em}
\begin{subfigure}[b]{0.48\textwidth}
\begin{tikzpicture}
\small
\begin{axis}[
        axis on top,
        width=0.75\textwidth,
        scale only axis,
        enlargelimits=false, 
        ytick={0,0.000001,0.000002,0.000003,0.000004},     
        xtick={0,500,1000,1500},    
        ylabel={${Z}_3$},
        xlabel={$S$},
        xmin=0,
        xmax=1500,
        ymin=0,
        ymax=0.000004,
        ]      
	\addplot graphics[xmin=0,ymin=0,xmax=1500,ymax=0.000004] {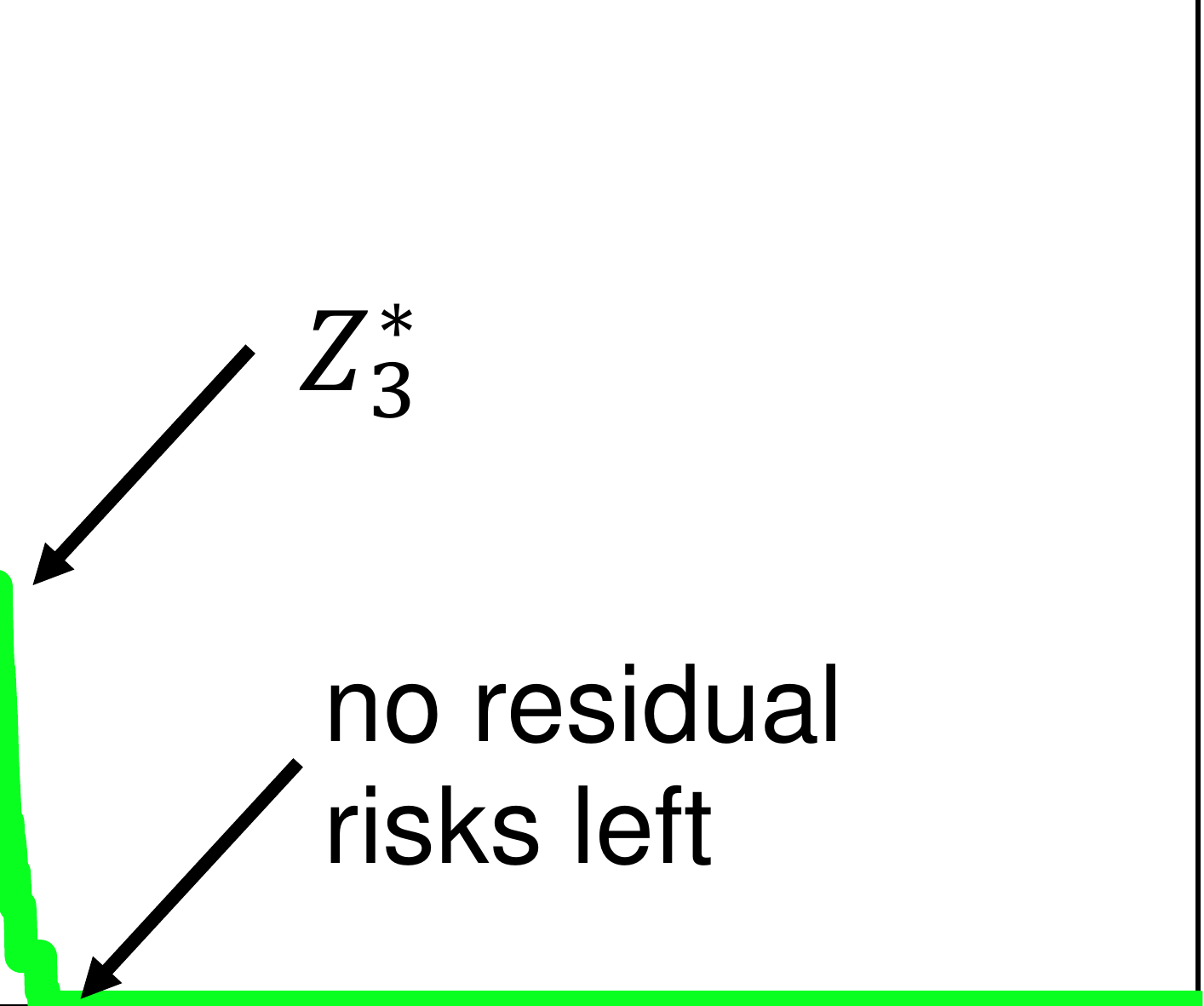};
  \end{axis}  
\end{tikzpicture}
   \caption{The residual risk (\includegraphics[height=0.5em]{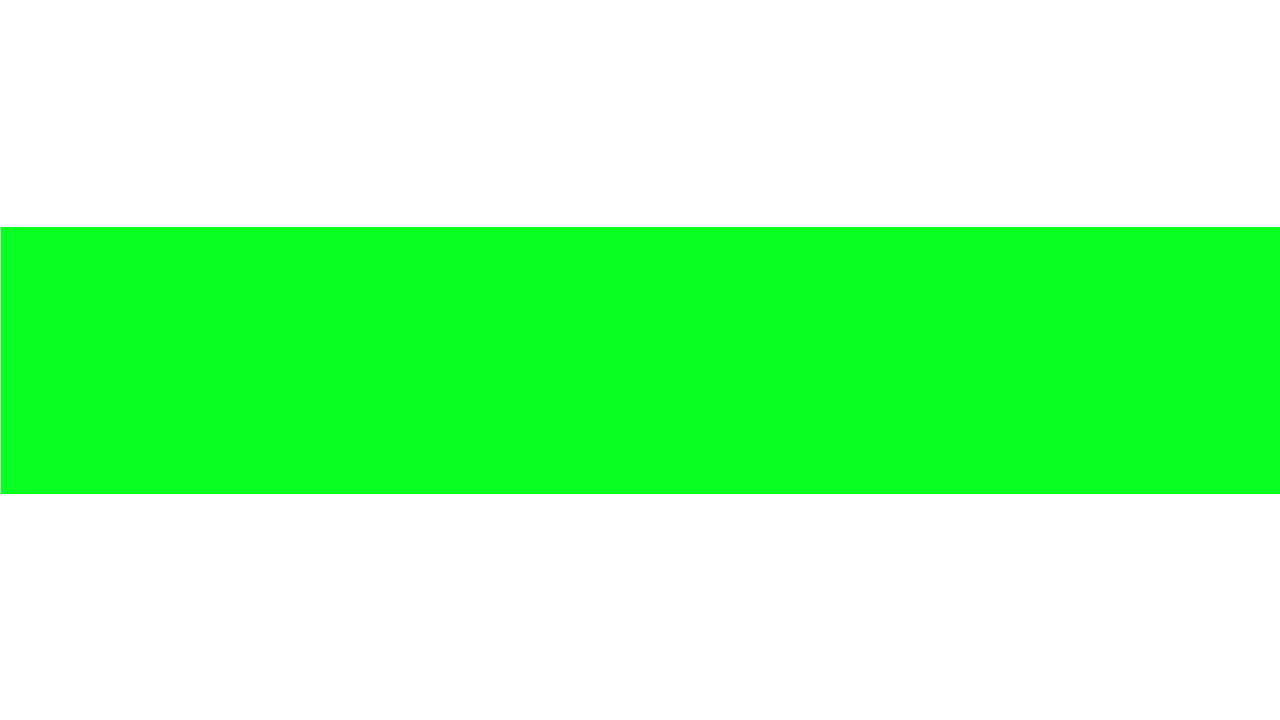}).}
   \label{fig:proper1_risk}
\end{subfigure}
\caption{{Combining the machine learning model and conventional security assessment for a single contingency. In (a) and (b), a standard classifier is used for identifying prioritising to identify missed alarms over false alarms, and (c) and (d) is the proposed probabilistic approach.} \vspace{-2.5em}}\label{fig:plot_proper2}
\end{figure}

\subsection{Probabilistic security assessment with machine learning}
In this study, the operator has a limited computational budget to analyse $S$ scenarios with conventional security assessment close to real-time operation. Contingency $c=3$ (connecting buses $1$ and $5$) was studied from the IEEE 6-bus system with cost of $\mathcal{C}_3= \frac{10000}{10001}$ and probability $p_3^C = 0.0002$. {The testing set was $|\Omega^P| = 1500$. Three different approaches of how to combine machine learning with probabilistic security assessment were studied as follows.}

In the first approach, no machine learning was used. {The operator selects randomly $S$ scenarios to assess with conventional security assessment. The only information available to decide without assessment on the security of an operating condition is the class distribution. Therefore, the operator can only assume the security label from this class distribution and then select randomly a subset of $S$ operating conditions to verify with security assessment the assumption on the security.}
The inaccurately assigned classes ${N}^1_3 +{N}^0_3$ were computed for the scenarios and the risk was calculated from Eq. \eqref{eq:costpred}. The results in Fig. \ref{fig:noml_alarms} and Fig. \ref{fig:noml_risk} show the linear reduction in inaccurately assigned labels and the risk {the more computational budget for security assessments $S$ the operator has available.} A traditional operator following the N-1 criterion needs to assess all operating conditions.

In the second approach, a standard classifier was used to focus on missed alarms {as they typically entail higher costs than false alarms $C^{F1}_c \gg C^{F0}_c$. No decision threshold was applied in the machine learning model and the security assessment were performed on the operating conditions that were classified as secure as the objective is to reduce missed alarms ${N}^1_c$. The results for inaccurate predictions and risks are shown in Fig. \ref{fig:proper1_alarms_alt} and Fig. \ref{fig:proper1_risk_alt}, respectively. First ${N}^1_3$ dropped to zero at $1205$ security assessments, then ${N}^0_3$ approached zero at $1499$ security assessments. However, a residual risk remained at any $S$ unless all operating conditions are assessed with conventional security assessment.}

{The third is the proposed approach to use the probability estimate of machine learning in the probabilistic security assessment.} The calibrated AdaBoost classifier was used to estimate the risk of relying on these predictions and security assessment were performed accordingly {as described in Sec. \ref{sec:combML}.} The result of reductions in inaccurate predictions is shown in Fig. \ref{fig:proper1_alarms}. Initially ($S=0$), $20$ predictions out of $|\Omega^P| = 1500$ were inaccurate. Then, the proposed approach identified these $20$ inaccurate predictions after only $59$ security assessments. Any further of the $1441$ security assessments were not needed as not reducing inaccuracies or risks. The proposed approach reduced the number of security assessments needed in comparison to the first approach by $\SI{96}{\percent}$. The proposed approach reduced risks with a steep slope and approached zero after these $59$ security assessments as per Fig. \ref{fig:proper1_risk}. This was a key finding as a traditional operator using the N-1 criterion could significantly reduce the computational effort to analyse each fault while ensuring the same risk tolerance level.

\subsection{Considering several contingencies}
{In this study, the performance of the proposed approach was tested for two test systems. First, the IEEE 6-bus system was used for two contingencies to illustrate the approach. Then, the French Transmission grid was used to showcase the benefits for multiple contingencies. }

\begin{figure}
\centering
\begin{subfigure}[b]{0.48\textwidth}
\begin{tikzpicture}
\small
\begin{axis}[
        axis on top,
        width=0.75\textwidth,
        scale only axis,
        enlargelimits=false, 
        ytick={0,10,20,30},     
        xtick={0,50,100,150},    
        ylabel={${N}^1_3 + {N}^0_3 + {N}^1_5 + {N}^0_5$},
        xlabel={$S$},
        xmin=0,
        xmax=150,
        ymin=0,
        ymax=30,
        ]          
	\addplot graphics[xmin=0,ymin=0,xmax=150,ymax=30] {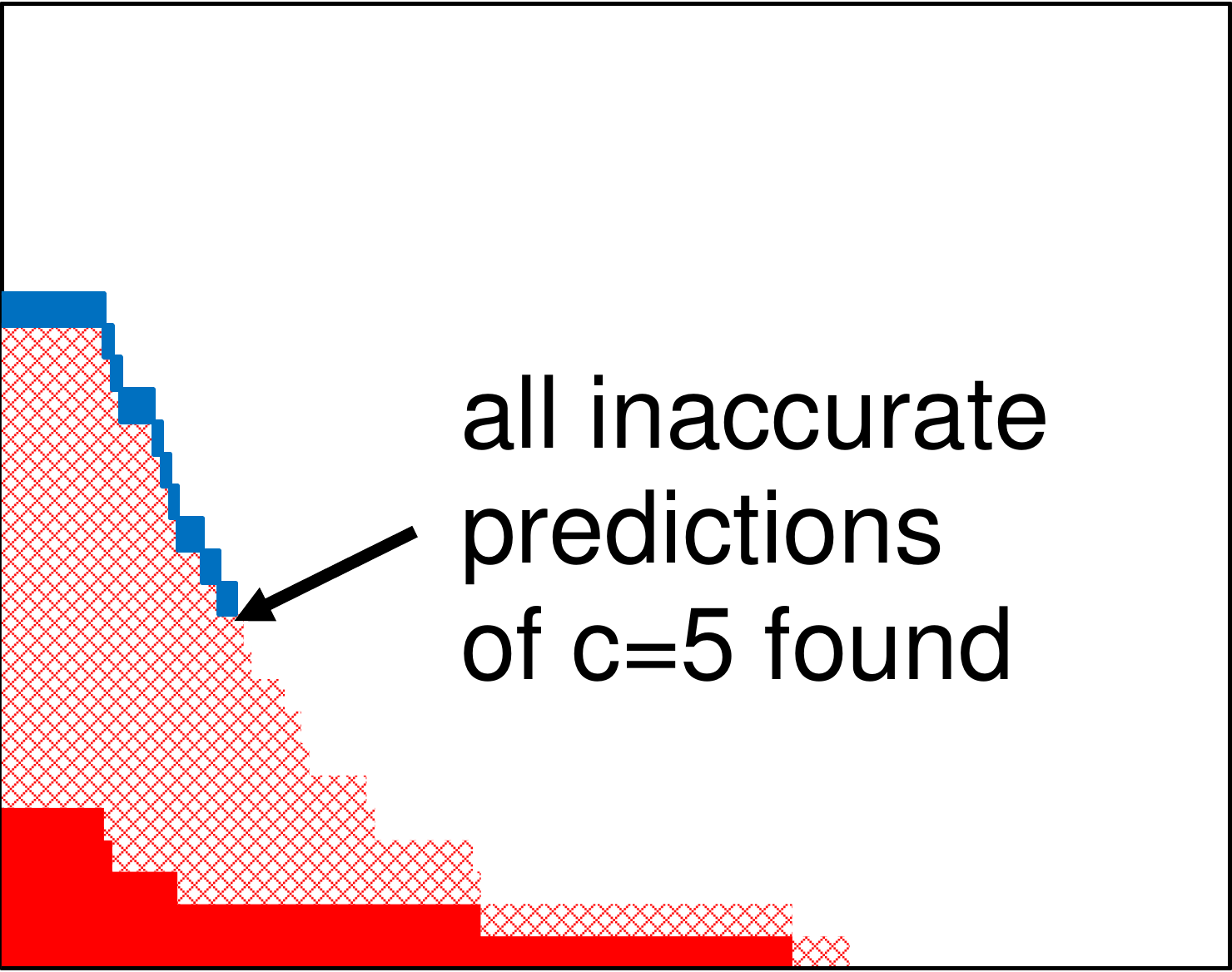};
  \end{axis}
\end{tikzpicture}
   \caption{Missed (\includegraphics[height=0.5em]{ma.png}, \includegraphics[height=0.5em]{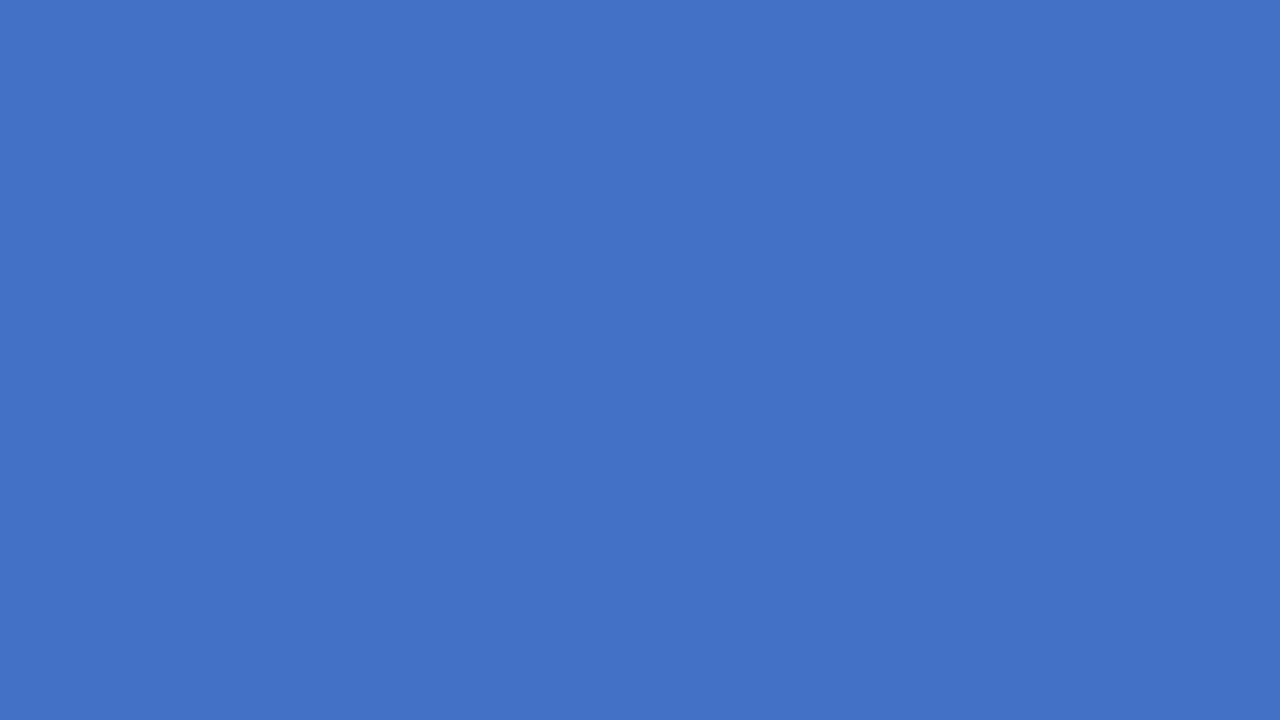}) and false alarms (\includegraphics[height=0.5em]{fa.png}, \includegraphics[height=0.5em]{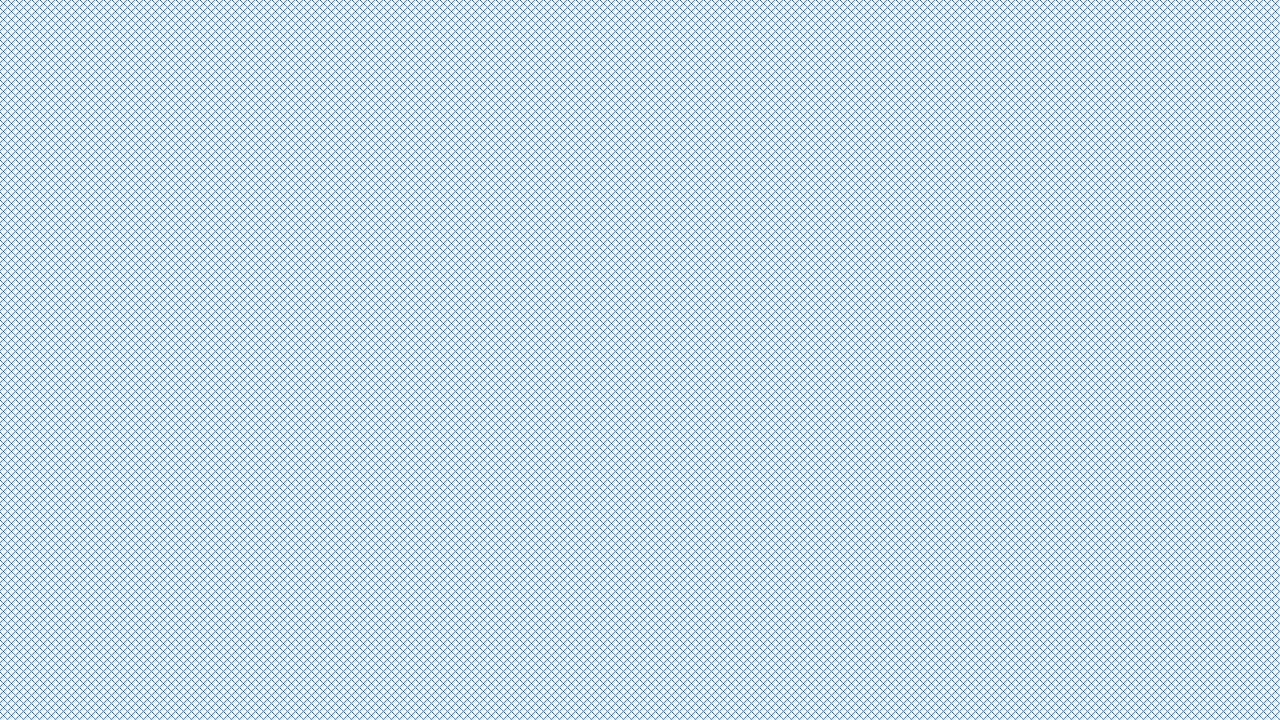}) of contingencies $c=3,5$.  
   }
   \label{fig:properbalance_alarms}
\end{subfigure}\hspace{1em}
\begin{subfigure}[b]{0.48\textwidth}
\begin{tikzpicture}
\small
\begin{axis}[
        axis on top,
        width=0.75\textwidth,
        scale only axis,
        enlargelimits=false, 
        ytick={0,0.000001,0.000002,0.000003,0.000004},     
        xtick={0,50,100,150},    
        ylabel={${Z}_3$ + ${Z}_5$},
        xlabel={S},
        xmin=0,
        xmax=150,
        ymin=0,
        ymax=0.000004,
        ]      
	\addplot graphics[xmin=0,ymin=0,xmax=150,ymax=0.000004] {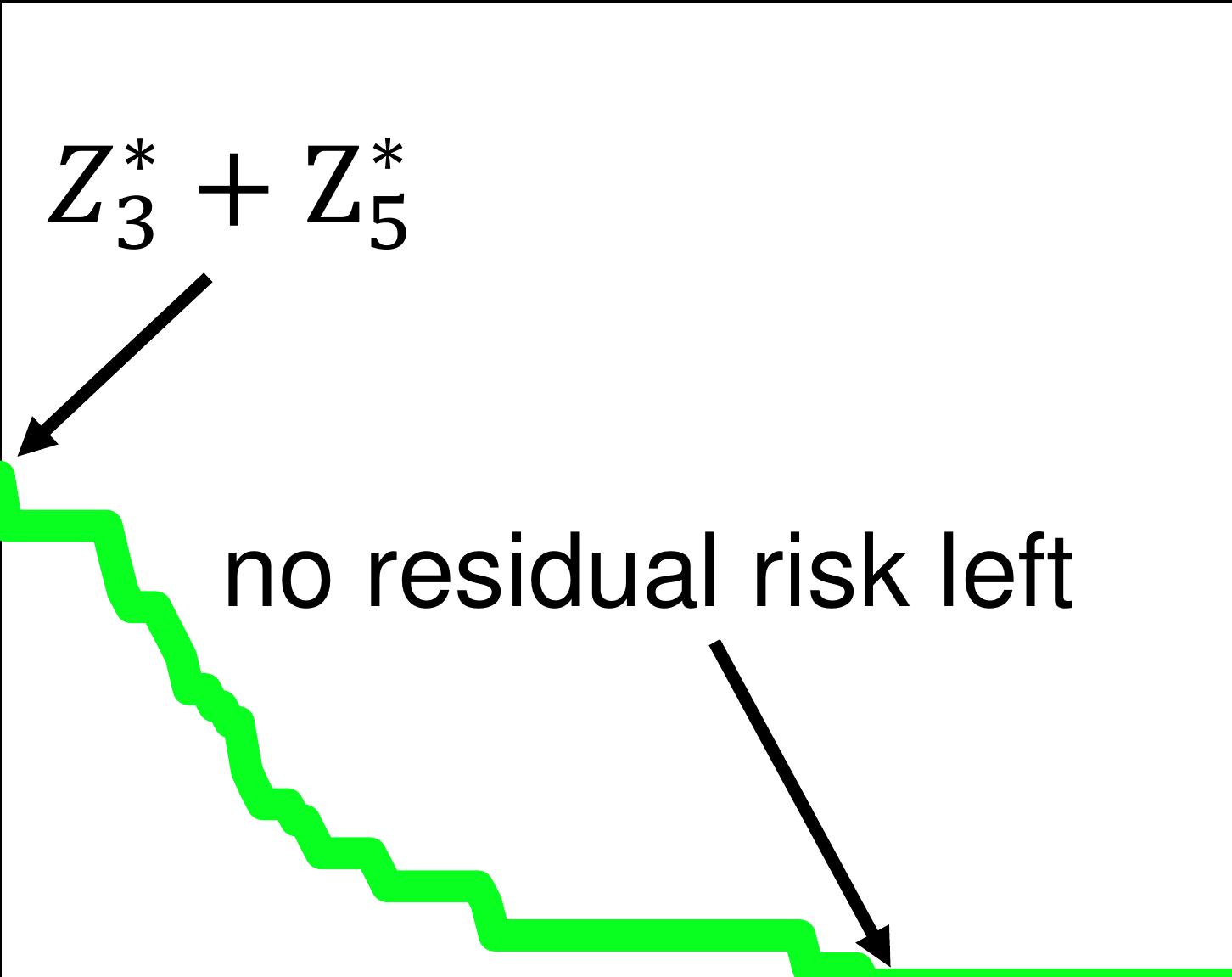};
  \end{axis}  
\end{tikzpicture}
   \caption{The residual risk (\includegraphics[height=0.5em]{grl.png}).\vspace{0.5em}}
   \label{fig:properbalance_risks}
\end{subfigure}
\caption{The proposed approach applied to many contingencies and operating conditions. Inaccurate predictions are in (a) and residual risk is in (b).
}\label{fig:plot_balance}
\end{figure}

In the IEEE test system, two contingencies $\Omega^C = \{3,5\}$ were used; parameters from $c=3$ as before and parameters for $c=5$ were $p_5^C = 0.0003$ and $\mathcal{C}_5=\frac{500}{501}$. One AdaBoost classifier was trained for each contingency and tested with $|\Omega^P|=1500$ operating conditions that resulted in $|\Omega^S|=3000$ testing scenarios. After only $104$ security assessments out of these $3000$, all inaccurate predictions were found and the risk drops to zero as presented in Fig. \ref{fig:plot_balance}. In the conventional approach, where machine-learning is not involved, all of these $3000$ scenarios needs assessments; hence the reduction in computations is $\SI{95}{\percent}$ that demonstrates the significance of this proposed approach.

\begin{figure}
\centering
\begin{subfigure}[b]{0.48\textwidth}
\begin{tikzpicture}
\small
\begin{axis}[
        axis on top,
        width=0.75\textwidth,
        scale only axis,
        enlargelimits=false, 
        ytick={0,400,800,1200},     
        xtick={0,6000,12000,18000},    
        ylabel={$\mathop{\sum} N^1_c + N^0_c$},
        xlabel={$S$},
        xmin=0,
        xmax=18000,
        ymin=0,
        ymax=1200,
        ]          
	\addplot graphics[xmin=0,ymin=0,xmax=18000,ymax=1200] {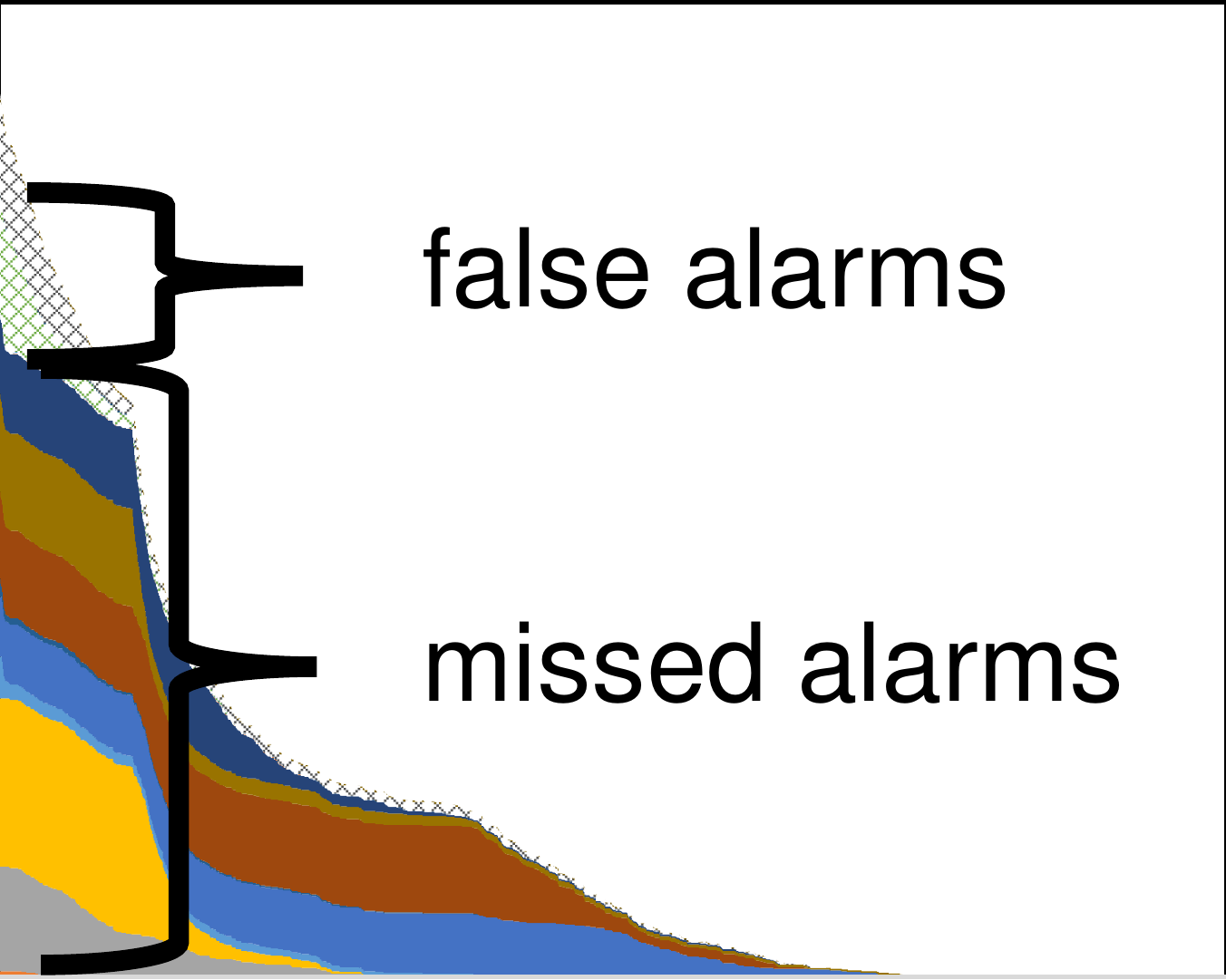};
  \end{axis}
\end{tikzpicture}
   \caption{Missed (dark) and false alarms (bright). One colour each contingency.}
   \label{fig:Frenchalarms}
\end{subfigure}\hspace{1em}
\begin{subfigure}[b]{0.48\textwidth}
\begin{tikzpicture}
\small
\begin{axis}[
        axis on top,
        width=0.75\textwidth,
        scale only axis,
        enlargelimits=false, 
        ytick={0,0.00005,0.0001,0.00015,0.0002},     
        xtick={0,6000,12000,18000},    
        ylabel={$\mathop{\sum} {Z_c}$},
        xlabel={$S$},
        xmin=0,
        xmax=18000,
        ymin=0,
        ymax=0.0002,
        ]      
	\addplot graphics[xmin=0,ymin=0,xmax=18000,ymax=0.0002] {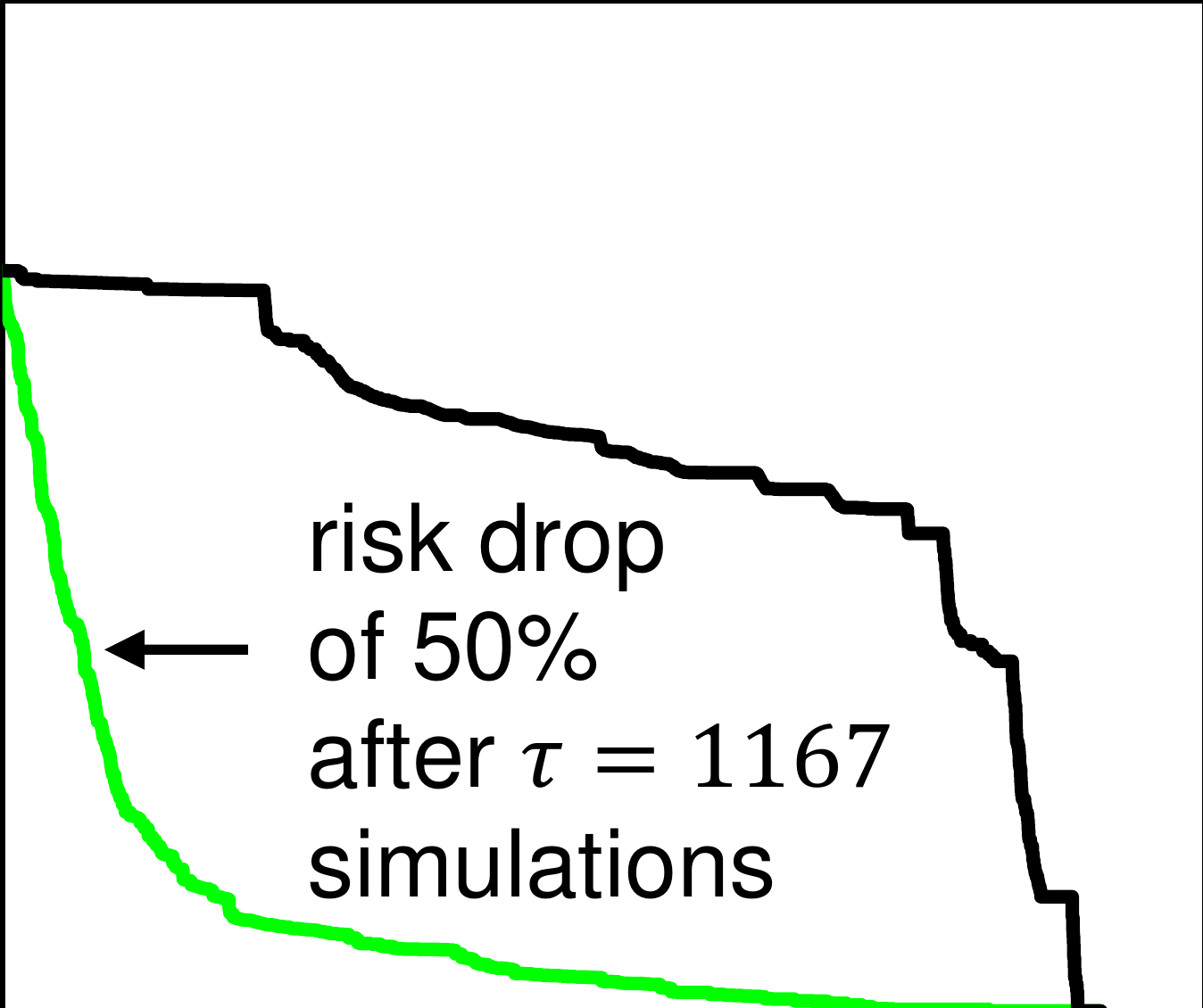};
  \end{axis}  
\end{tikzpicture}
   \caption{Residual risks: proposed approach (\includegraphics[height=0.5em]{grl.png}), standard classifier (\includegraphics[height=0.5em]{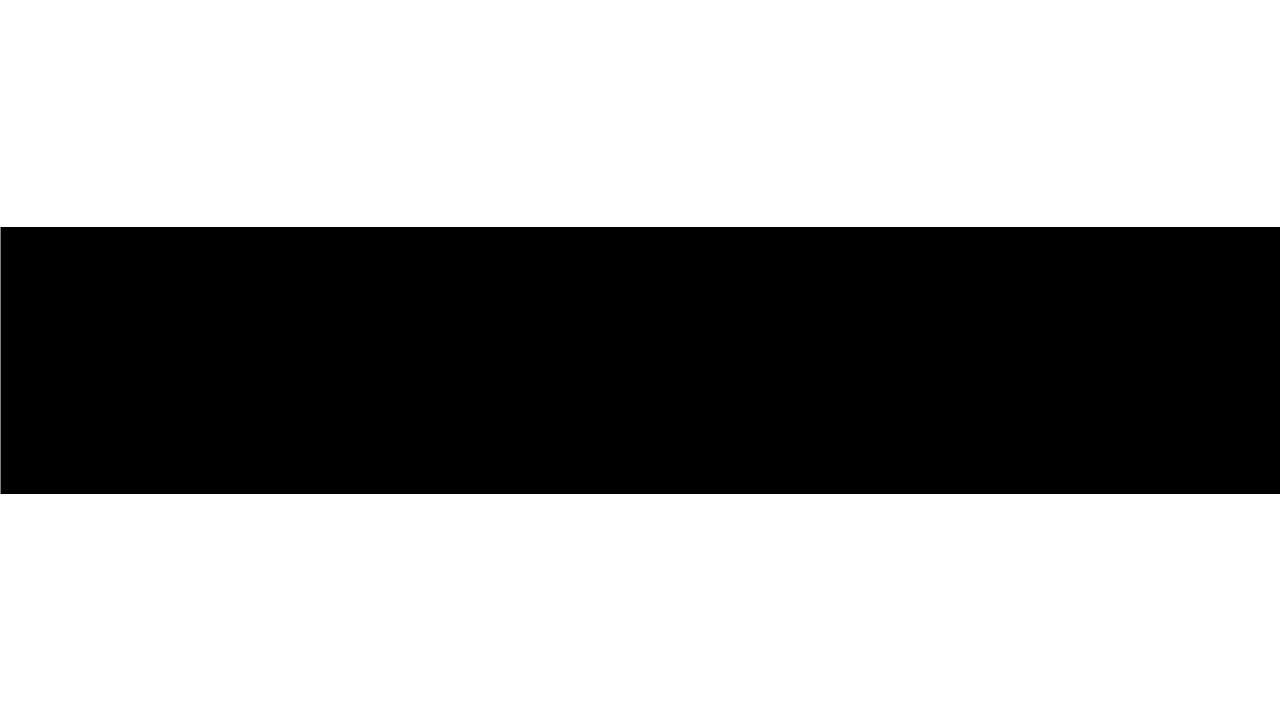}).}
   \label{fig:Frenchrisks}
\end{subfigure}
\caption{The proposed approach was applied to multiple contingencies in the French transmission grid. Inaccurate predictions are in (a) and the residual risk in (b). 
}\label{fig:French}
\end{figure}

In the French system, $11$ contingencies were considered resulting in $16500$ testing scenarios ($|\Omega^S|=1500 \times 11 = 16500$). $11$ AdaBoost classifiers were trained and used as described in {in Sec. \ref{sec:combML}.} In Fig. \ref{fig:Frenchalarms} the reductions of missed alarms of the proposed approach are presented. $1093$ of the $16500$ scenarios were inaccurately predicted when relying only on machine learning. Hence, the test error was $\SI{7}{\percent}$. the proposed approach reduced these inaccurate predictions by $\SI{50}{\percent}$ ($547$) within the first $2215$ security assessments. {In terms of risks and in comparison with a machine learning approach applying a standard classifier, the proposed approach reduced the risks much faster (Fig. \ref{fig:Frenchrisks}).} In the standard classifier, the risk decreased slowly and dropped sharply around $15000$ security assessments. This result implied that the risk of false alarms was higher than of missing alarms. The risk reduced quickly within the first security assessments in the proposed approach: e.g., the risk reduced by $\SI{50}{\percent}$ after $1167$ assessments ($\SI{7}{\percent}$ of the $16500$). This monotonic decrease in the overall residual risk $\mathop{\sum} {Z_c}$ showed that the proposed approach identified high-risk operating conditions in a real data-set and for a large grid.

\subsection{Sensitivity of parameter estimations}
The sensitivity to inaccurate parameter estimations of the severity/costs of contingencies $C^{F1}_c$ and the likelihood of contingencies $p_c$ was investigated in two studies on the French system. The first study focused on inaccurate estimations in a single contingency and the second on a systematic inaccuracy in all contingencies. 

\begin{figure}
\centering
\begin{subfigure}[b]{0.48\textwidth}
\begin{tikzpicture}
\small
\begin{axis}[
        axis on top,
        width=0.75\textwidth,
        scale only axis,
        enlargelimits=false, 
        ytick={0,0.00005,0.0001,0.00015,0.0002},     
        xtick={0,6000,12000,18000},    
        ylabel={$\mathop{\sum} {Z_c}$},
        xlabel={$S$},
        xmin=0,
        xmax=18000,
        ymin=0,
        ymax=0.0002,
        ]          
	\addplot graphics[xmin=0,ymin=0,xmax=18000,ymax=0.0002] {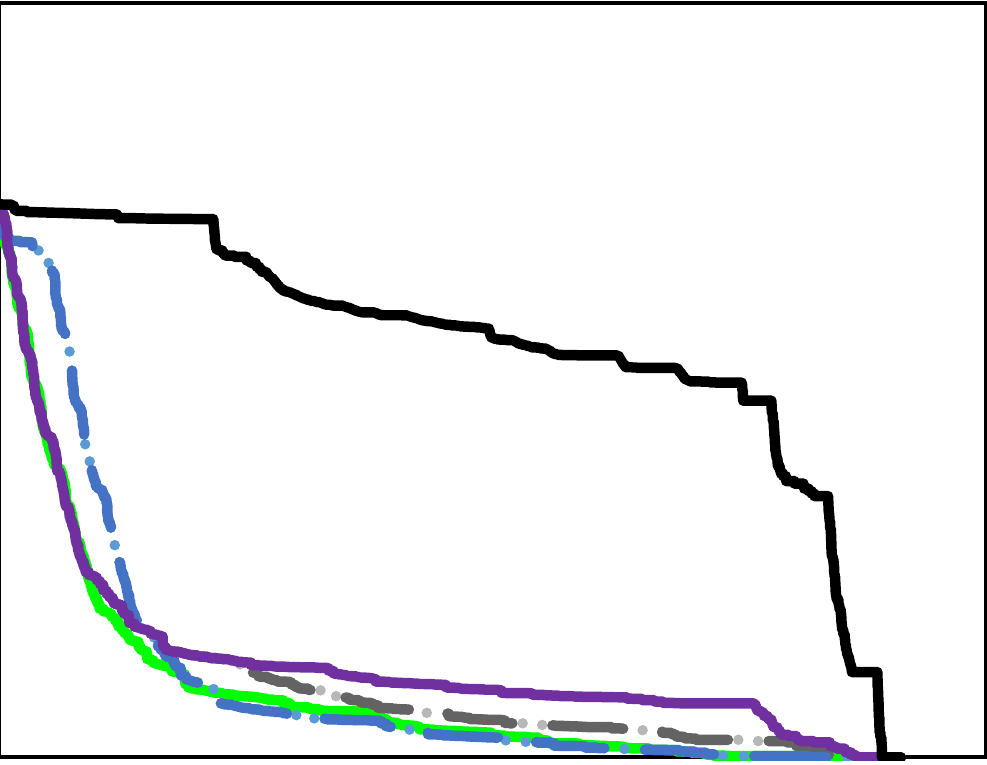};
  \end{axis}
\end{tikzpicture}
   \caption{$\alpha =100$ at a single contingency.}
   \label{fig:Frenchrisks6}
\end{subfigure}\hspace{1em}
\begin{subfigure}[b]{0.48\textwidth}
\begin{tikzpicture}
\small
\begin{axis}[
        axis on top,
        width=0.75\textwidth,
        scale only axis,
        enlargelimits=false, 
        ytick={0,0.00005,0.0001,0.00015,0.0002},     
        xtick={0,6000,12000,18000},    
        ylabel={$\mathop{\sum} {Z_c}$},
        xlabel={$S$},
        xmin=0,
        xmax=18000,
        ymin=0,
        ymax=0.0002,
        ]      
	\addplot graphics[xmin=0,ymin=0,xmax=18000,ymax=0.0002] {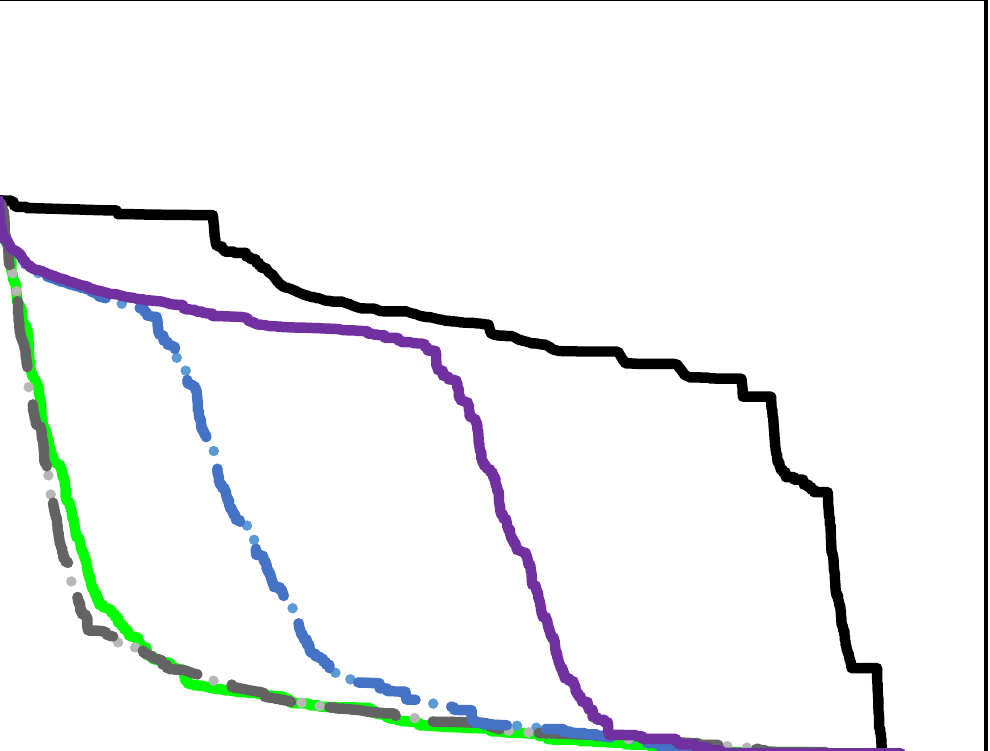};
  \end{axis}  
\end{tikzpicture}
   \caption{$\alpha =10$ at all contingencies.}
   \label{fig:Frenchrisks2}
\end{subfigure}
\caption{The risk-sensitivity on inaccurate parameter estimations for (a) a single contingency and at (b) all contingencies. The inaccuracies $\alpha$ and $\frac{1}{\alpha}$ are either in costs (\protect\includegraphics[height=0.5em]{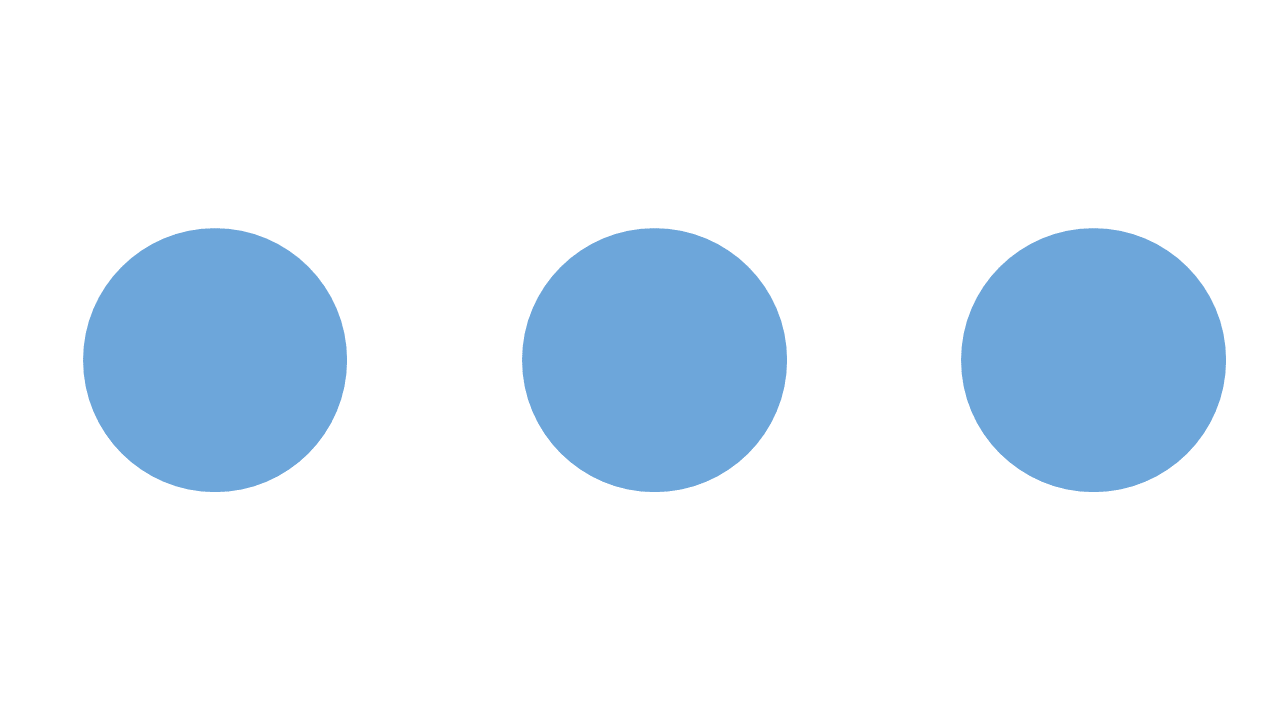}, \protect\includegraphics[height=0.5em]{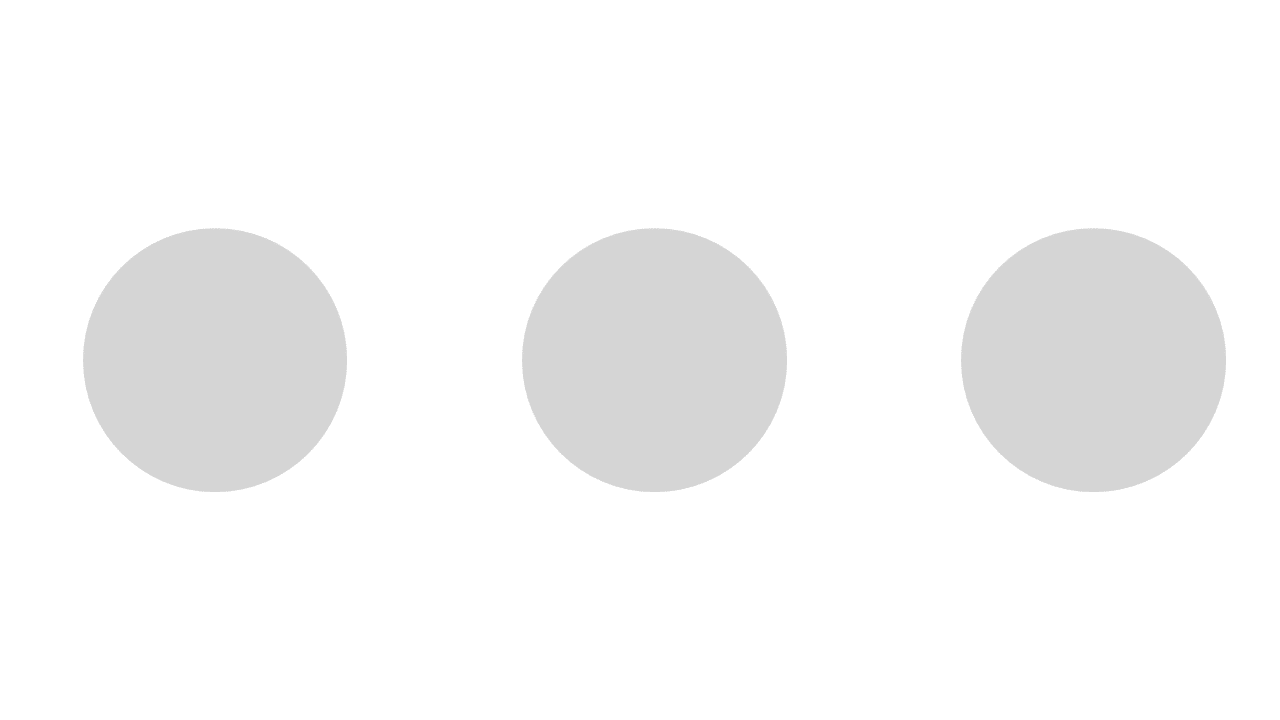}) or in probabilities (\protect\includegraphics[height=0.5em]{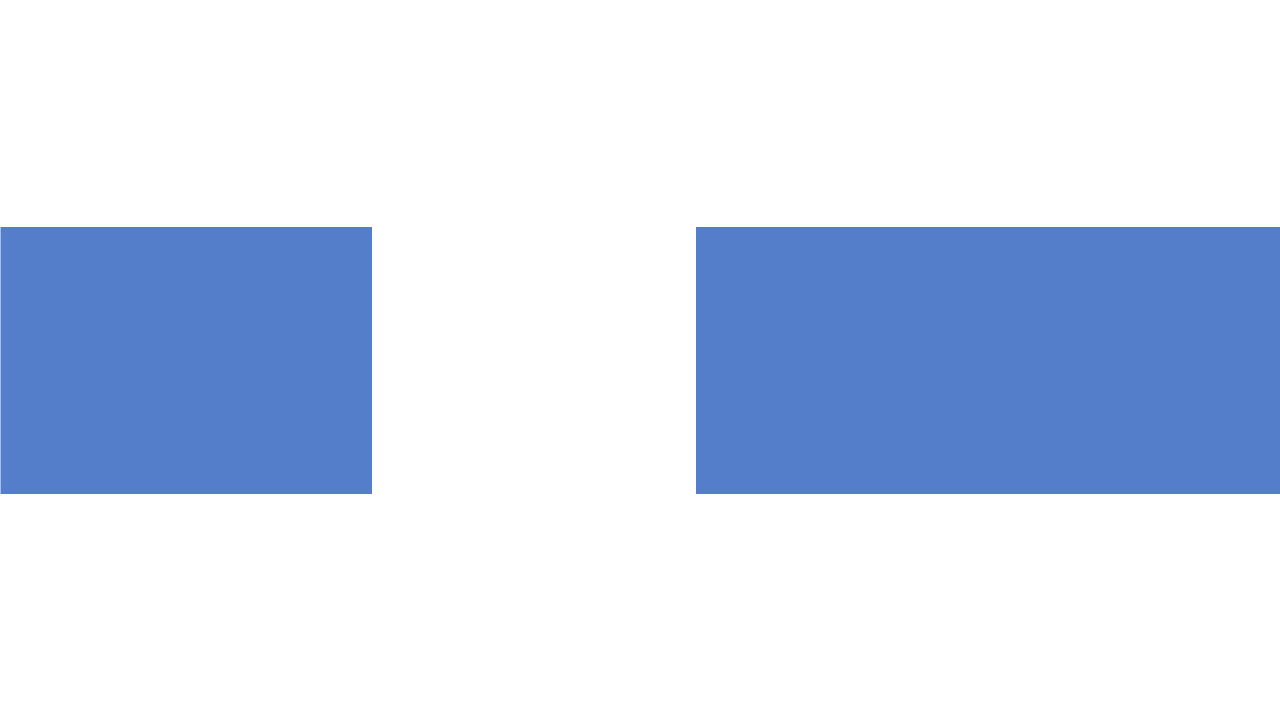}, \protect\includegraphics[height=0.5em]{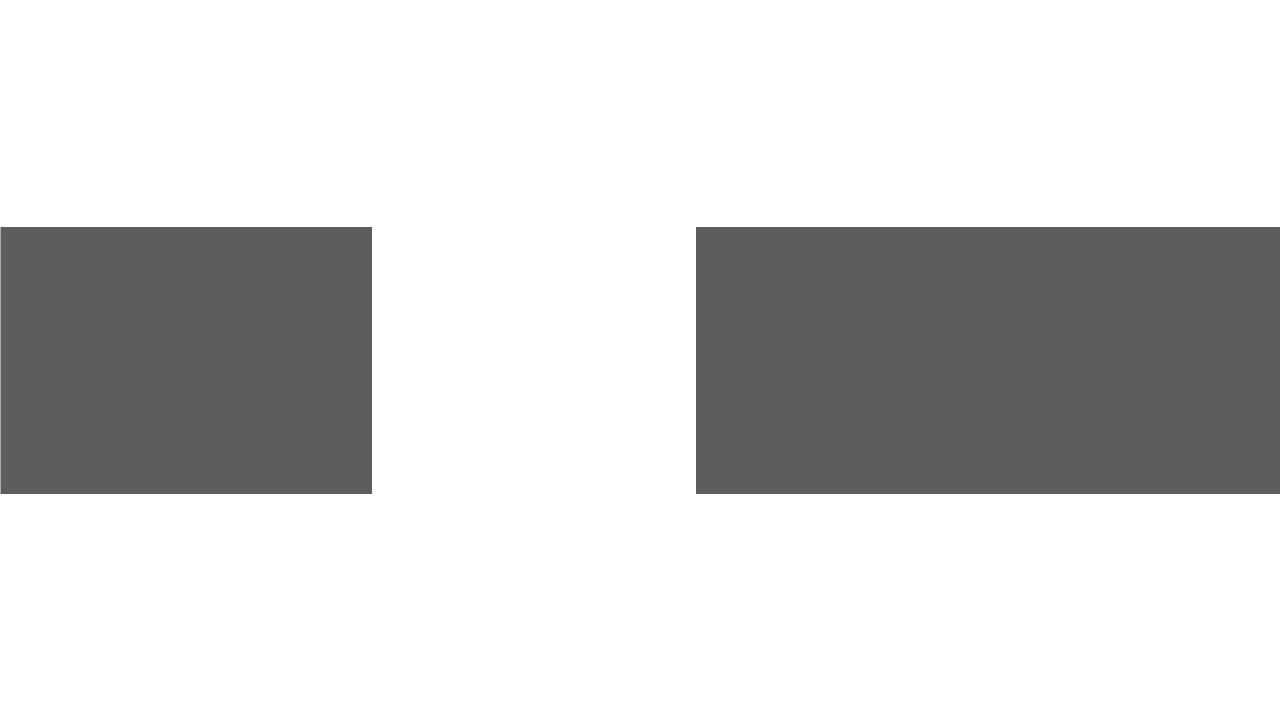}), or superposed (\protect\includegraphics[height=0.5em]{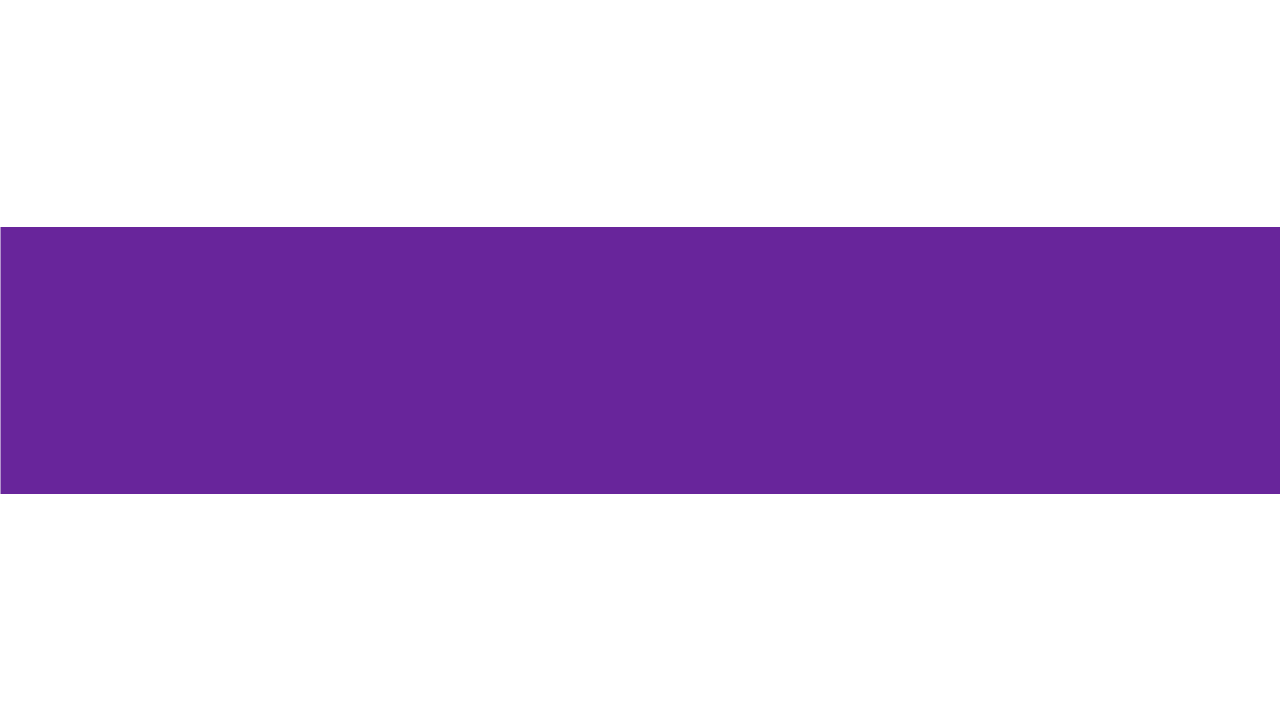}). Accurate estimation in the proposed approach (\protect\includegraphics[height=0.5em]{grl.png}) and standard classifier (\protect\includegraphics[height=0.5em]{sole.png}).
}\label{fig:French2}
\end{figure}

In the first study, the estimated cost or likelihood of a single and randomly selected contingency was $\alpha=100$ times higher/lower than the actual values $\mathcal{C}_c=\frac{1000}{1001}$ and $p_c=0.0005$. The estimates were used to guide security assessments and the actual parameters were used to compute the residual risk $\mathop{\sum} {Z_c}$. Then, the analysis of ${Z_c}$ allowed to judge the impact of inaccurate estimations. Slightly slower reductions in risks were observed in the results in Fig. \ref{fig:Frenchrisks6} as one can expect. However, the proposed approach that used these (inaccurate) parameters still significantly outperformed a standard classifier.

In the second study, the inaccuracy was systematic at $\alpha=10$ in all $11$ contingencies either in costs $C^{F1}_c$ or likelihoods $p_c$ or superposed. The results in Fig. \ref{fig:Frenchrisks2} showed risks increased significantly at $\alpha=10$ and reduced slightly at the inverse $\frac{1}{\alpha}=0.1$. When superposed where the inaccuracy of all probabilities and costs were at $\alpha=10$ the risks increased drastically. Also, here, although the proposed approach was strongly affected it still outperformed a standard classifier.

\subsection{Computations}

{The generation of the training database requires computational time for security assessments. In the case of dynamic security, this would require analysing multiple stability phenomena and some involve time-domain simulations. A time-domain simulation in the French system takes around $\SI{57}{\second}$ on a computing cluster \citep{Kon16}.} The training and calibration of classifiers took on average $\SI{445}{\second}$ on a standard notebook. However, this training is required only once. A significant advantage of the proposed approach was that re-training was not required when the input parameters change. {For instance, if the likelihoods of contingencies change with time, they can instantly be updated and considered in the decision threshold. This update requires no computational time. This is one of the key advantages of this approach. }

The prediction of an operating condition in the short-term operation took $\ll\SI{0.1}{\second}$. {In comparison, the assessment of dynamic security requires analysing the stability in the time-domain and this requires time-domain simulation of more $\SI{57}{\second}$ for a single operating condition in the French system. This difference illustrates the main benefit of using machine learning in the probabilistic security assessment close to real-time. The proposed approach combines this advantage with the advantage of conventional security assessment being always accurate, and hence can correct the machine learning models.}

\subsection{Discussion}
The probabilistic security assessment with using machine learning showed promising results on addressing cost skewness and class imbalances in the training database. Inaccurate predictions are effectively reduced by identifying operating conditions and contingencies entailing high risks. $\SI{25}{\percent}$ of security assessments identified inaccurate predictions from the machine learning model in the French study. The risk of relying on machine learning reduces, even when parameter estimations are inaccurate up to two orders of magnitudes. A key finding is that computations reduced by around $\SI{90}{\percent}$ in the proposed online workflow. This finding enables the opportunity to account for many more contingency scenarios. When moving toward a probabilistic security assessment, the high number of scenarios may become the bottleneck as many more (flexible) devices will be deployed in the future power system and all these devices can result in contingencies. Another key benefit is that frequent changes in the likelihood of contingencies can be effectively considered. In addition, the proposed approach can be used when a machine learning model is trained for multiple contingencies to reduce training times, and frequent changes can still be taken into account as these changes are considered as the last step of the proposed workflow.

The limits of the proposed probabilistic security assessment with machine learning is to not consider all risks. Additional risks of the proposed approach may relate to the limited number of operating conditions used, modelling errors, inaccuracies in the security assessments, inaccurate estimations of other input parameters, etc. Specifically, the estimations of costs/impact and the likelihood of contingencies are not straightforward, and work is for instance done in \citep{Xia06,Fan16,Jam20}. More general limitations exist for machine learning-based approaches, such as that the number of security assessments needed in the offline workflow may be large as well, as the training database and classifiers must be updated as the underlying probability distributions of the data changes and the network topology changes over time. The proposed workflow assumes a single machine learning model for each contingency and whether this workflow is preferable over a workflow considering one model for multiple contingencies depends on the specific application. The one-time training time may be lower when a single model is used, however, the model may be more complex, the accuracy may be lower, and more frequent updating of the model may be needed as the entire model requires updating when a system or data change occurs. In comparison to the proposed workflow with one model per contingency, only the affected model would require updating. Specifically, changes in the topology have a strong impact on the stability analysis and with that on the assessment of security and it was in the focus of several research efforts \citep{Duc20}. An additional limitation is the learning approach Adaboost may be sensitive to data that contains noise and outliers \citep{Lon10}. If the training data is noisy and contains outliers than other learning approaches based on randomisation (e.g., Random Forest or Extremely Randomised Trees) can be considered to replace Adaboost. All other proposed steps can further be used. A final limitation is that no theoretical guarantees can be provided for computing the risks, the proposed approach allows, however, for estimating the level of risk.

\section{Conclusion}\label{sec:conc}
This work focused on combining machine learning within a probabilistic security assessment with the vision to consider dynamic security besides static security assessments in real-time operations. Machine learning was trained to predict the output of a security assessment. Subsequently, the probabilistic output of the trained classifier was calibrated to be accurate. Then, risk-minimised predicting decisions were obtained from the trained classifiers. These predictions across many contingencies and operating conditions provide the operator with an initial risk assessment of many possible future scenarios. Subsequently, the operators assess the high-risk scenarios with conventional security assessments. Hence, the risk of using machine learning is significantly reduced as inaccurate predictions can be corrected. Case studies demonstrated that conventional security assessments can be reduced by up-to $ \SI{95}{\percent}$ but risk-tolerance level stays the same. The proposed workflow identified the most critical scenarios quickly and is robust against inaccurate predictions of the likelihood of contingencies and can account for changes in these within real-time. This work supports the transition from a deterministic security assessment from the past that resulted in highly conservative system operations that are cost-inefficient. This proposed workflow maximises the effectiveness of probabilistic security assessment by supporting to include dynamic security assessments. In the future, the proposed workflow will be tested considering dynamic metrics, multiple asset failures, the cost benefits when moving to this probabilistic operating scheme can be quantified, and training various machine learning models for different stability phenomena. 

\section{Acknowledgements}
This work was supported by a studentship funded by the Engineering and Physical Sciences Research Council. We are thankful to colleagues from Reseau de Transport d'Electricite who provided expertise that greatly assisted the research. {We also thank the reviewers for their insightful thoughts and discussion in the review process.}

\bibliographystyle{elsarticle-num}

\bibliography{MyLib}

\begin{thebibliography}{10}
\expandafter\ifx\csname url\endcsname\relax
  \def\url#1{\texttt{#1}}\fi
\expandafter\ifx\csname urlprefix\endcsname\relax\def\urlprefix{URL }\fi
\expandafter\ifx\csname href\endcsname\relax
  \def\href#1#2{#2} \def\path#1{#1}\fi

\bibitem{Pan12}
P.~Panciatici, G.~Bareux, L.~Wehenkel, Operating in the fog: Security
  management under uncertainty, IEEE Power and Energy Magazine 10~(5) (2012)
  40--49.

\bibitem{Fou88}
A.~A. {Fouad}, F.~{Aboytes}, V.~F. {Carvalho}, S.~L. {Corey}, K.~J. {Dhir},
  R.~{Vierra}, Dynamic security assessment practices in north america, IEEE
  Transactions on Power Systems 3~(3) (1988) 1310--1321.

\bibitem{Bil96}
R.~Billinton, R.~N. Allan, Reliability Evaluation of Power Systems, Springer
  Science \& Business Media, 1996.

\bibitem{Kun04}
P.~Kundur, J.~Paserba, V.~Ajjarapu, G.~Andersson, A.~Bose, C.~Canizares,
  N.~Hatziargyriou, D.~Hill, A.~Stankovic, C.~Taylor, et~al., Definition and
  classification of power system stability ieee/cigre joint task force on
  stability terms and definitions, IEEE Transactions on Power Systems 19~(3)
  (2004) 1387--1401.

\bibitem{Lia67}
T.~E.~D. Liacco, The adaptive reliability control system, IEEE Transactions on
  Power Apparatus and Systems PAS-86~(5) (1967) 517--531.

\bibitem{Cap11}
F.~Capitanescu, J.~M. Ramos, P.~Panciatici, D.~Kirschen, A.~M. Marcolini,
  L.~Platbrood, L.~Wehenkel, State-of-the-art, challenges, and future trends in
  security constrained optimal power flow, Electric Power Systems Research
  81~(8) (2011) 1731--1741.

\bibitem{Cap16}
F.~Capitanescu, Critical review of recent advances and further developments
  needed in ac optimal power flow, Electric Power Systems Research 136 (2016)
  57--68.

\bibitem{Kun94}
P.~Kundur, N.~J. Balu, M.~G. Lauby, Power system stability and control, Vol.~7,
  McGraw-hill New York, 1994.

\bibitem{Blo00}
V.~D. Blondel, J.~N. Tsitsiklis, A survey of computational complexity results
  in systems and control, Automatica 36~(9) (2000) 1249--1274.

\bibitem{Chi01}
H.-D. Chiang, Power System Stability, Wiley Encyclopedia of Electrical and
  Electronics Engineering, 1999.

\bibitem{McC99}
J.~D. McCalley, V.~Vittal, N.~Abi-Samra, An overview of risk based security
  assessment, in: IEEE Power Engineering Society Summer Meeting, Vol.~1, 1999,
  pp. 173--178.

\bibitem{McC04}
J.~McCalley, S.~Asgarpoor, L.~Bertling, R.~Billinion, H.~Chao, J.~Chen,
  J.~Endrenyi, R.~Fletcher, A.~Ford, C.~Grigg, Probabilistic security
  assessment for power system operations, in: IEEE Power Engineering Society
  General Meeting, 2004, pp. 212--220.

\bibitem{All00}
R.~Allan, R.~Billinton, Probabilistic assessment of power systems, Proceedings
  of the IEEE 88~(2) (2000) 140--162.

\bibitem{Hey19}
E.~Heylen, M.~Ovaere, S.~Proost, G.~Deconinck, D.~Van~Hertem, A
  multi-dimensional analysis of reliability criteria: From deterministic n−1
  to a probabilistic approach, Electric Power Systems Research 167 (2019)
  290--300.

\bibitem{Kir04}
D.~Kirschen, D.~Jayaweera, D.~Nedic, R.~Allan, A probabilistic indicator of
  system stress, IEEE Transactions on Power Systems 19~(3) (2004) 1650--1657.

\bibitem{Mar04}
N.~Maruejouls, V.~Sermanson, S.~Lee, P.~Zhang, A practical probabilistic
  reliability assessment using contingency simulation, IEEE PES Power Systems
  Conference and Exposition 3 (2004) 1312--1318.

\bibitem{Kir07}
D.~S. Kirschen, D.~Jayaweera, Comparison of risk-based and deterministic
  security assessments, IET Generation, Transmission \& Distribution 1~(4)
  (2007) 527--533.

\bibitem{Kir03}
D.~S. {Kirschen}, K.~R.~W. {Bell}, D.~P. {Nedic}, D.~{Jayaweera}, R.~N.
  {Allan}, Computing the value of security, IEEE Proceedings - Generation,
  Transmission and Distribution 150~(6) (2003) 673--678.

\bibitem{Ni03}
N.~Ming, J.~D. McCalley, V.~Vittal, T.~Tayyib, Online risk-based security
  assessment, IEEE Transactions on Power Systems 18~(1) (2003) 258--265.

\bibitem{Dis11}
A.~Dissanayaka, U.~D. Annakkage, B.~Jayasekara, B.~Bagen, Risk-based dynamic
  security assessment, IEEE Transactions on Power Systems 26~(3) (2011)
  1302--1308.

\bibitem{Pre15}
R.~Preece, J.~V. Milanovi{\'c}, Probabilistic risk assessment of rotor angle
  instability using fuzzy inference systems, IEEE Transactions on Power Systems
  30~(4) (2014) 1747--1757.

\bibitem{Cia17}
E.~Ciapessoni, D.~Cirio, S.~Massucco, A.~Morini, A.~Pitto, F.~Silvestro,
  Risk-based dynamic security assessment for power system operation and
  operational planning, Energies 10~(4) (2017) 475.

\bibitem{Str16}
G.~Strbac, D.~Kirschen, R.~Moreno, Reliability standards for the operation and
  planning of future electricity networks, Foundations and Trends in Electric
  Energy Systems 1~(3) (2016) 143--219.

\bibitem{Gar14}
{GARPUR Consortium and others}, Current practices, drivers and barriers for new
  reliability standards, 7th framework programme, EU Commission grant agreement
  608540 (2014).

\bibitem{Hey18}
E.~Heylen, G.~Deconinck, D.~{Van Hertem}, Review and classification of
  reliability indicators for power systems with a high share of renewable
  energy sources, Renewable and Sustainable Energy Reviews 97 (2018) 554 --
  568.

\bibitem{Xia06}
F.~Xiao, J.~D. McCalley, Y.~Ou, J.~Adams, S.~Myers, Contingency probability
  estimation using weather and geographical data for on-line security
  assessment, in: IEEE International Conference on Probabilistic Methods
  Applied to Power Systems, 2006, pp. 1--7.

\bibitem{Fan16}
R.~Z. Fanucchi, M.~Bessani, M.~H.~M. Camillo, J.~B.~A. London, C.~D. Maciel,
  Failure rate prediction under adverse weather conditions in an electric
  distribution system using negative binomial regression, in: IEEE
  International Conference on Harmonics and Quality of Power, 2016, pp.
  478--483.

\bibitem{Jam20}
M.~R. {Jamieson}, G.~{Strbac}, K.~R.~W. {Bell}, Quantification and
  visualisation of extreme wind effects on transmission network outage
  probability and wind generation output, IET Smart Grid 3~(2) (2020) 112--122.

\bibitem{Don18}
B.~{Donnot}, I.~{Guyon}, A.~{Marot}, M.~{Schoenauer}, P.~{Panciatici},
  Optimization of computational budget for power system risk assessment, in:
  IEEE PES Innovative Smart Grid Technologies Conference Europe, 2018, pp.
  1--6.

\bibitem{Duc20}
L.~Duchesne, E.~Karangelos, L.~Wehenkel, Recent developments in machine
  learning for energy systems reliability management, Proceedings of the IEEE
  (2020).

\bibitem{Weh98}
L.~A. Wehenkel, Automatic Learning Techniques in Power Systems, Kluwer Academic
  Publishers, 1998.

\bibitem{Lou10}
L.~Loud, S.~Guillon, G.~Vanier, J.~Huang, L.~Riverin, D.~Lefebvre, J.~Rizzi,
  Hydro-qu{\'e}bec's challenges and experiences in on-line dsa applications,
  in: IEEE PES General Meeting, 2010, pp. 1--8.

\bibitem{Sam10}
S.~R. Samantara, I.~Kamwa, G.~Joos, Ensemble decision trees for phasor
  measurement unit-based wide-area security assessment in the operations time
  frame, IET Generation, Transmission \& Distribution 4~(12) (2010) 1334--1348.

\bibitem{Vas16}
M.~H. Vasconcelos, L.~M. Carvalho, J.~Meirinhos, N.~Omont, P.~Gambier-Morel,
  G.~Jamgotchian, D.~Cirio, E.~Ciapessoni, A.~Pitto, I.~Konstantelos,
  G.~Strbac, M.~Ferraro, C.~Biasuzzi, Online security assessment with load and
  renewable generation uncertainty: The itesla project approach, in:
  IEEEational Conference on Probabilistic Methods Applied to Power Systems,
  2016, pp. 1--8.

\bibitem{Kon16}
I.~{Konstantelos}, G.~{Jamgotchian}, S.~H. {Tindemans}, P.~{Duchesne},
  S.~{Cole}, C.~{Merckx}, G.~{Strbac}, P.~{Panciatici}, Implementation of a
  massively parallel dynamic security assessment platform for large-scale
  grids, IEEE Transactions on Smart Grid 8~(3) (2017) 1417--1426.

\bibitem{Kon18}
I.~Konstantelos, M.~Sun, S.~H. Tindemans, S.~Issad, P.~Panciatici, G.~Strbac,
  Using vine copulas to generate representative system states for machine
  learning, IEEE Transactions on Power Systems 34~(1) (2018) 225--235.

\bibitem{Van14}
L.~Vanfretti, F.~R.~S. Sevilla, A three-layer severity index for power system
  voltage stability assessment using time-series from dynamic simulations, in:
  IEEE PES Innovative Smart Grid Technologies, Europe, 2014, pp. 1--5.

\bibitem{LiuA14}
C.~Liu, K.~Sun, Z.~H. Rather, Z.~Chen, C.~L. Bak, P.~Thogersen, P.~Lund, A
  systematic approach for dynamic security assessment and the corresponding
  preventive control scheme based on decision trees, in: IEEE PES General
  Meeting, 2014, pp. 1--1.

\bibitem{Sev15}
F.~R.~S. Sevilla, L.~Vanfretti, Static stability indexes for classification of
  power system time-domain simulations, in: IEEE PES Innovative Smart Grid
  Technologies Conference, 2015, pp. 1--5.

\bibitem{Gen10}
I.~Genc, R.~Diao, V.~Vittal, S.~Kolluri, S.~Mandal, Decision tree-based
  preventive and corrective control applications for dynamic security
  enhancement in power systems, IEEE Transactions on Power Systems 25~(3)
  (2010) 1611--1619.

\bibitem{Jaf18}
S.~Jafarzadeh, V.~M.~I. Genc, Probabilistic dynamic security assessment of
  large power systems using machine learning algorithms, Turkish Journal of
  Electrical Engineering \& Computer Sciences 26~(3) (2018) 1479--1490.

\bibitem{Kri11}
V.~Krishnan, J.~D. McCalley, S.~Henry, S.~Issad, Efficient database generation
  for decision tree based power system security assessment, IEEE Transactions
  on Power Systems 26~(4) (2011) 2319--2327.

\bibitem{Tha18}
F.~{Thams}, A.~{Venzke}, R.~{Eriksson}, S.~{Chatzivasileiadis}, Efficient
  database generation for data-driven security assessment of power systems,
  IEEE Transactions on Power Systems 35~(1) (2020) 30--41.

\bibitem{He12}
M.~He, J.~Zhang, V.~Vittal, {A Data Mining Framework for Online Dynamic
  Security Assessment: Decision Trees, Boosting, and Complexity Analysis}, in:
  IEEE PES Innovative Smart Grid Technologies, 2012, pp. 1--8.

\bibitem{He13}
M.~He, J.~Zhang, V.~Vittal, {Robust Online Dynamic Security Assessment Using
  Adaptive Ensemble Decision-Tree Learning}, IEEE Transactions on Power Systems
  28~(4) (2013) 4089--4098.

\bibitem{Guy03}
I.~Guyon, A.~Elisseeff, {An Introduction to Variable and Feature Selection},
  Journal of Machine Learning Research 3 (2003) 1157--1182.

\bibitem{Li16}
J.~Li, K.~Cheng, S.~Wang, F.~Morstatter, R.~P.~Trevino, J.~Tang, H.~Liu,
  {Feature Selection: A Data Perspective}, ACM Computing Surveys 50 (2016).

\bibitem{Moh15}
H.~Mohammadi, M.~Dehghani, {PMU} based voltage security assessment of power
  systems exploiting principal component analysis and decision trees,
  International Journal of Electrical Power \& Energy Systems 64 (2015)
  655--663.

\bibitem{Sun18}
M.~Sun, I.~Konstantelos, G.~Strbac, A deep learning-based feature extraction
  framework for system security assessment, IEEE Transactions on Smart Grid
  10~(5) (2018) 5007--5020.

\bibitem{Cha02}
N.~V. Chawla, K.~W. Bowyer, L.~O. Hall, W.~P. Kegelmeyer, Smote: synthetic
  minority over-sampling technique, Journal of Artificial Intelligence Research
  16 (2002) 321--357.

\bibitem{Dru03}
C.~Drummond, R.~C. Holte, et~al., C4. 5, class imbalance, and cost sensitivity:
  why under-sampling beats over-sampling, in: Workshop on learning from
  imbalanced datasets II, Vol.~11, Citeseer, 2003, pp. 1--8.

\bibitem{Zhu17}
L.~Zhu, C.~Lu, Z.~Y. Dong, C.~Hong, Imbalance learning machine-based power
  system short-term voltage stability assessment, IEEE Transactions on
  Industrial Informatics 13~(5) (2017) 2533--2543.

\bibitem{Zho19}
Y.~Zhou, Q.~Guo, H.~Sun, Z.~Yu, J.~Wu, L.~Hao, A novel data-driven approach for
  transient stability prediction of power systems considering the operational
  variability, International Journal of Electrical Power \& Energy Systems 107
  (2019) 379--394.

\bibitem{Zha18}
Y.~Zhang, Y.~Xu, Z.~Y. Dong, R.~Zhang, {A Hierarchical Self-Adaptive
  Data-Analytics Method for Real-Time Power System Short-Term Voltage Stability
  Assessment}, IEEE Transactions on Industrial Informatics 15~(1) (2018)
  74--84.

\bibitem{Bal18}
N.~G. Baltas, P.~Mazidi, J.~Ma, F.~de~Asis~Fernandez, P.~Rodriguez, A
  comparative analysis of decision trees, support vector machines and
  artificial neural networks for on-line transient stability assessment, in:
  IEEE International Conference on Smart Energy Systems and Technologies, 2018,
  pp. 1--6.

\bibitem{Tan17}
B.~{Tan}, J.~{Yang}, X.~{Pan}, J.~{Li}, P.~{Xie}, C.~{Zeng}, Representational
  learning approach for power system transient stability assessment based on
  convolutional neural network, The Journal of Engineering 2017~(13) (2017)
  1847--1850.

\bibitem{Wan16}
B.~Wang, B.~Fang, Y.~Wang, H.~Liu, Y.~Liu, Power system transient stability
  assessment based on big data and the core vector machine, IEEE Transactions
  on Smart Grid 7~(5) (2016) 2561--2570.

\bibitem{Roz17}
R.~Eskandarpour, A.~Khodaei, Leveraging accuracy-uncertainty tradeoff in svm to
  achieve highly accurate outage predictions, IEEE Transactions on Power
  Systems 33~(1) (2017) 1139--1141.

\bibitem{Cre19}
J.~L. Cremer, I.~Konstantelos, G.~Strbac, From optimization-based machine
  learning to interpretable security rules for operation, IEEE Transactions on
  Power Systems 34~(5) (2019) 3826--3836.

\bibitem{Cre18b}
J.~L. Cremer, I.~Konstantelos, S.~H. Tindemans, G.~Strbac, Data-driven power
  system operation: Exploring the balance between cost and risk, IEEE
  Transactions on Power Systems 34~(1) (2018) 791--801.

\bibitem{Liu18}
C.~Liu, F.~Tang, C.~Leth~Bak, An accurate online dynamic security assessment
  scheme based on random forest, Energies 11~(7) (2018) 1914.

\bibitem{Maj18}
S.~S. {Maaji}, G.~{Cosma}, A.~{Taherkhani}, A.~A. {Alani}, T.~M. {McGinnity},
  On-line voltage stability monitoring using an ensemble adaboost classifier,
  in: International Conference on Information Management, 2018, pp. 253--259.

\bibitem{Hou95}
I.~Houben, L.~Wehenkel, M.~Pavella, Coupling of k-nn with decision trees for
  power system transient stability assessment, in: IEEE Conference on Control
  Applications, 1995, pp. 825--832.

\bibitem{Fan18}
H.~Fan, Y.~Chen, S.~Huang, X.~Zhang, H.~Guan, D.~Shi, Post-fault transient
  stability assessment based on k-nearest neighbor algorithm with mahalanobis
  distance, in: IEEE International Conference on Power System Technology, 2018,
  pp. 4417--4423.

\bibitem{Sun16}
H.~Sun, F.~Zhao, H.~Wang, K.~Wang, W.~Jiang, Q.~Guo, B.~Zhang, L.~Wehenkel,
  Automatic learning of fine operating rules for online power system security
  control, IEEE Transactions on Neural Networks and Learning Systems 27~(8)
  (2016) 1708--1719.

\bibitem{Cos16}
D.~C.~L. Costa, M.~V.~A. Nunes, J.~P.~A. Vieira, U.~H. Bezerra, Decision
  tree-based security dispatch application in integrated electric power and
  natural-gas networks, Electric Power Systems Research 141 (2016) 442--449.

\bibitem{Cre18a}
J.~L. {Cremer}, I.~{Konstantelos}, G.~{Strbac}, S.~H. {Tindemans},
  Sample-derived disjunctive rules for secure power system operation, in: IEEE
  International Conference on Probabilistic Methods Applied to Power Systems,
  2018, pp. 1--6.

\bibitem{Pow11}
D.~Powers, Evaluation: From precision, recall and f-factor to roc,
  informedness, markedness \& correlation, Journal of Machine Learning
  Technologies 2 (2011) 37--63.

\bibitem{Elk01}
C.~Elkan, The foundations of cost-sensitive learning, in: International Joint
  Conference on Artificial Intelligence, Vol.~17, Lawrence Erlbaum Associates
  Ltd, 2001, pp. 973--978.

\bibitem{Han17}
F.~Hang, S.~Huang, Y.~Chen, S.~Mei, Power system transient stability assessment
  based on dimension reduction and cost-sensitive ensemble learning, in: IEEE
  Conference on Energy Internet and Energy System Integration), 2017, pp. 1--6.

\bibitem{Zho18}
Y.~Zhou, W.~Zhao, Q.~Guo, H.~Sun, L.~Hao, Transient stability assessment of
  power systems using cost-sensitive deep learning approach, in: IEEE
  Conference on Energy Internet and Energy System Integration, 2018, pp. 1--6.

\bibitem{Fre97}
Y.~Freund, R.~E. Schapire, A decision-theoretic generalization of on-line
  learning and an application to boosting, Journal of Computer and System
  Sciences 55~(1) (1997) 119--139.

\bibitem{Has09}
T.~Hastie, S.~Rosset, J.~Zhu, H.~Zou, Multi-class adaboost, Statistics and its
  Interface 2~(3) (2009) 349--360.

\bibitem{Die00}
T.~G. Dietterich, Ensemble methods in machine learning, in: International
  workshop on multiple classifier systems, Springer, 2000, pp. 1--15.

\bibitem{He09}
H.~{He}, E.~A. {Garcia}, Learning from imbalanced data, IEEE Transactions on
  Knowledge and Data Engineering 21~(9) (2009) 1263--1284.

\bibitem{Car05}
K.~McCarthy, B.~Zabar, G.~Weiss, Does cost-sensitive learning beat sampling for
  classifying rare classes?, in: Proceedings of the 1st International Workshop
  on Utility-Based Data Mining, Association for Computing Machinery, New York,
  NY, USA, 2005, p. 69–77.

\bibitem{Pla99}
J.~Platt, Probabilistic outputs for support vector machines and comparisons to
  regularized likelihood methods, Advances in large margin classifiers 10~(3)
  (1999) 61--74.

\bibitem{Nik16}
N.~Nikolaou, N.~Edakunni, M.~Kull, P.~Flach, G.~Brown, Cost-sensitive boosting
  algorithms: Do we really need them?, Machine Learning 104~(2-3) (2016)
  359--384.

\bibitem{Fla12}
P.~Flach, Machine Learning: The Art and Science of Algorithms that Make Sense
  of Data, Cambridge University Press, 2012.

\bibitem{Nic052}
A.~Niculescu-Mizil, R.~Caruana, Obtaining calibrated probabilities from
  boosting, in: Conference on Uncertainty in Artificial Intelligence, 2005, p.
  413.

\bibitem{Wod84}
A.~J. Wood, B.~Wollenberg, Power generation, operation and control, new york:
  John willey \& sons (1984).

\bibitem{Bre84}
L.~Breiman, J.~H. Friedman, R.~A. Olshen, C.~J. Stone, Classification and
  regression trees, Wadsworth \& Brooks Monterey, CA (1984).

\bibitem{Ped11}
F.~Pedregosa, G.~Varoquaux, A.~Gramfort, V.~Michel, B.~Thirion, O.~Grisel,
  M.~Blondel, P.~Prettenhofer, R.~Weiss, V.~Dubourg, Scikit-learn: Machine
  learning in python, Journal of Machine Learning Research 12~(Oct) (2011)
  2825--2830.

\bibitem{Lon10}
P.~M. Long, R.~A. Servedio, Random classification noise defeats all convex
  potential boosters, Machine learning 78~(3) (2010) 287--304.

\end{thebibliography}


\end{document}